\documentclass[5p,times,preprint]{elsarticle}

\usepackage{amsmath}
\usepackage{amssymb}
\usepackage{soul}

\usepackage[]{lineno}

\usepackage{enumerate}
\usepackage{graphicx}
\usepackage{dcolumn}
\usepackage{bm}
\usepackage{color}

\usepackage{epstopdf}
\usepackage{subcaption}
\usepackage{tcolorbox}
\usepackage{url}
\usepackage{setspace}
\journal{Nuclear Instruments and Methods in Physics Research A}
\begin{document}
\begin{frontmatter}
\title{Characterization of Multianode Photomultiplier Tubes for use in the CLAS12 RICH Detector}

\author[A]{P.~Degtiarenko }
\author[B]{A. Kim \corref{cor1}} 
\ead{kenjo@jlab.org}
\cortext[cor1]{ Corresponding author Tel: +1 757 269 6356}
\author[A]{V. Kubarovsky }
\author[A]{B. Raydo}
\author[C]{A. Smith}
\author[D]{F. Benmokhtar}

\address[A]{Thomas Jefferson National Accelerator Facility, Newport News, VA 23606, USA}
\address[B]{University of Connecticut, Storrs, CT 06269, USA}
\address[C]{Duke University, Durham, NC 27705, USA}
\address[D]{Duquesne University, Pittsburgh, PA, 15282, USA}

\begin{abstract}
We present results of the detailed study of several hundred Hamamatsu H12700 Multianode Photomultiplier Tubes (MaPMTs), characterizing their response to the Cherenkov light photons in the second Ring Imaging Cherenkov detector, a part of the CLAS12 upgrade at Jefferson Lab.
The total number of pixels studied was 25536.
The single photoelectron spectra were measured for each pixel at different high voltages and  light intensities of the laser test setup. Using the same dedicated front-end electronics as in the first RICH detector, the setup allowed us to characterize each pixel's properties such as gain, quantum efficiency, signal crosstalk between neighboring pixels,
and determine the signal threshold values to optimize their efficiency to detect Cherenkov photons.
A recently published state-of-the-art mathematical model, describing photon detector response functions measured in low light conditions, was extended to include the description of the crosstalk contributions to the spectra.
The database of extracted parameters will be used for the final selection of the MaPMTs, their arrangement in the new RICH detector, and the optimization of the operational settings of the front-end electronics.
The results show that the characteristics of the H12700 MaPMTs satisfy our requirements for the position-sensitive single photoelectron detectors.
\end{abstract}

\begin{keyword}
Ring Imaging Cherenkov detector \sep
Hamamatsu Multianode Photomultiplier tubes H8500 and H12700 \sep
Photon detector \sep Photomultiplier \sep
Photoelectron \sep  Signal amplitude spectra \sep Signal crosstalk \sep
Photon detection efficiency
\end{keyword}

\end{frontmatter}


\section{Introduction}
As part of the ongoing study of the structure of nucleons \cite{Avakian:2010ae}  in Hall B at the Thomas Jefferson National Accelerator Facility (JLab), the CEBAF Large Acceptance Spectrometer (CLAS12) \cite{Burkert:2020akg} is being used to accurately identify the secondary particles of high energy reactions, to assist in probing the strangeness frontier, and to aid in characterizing the transverse momentum distribution (TMD) and generalized parton distribution (GPD) functions of the nucleon. Indispensable to this task is the ability to identify kaons, pions, and protons.  With the CLAS12 spectrometer providing accurate momentum measurements, the Ring Imaging Cherenkov detector (RICH) \cite{Contalbrigo:2020,Contalbrigo:2020snw,Mirazita:2017vav,Contalbrigo:2014rqa} provides tandem Cherenkov light-cone radius measurements that yield the velocities of near light-speed particles, thus facilitating mass-dependent particle identification.

\begin{figure}[h!bt]
	\centering
	\includegraphics[width=0.95\linewidth]{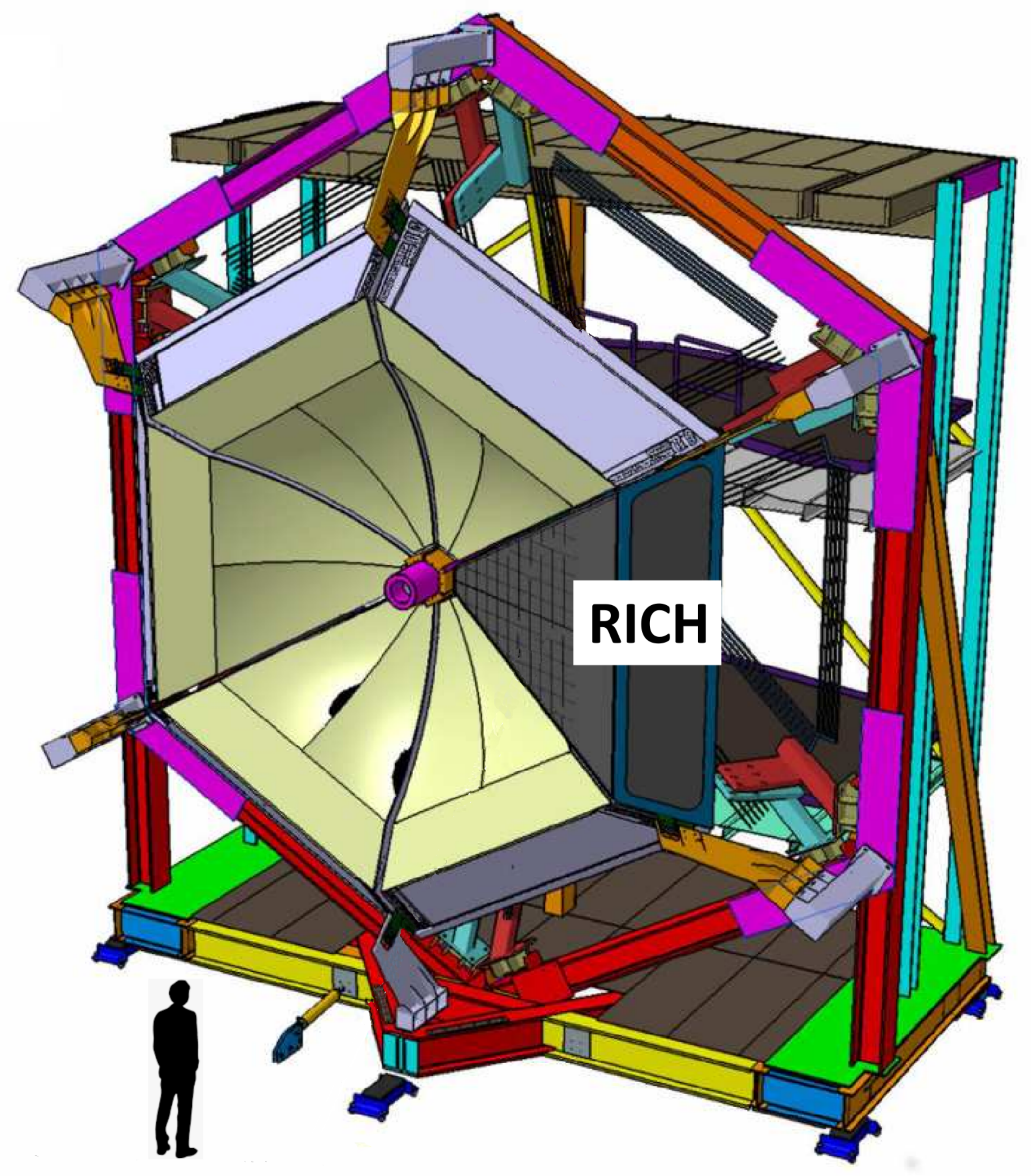}
	\includegraphics[width=0.95\linewidth]{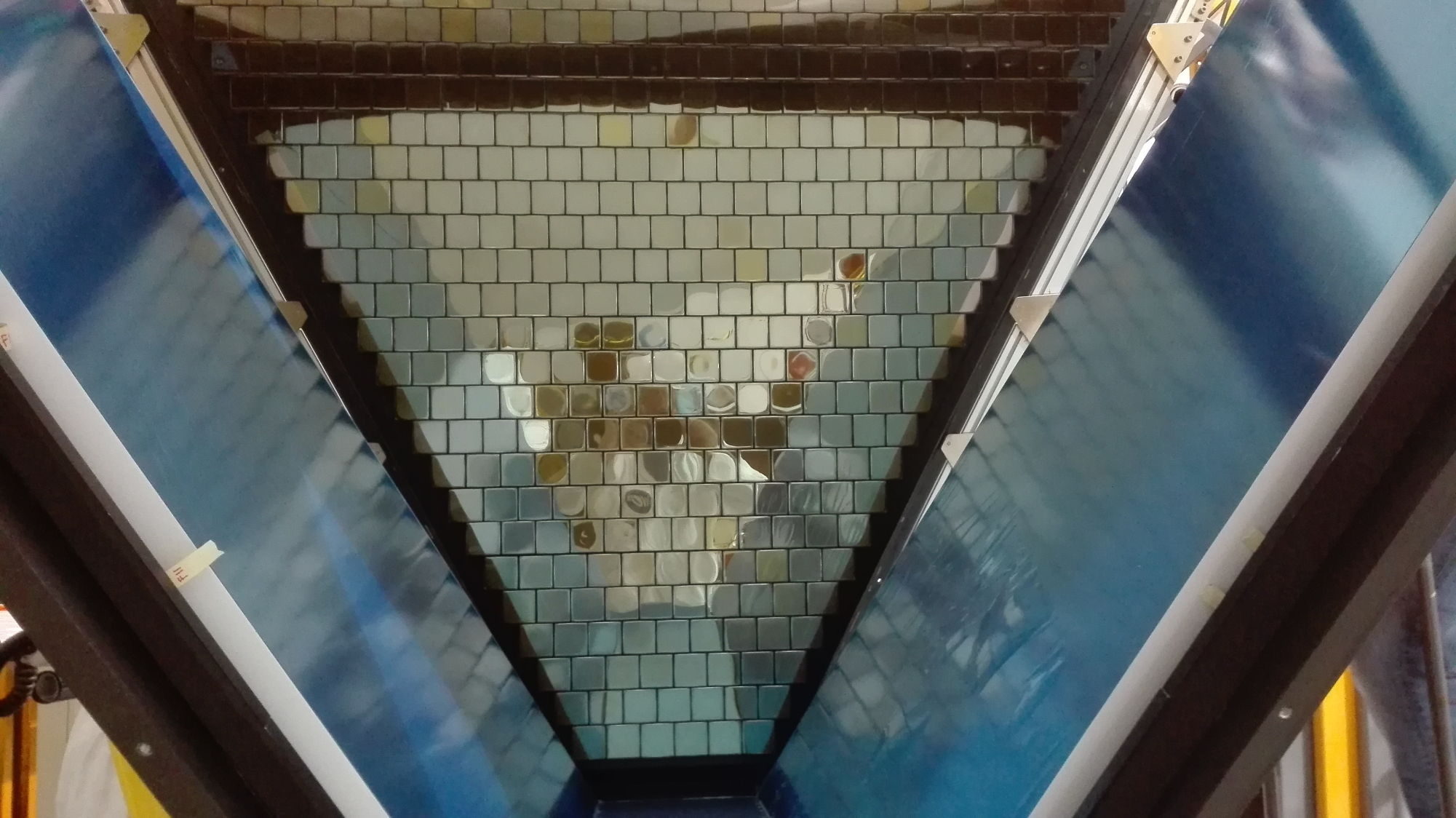}
	\caption{Top: The part of the CLAS12 detector with the RICH covering one out of six sectors. Bottom: the photomatrix of multianode photomultipliers and the mirror system.}
	\label{fig:RICHdetector}
\end{figure}

The photomatrix  wall is a crucial component of the RICH detector (see Fig.~\ref{fig:RICHdetector}). It is relatively large (area about 1 m$^2$) and should be comprised of many photon detection devices such as photomultiplier tubes.
Due to the imaging aspect of the RICH they must provide a spatial resolution of less than 1 cm.
Since multiple photon detectors are tiled into large arrays, they should have large active area with minimal dead-space.
The photon detectors must also efficiently detect single photon level signals and should be sensitive to visible light due to the aerogel radiator material.
Multianode Photomultiplier Tubes from Hamamatsu are perfect candidates for the CLAS12 RICH detector, as they are flat-panel PMTs offering an adequate compromise between detector performance and cost.
Each MaPMT consists of an 8 by 8 array of pixels, each with dimension of 6~mm x 6~mm.
The pixel numbers increment from left to right, top to bottom, with pixel \#1 in the top left corner.
Furthermore, the device has a very high packing fraction of 89\% with a high quantum efficiency of 20-30\% in the visible light region.
The tubes also have excellent immunity to magnetic fields because all internal parts are housed in a metal package and the distance between dynode electrodes is very short.

\begin{figure}[htb]
  \centering
  \includegraphics[width=0.8\linewidth]{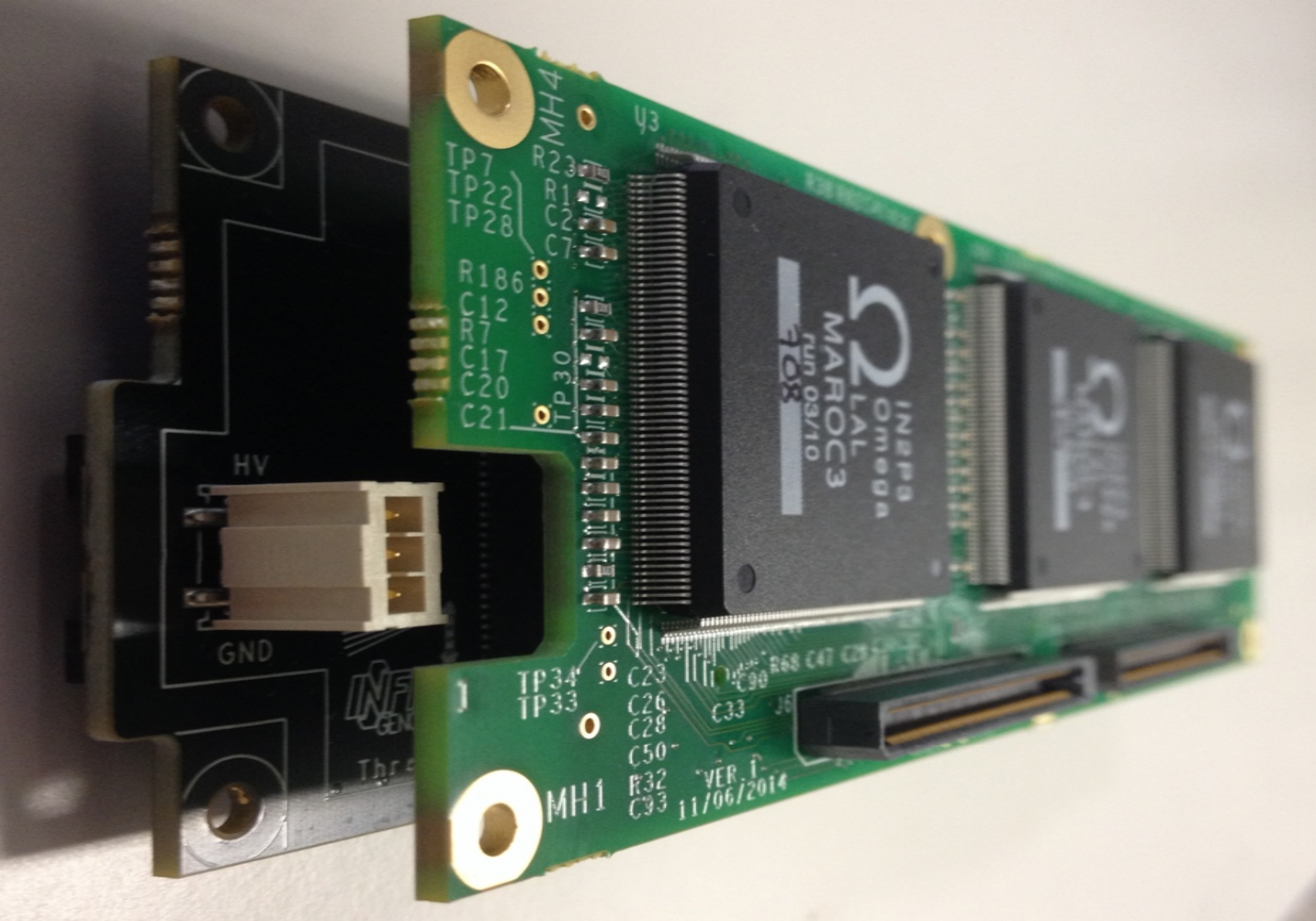}
  \includegraphics[width=0.8\linewidth]{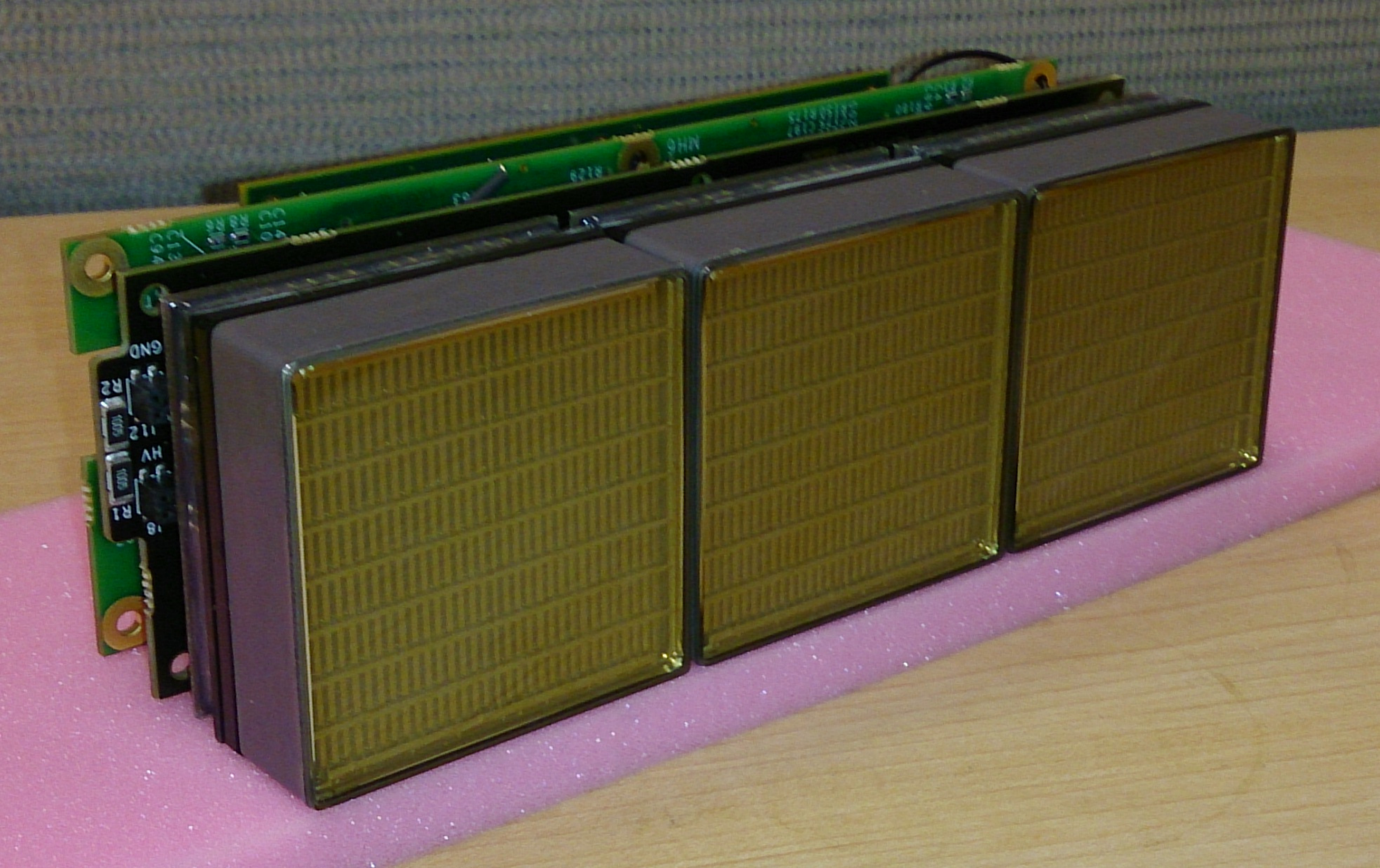}
  \caption{Front-end electronics readout board and mounted MaPMTs.}
  \label{fig:feboards}
\end{figure}

Initially, the Hamamatsu H8500 MaPMT model \cite{H8500} was chosen as the best option because they provide high quantum efficiency for visible light and sufficient spatial resolution (6x6 mm$^2$) at a limited cost.
However, Hamamatsu has released the new H12700 MaPMT model  \cite{H12700} that shows enhanced single photoelectron (SPE) detection, reduced crosstalk between pixels, and is otherwise similar in spatial resolution and  cost to the H8500 MaPMTs.
The first RICH detector was installed in sector 4 of the CLAS12 detector in 2018.
There are 391 Hamamatsu MaPMTs  in the photodector matrix, 76 of them are H8500 and 315 H12700. 
The second RICH detector is almost identical to the first one, fully equipped with H12700 MaPMTs. It has been installed in CLAS12 and is presently taking data.
The characterization of MaPMTs for both detectors was done using a laser stand equipped with custom front-end electronics boards which have much better parameters than the FADCs~\cite{FADC250} used for preliminary studies and installed in the most of the CLAS12 subsystems.
This highly integrated front-end (FE) electronics with modular design~\cite{RICH_FE} was developed for a large array of Hamamatsu H8500 and H12700 MaPMTs to minimize the impact of the electronics material on the CLAS12 subsystems downstream of the RICH detector.
The architecture of the readout electronics consists of front-end cards with dedicated Application Specific Integrated Circuits (ASICs), configured, controlled, and read out by Field Programmable Gate Arrays (FPGAs) \cite{RICH_FE}.
The ASIC board is based on the MAROC3  integrated circuit \cite{MAROC} whose excellent single photon capabilities both in analog and binary mode have been confirmed.
The three-tile electronics module with and without the three H12700 MaPMTs installed is shown in~Fig.~\ref{fig:feboards}.
The performance of the MAROC chips was tested and was found suitable for the RICH requirements:
\begin{itemize}
	\item 100\% efficiency at 1/3 of the single photoelectron signal (50~fC)
	\item time resolution of 1~ns
	\item short deadtime to sustain a trigger rate of 30~kHz
	\item latency of 8~$\mu$s
\end{itemize}
We made detailed characterization of around 400 H12700 MaPMTs, as well as several H8500 to make a comparison of the two models.
These data turned out to be useful for evaluating the performance of the first CLAS12 RICH detector where both MaPMT models are used.
The single photoelectron spectra were measured for each pixel at different high voltages and light intensities of the laser test setup.
Using the dedicated front-end electronics, standard for the RICH detectors, the setup allowed us to characterize each pixel’s properties such as gain, quantum efficiency, signal crosstalk between neighboring pixels, and determine the signal threshold values to optimize their efficiency to detect Cherenkov photons.
These parameters were determined for each pixel in the set of 400 MaPMTs, giving us the opportunity to select the best MaPMTs and determine the working parameters of the front-end electronics in the real experiment.
The results of this study are presented in this paper.

The remaining structure of this paper is laid out as follows. 
\begin{itemize}
	\item Section 2 presents the design of the laser test stand for the MaPMT automated characterization, allowing illumination of every pixel by the precisely calibrated low light pulses in the controlled stable environment, and collecting the response data. 
	\item Section 3 describes the procedures for the absolute calibration of the readout electronics converting the output signal amplitudes to linear charge scale in pC for every pixel. 
	\item Section 4 illustrates the techniques for the pixel-to-pixel crosstalk measurements, and possible algorithms for the separation of the crosstalk from real signals. 
	\item Section 5 describes the technique of absolute calibration of the light source, as a prerequisite for the measurement of quantum efficiency in every pixel. 
	\item Section 6 describes the computational model used in the data analysis to extract such critical parameters for each anode, as its quantum efficiency, gain, the shape of the single photoelectron amplitude response function, and contribution of the crosstalk signals from the neighboring pixels, and introducing the novel technique of characterizing the crosstalk contributions in the model. 
	\item Section 7 illustrates the self-consistency of the algorithm for the parameters' extraction using the measurements at different light intensities and different high voltages applied. 
	\item Section 8 presents the results of the full characterization and study of all 399 MaPMTs, showing the spread of the extracted parameters and evaluating the systematic errors from the independent redundant measurements. The results make possible the evaluation of average and individual pixel characteristics of the full MaPMT array for the purposes of selection and arrangement of the MaPMTs in the RICH detector, and for use in the experimental data analysis.
\end{itemize}

\section{Laser stand for the MaPMT characterization}
\begin{figure}[hbt]
	\centering
	\includegraphics[width=0.95\linewidth]{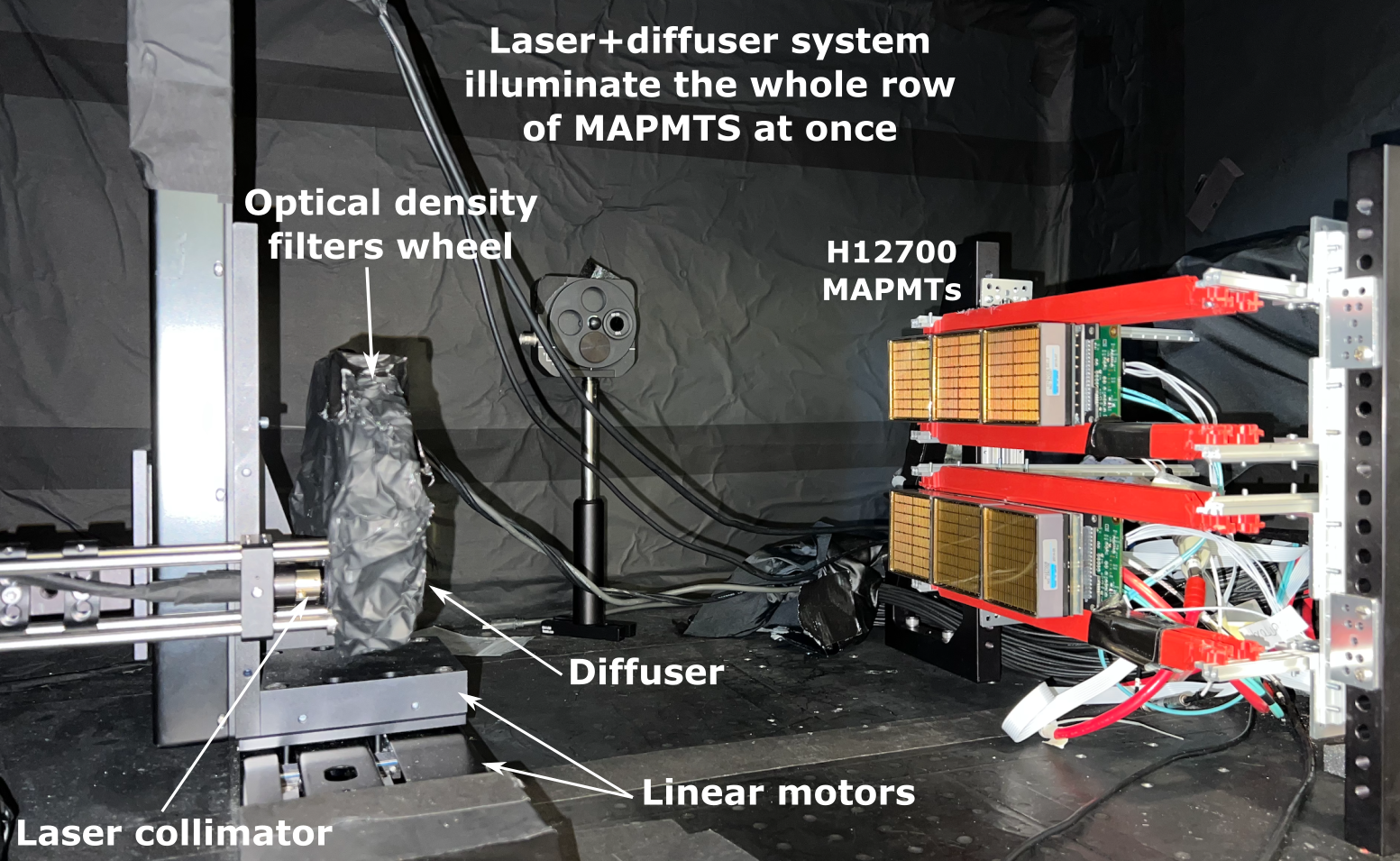}
	\caption{Inner view of the laser test stand.}
	\label{fig:MAPMTtest}
\end{figure}

The large number of the channels in the RICH detector  poses a challenging problem for the MaPMT testing and calibration.
The RICH consists of 391 MaPMTs, resulting in a total of 25024 channels. In order to test them efficiently within a reasonable timeframe, the fully automated test stand was built to evaluate 6 MaPMTs at once, as shown in Fig.~\ref{fig:MAPMTtest}.

\begin{figure}[bt]
	\centering
	\begin{subfigure}[b]{0.628\linewidth}
		\includegraphics[width=\linewidth]{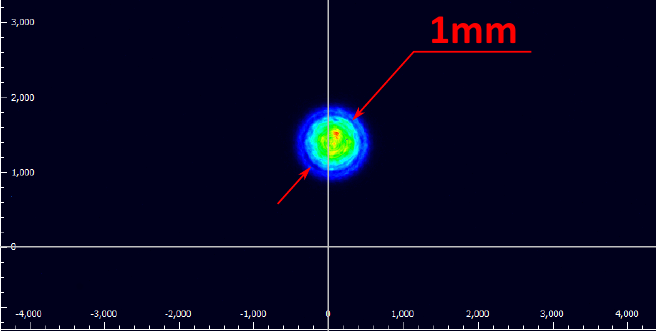}
		\caption{Focused laser beam with the dimension much less than the MaPMT pixel size.}
		\label{fig:beamopt1}
	\end{subfigure}
	\begin{subfigure}[b]{0.354\linewidth}
		\includegraphics[width=\linewidth]{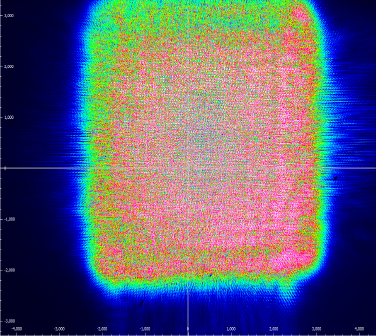}
		\caption{Square pattern illuminating the full MaPMT surface.}
		\label{fig:beamopt2}
	\end{subfigure}
	\caption{The laser light output options.}
\end{figure}

The test stand consists of a picosecond diode  laser PiL047X with a 470 nm wavelength, 2 long travel motorized stands to drive the laser fiber in two-dimensional space for individual pixel illumination, a motorized wheel with a neutral density filter system, and 2 adapter boards for the MaPMTs with JLab designed front-end electronics boards \cite{Contalbrigo:2020}.
The laser light is directed through the fiber and attenuated to the single photon level using neutral density filters to mimic the conditions of the RICH detector.
The remotely operated filter wheel has 6 positions allowing to switch the light attenuation and evaluate MaPMT at different light intensities. Ultra-low and high intensity settings were used for dedicated tests, and the mass MaPMT study was performed using the wheel positions 3, 4, and 6.
The motors can be controlled to move the focused laser beam (see Fig.~\ref{fig:beamopt1}) across the entire surface of the MaPMT entrance window and illuminate one by one all 64 pixels individually.
Alternatively, the Engineered Diffuser can be used to scatter the laser beam and produce a square pattern with a non-Gaussian intensity distribution (see Fig.~\ref{fig:beamopt2}). 
The second option is used to illuminate the full row of 3 MaPMTs at once.

All laser stand equipment is placed in a black box with non-reflective black material on the optical table. The laser interlock safety box automatically switches off the laser, as well as the front-end electronics low voltage and MaPMT high voltage, to prevent possible photomultiplier damage or human exposure to the laser light in case the front door of the black box is opened during measurements.

This configuration minimizes the routine workload and allows for the evaluation of 6 MaPMTs (equivalent to 384 conventional PMTs!) at different high voltages and different light intensities within 6 hours with less than 15 minutes of human interaction used to load the MaPMTs to the front-end boards.

The measurements of custom front-end electronics together with the installed MaPMTs in the RICH black box setup were crucial to understand their performance in the RICH detector.
To test and calibrate it, multiple tests with an internal onboard charge injector, an external charge injector, and a signal generator were performed.
As shown in Fig.~\ref{fig:MAPMTtest}, the RICH MaPMT test setup can house two FE boards inside the black box.
The communication between the FPGA board and the PC is performed using TCP/IP protocol over optical Ethernet (1000BASE-SX).
The data acquisition program executes on a remote PC running Linux OS, configures the FPGA and MAROC boards, and collects the data through a network interface.
The current setup allows fast evaluation of the FE modules with a highly automated procedure, which is important because the RICH panel consists of 115 tiles with 3-MaPMT and 23 tiles with 2-MaPMT FE modules.
\section{MAROC chip calibration}

\begin{figure}[hbt]
	\centering
	\includegraphics[width=\linewidth]{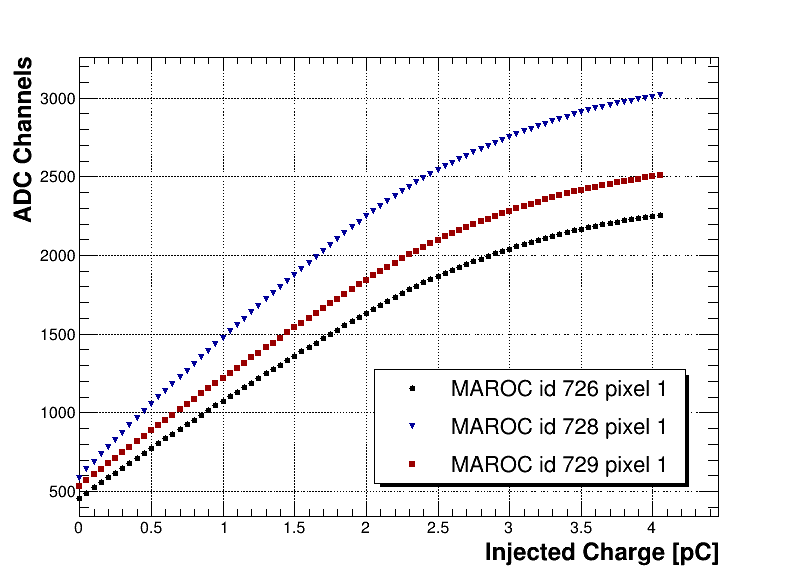}
	\caption{Response of the MAROC slow shaper in ADC channels as a function of the injected charge. The curves shown are for pixel \#1 in three different MAROC boards.}
	\label{fig:MAROCcalib}
\end{figure}

To allow the cross-comparison between different pixels and different MaPMTs  in universal units, and to correct for the non-linearity of the ADC readout at higher amplitudes, a procedure was developed to convert the amplitude of the MAROC slow shaper signal from ADC channels into charge. The MAROC has a built-in charge injection functionality consisting of a test input pin that is connected to the preamplifiers through a logic network of switches and 2 pF capacitors.
Together with an external step function generator, this can be used to inject a controllable amount of charge directly into the preamplifiers. We measured the output of the slow shaper in ADC channels for 82 different input charges ranging from 0 to 4 pC.
Figure~\ref{fig:MAROCcalib} shows the relationship between the injected charge and the measured amplitude in units of ADC channels for three different readout channels. The relationship between charge and ADC channels is linear up to about 1.5 pC.
This distribution was observed to vary between chips and pixels, and thus individual distributions were measured for all 64 pixels on each MAROC used in this study. 

\begin{figure}[hbt!]
	\centering
	\includegraphics[width=\linewidth]{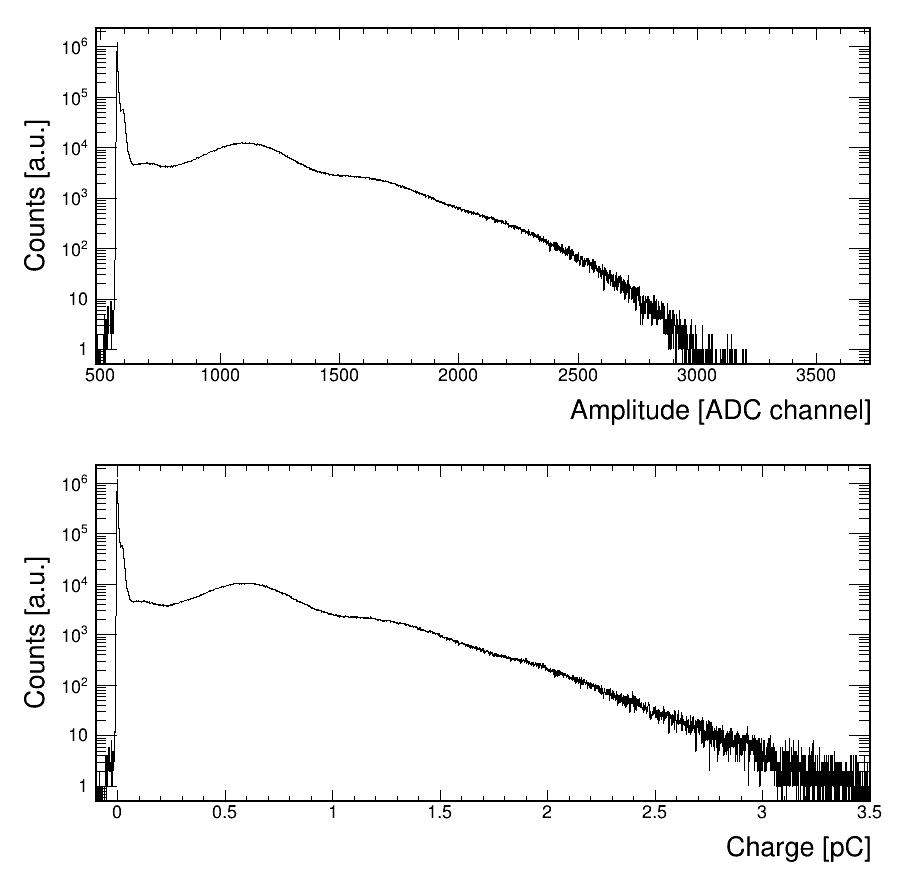}
	\caption{Top: A typical SPE spectrum for one H12700 pixel in units of ADC channel. Bottom: The same spectrum after converting the units into pC.}
	\label{fig:H12700calib}
\end{figure}

\begin{figure}[hbt!]
	\centering
	\includegraphics[width=\linewidth]{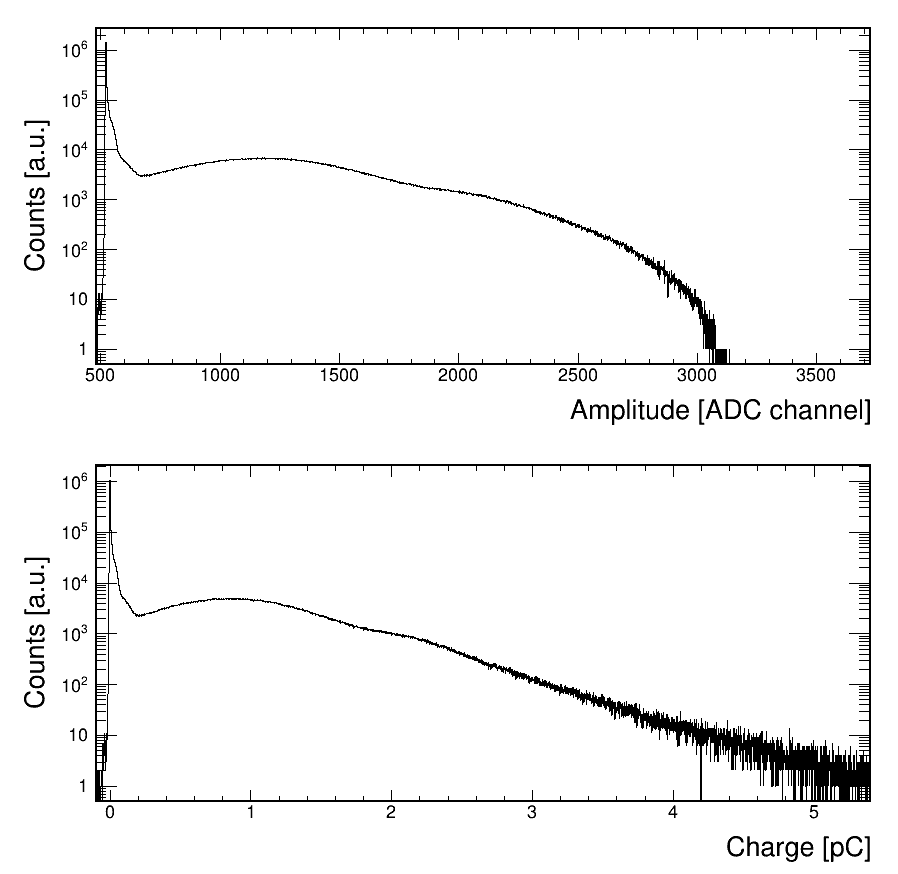}
	\caption{Top: A typical SPE spectrum for one H8500 pixel in units of ADC channels. Bottom: The same spectrum after converting the units into pC.}
	\label{fig:H8500calib}
\end{figure}

This calibration data was used to convert the measured amplitude in ADC channels into charge collected on an event-by-event basis. A local polynomial regression was used to provide a one-to-one mapping of adc channel to charge. Figure ~\ref{fig:H12700calib} and Fig.~\ref{fig:H8500calib} show typical amplitude distributions before and after this conversion was applied for one H12700 MaPMT pixel and one H8500 MaPMT pixel, respectively. For both, the conversion to charge extends the high-amplitude tails of the spectra due to the non-linearity of the ADC readout.

\section{Cross talk measurements}

\begin{figure*}[h!bt] 
    \centering 
    \includegraphics[width=.9\textwidth,height=0.5\textwidth]{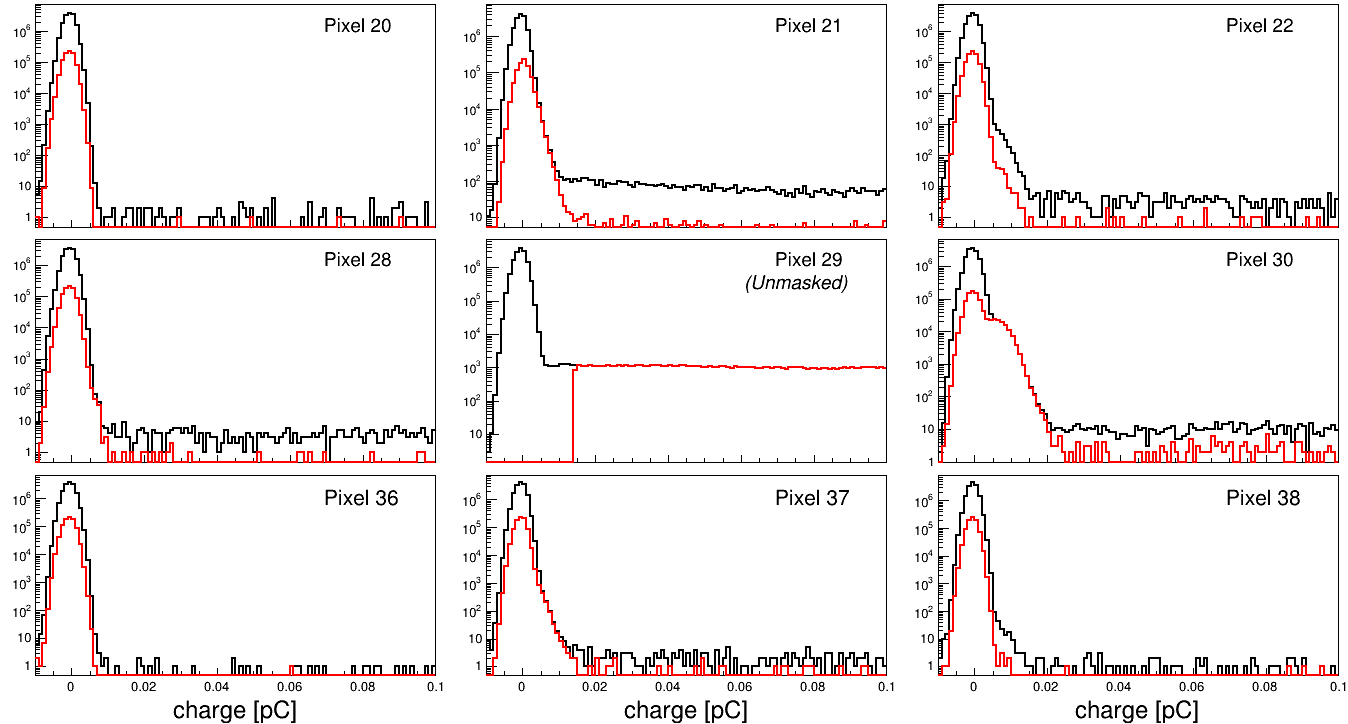}
    \caption{Black: the charge spectra for pixel 29 of a typical H12700 MaPMT and the surrounding pixels when only pixel 29 was illuminated by the laser light. Red: the same spectra with the cut that the signal in pixel 29 is 10$\sigma$ above pedestal.}
    \label{fig:H12700pinhole}
\end{figure*}

To demonstrate the crosstalk between adjacent pixels on the MaPMTs, we collected data where the whole PMT face was masked with a sheet of black paper, and a single 3 mm diameter hole was punctured over the center of one pixel. Despite the majority of the laser light being incident on the single unmasked pixel, we observed signals above pedestal in the surrounding pixels as well. Figure ~\ref{fig:H12700pinhole} shows the measured spectra for the central and neighboring pixels when the puncture hole was directly above pixel 29. There are two types of events we see in the surrounding pixels of this data set. The first is the electronic crosstalk resulting from the electron cascade in the central pixel. The signal measured in a neighboring pixel is directly proportional to that which is measured in the central pixel. In Fig.~\ref{fig:H12700pinhole}, these types of events are characterized by a shoulder attached to the right of the pedestal. This is most prominently seen in the spectrum for the pixel directly to the right of the central pixel of Fig.~\ref{fig:H12700pinhole} (pixel 30). Because of the strong correlation of the crosstalk to the central pixel, these types of events can be identified and removed from the data offline. More will be discussed on this later.

The second type of event observed in the neighboring pixels is 
the optical crosstalk due to
the displacement of the photoelectron emitted by the photocathode. When the incident photon hits the unmasked pixel, there is some probability that the emitted photoelectron is detected in one of the neighboring pixels instead. Because there is no correlation with the signal in the central pixel for these events, there is no way to identify these signals on an event-by-event basis. In Fig.~\ref{fig:H12700pinhole}, the spectra drawn in red have the additional cut applied that the signal in the central pixel should be greater than 10$\sigma$ above the pedestal. With this cut applied, the number of events beyond the crosstalk shoulder in the neighboring pixels is reduced by more than an order of magnitude.

\begin{figure}[h!bt]
	\centering
	\includegraphics[width=\linewidth]{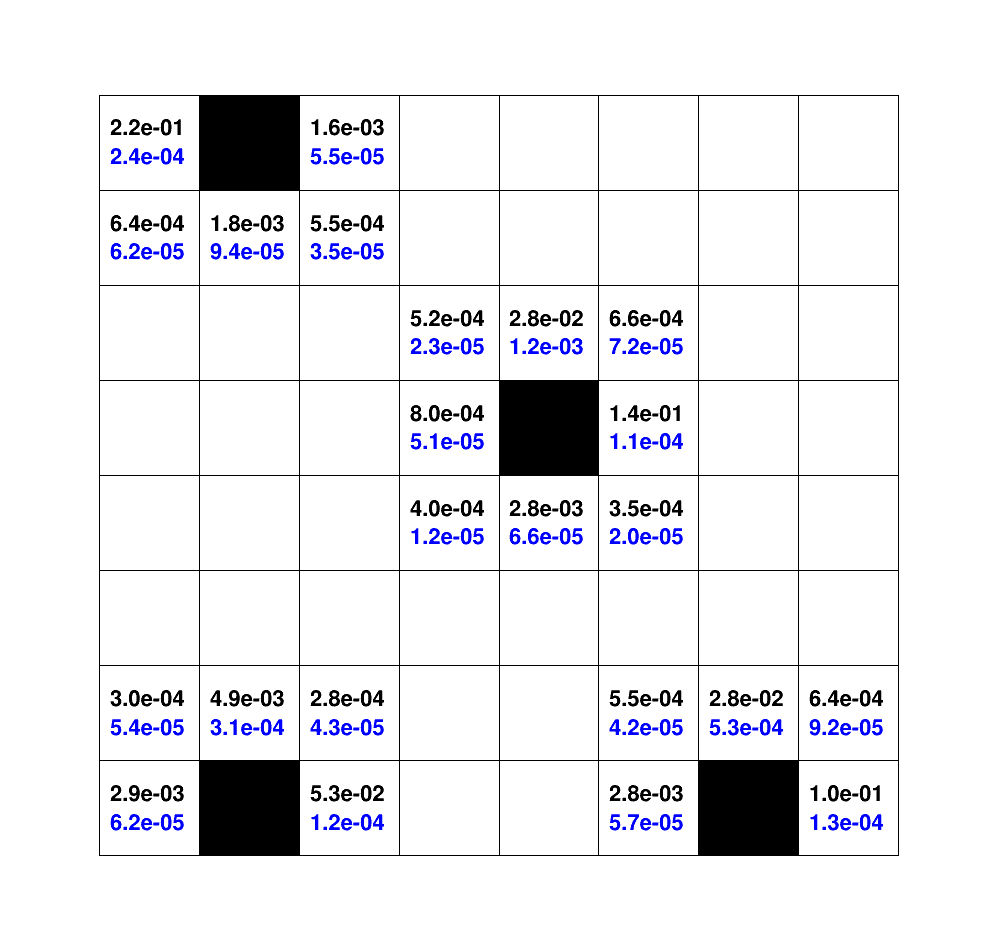}
	\caption{For each highlighted pixel a separate run was taken where only this pixel had a 3 mm hole punctured in the mask covering the whole PMT face. The numbers in black in the surrounding pixels represent the fraction of electronic crosstalk events in that pixel. The numbers in blue represent the fraction of optical crosstalk events where the photoelectron emitted from a photon incident on the photocathode of the unmasked pixel is detected in one of the neighboring anodes.}
	\label{fig:H12700_ct_ratio}
\end{figure}

\begin{figure*}[h!bt]
	\centering
	\includegraphics[width=0.9\linewidth]{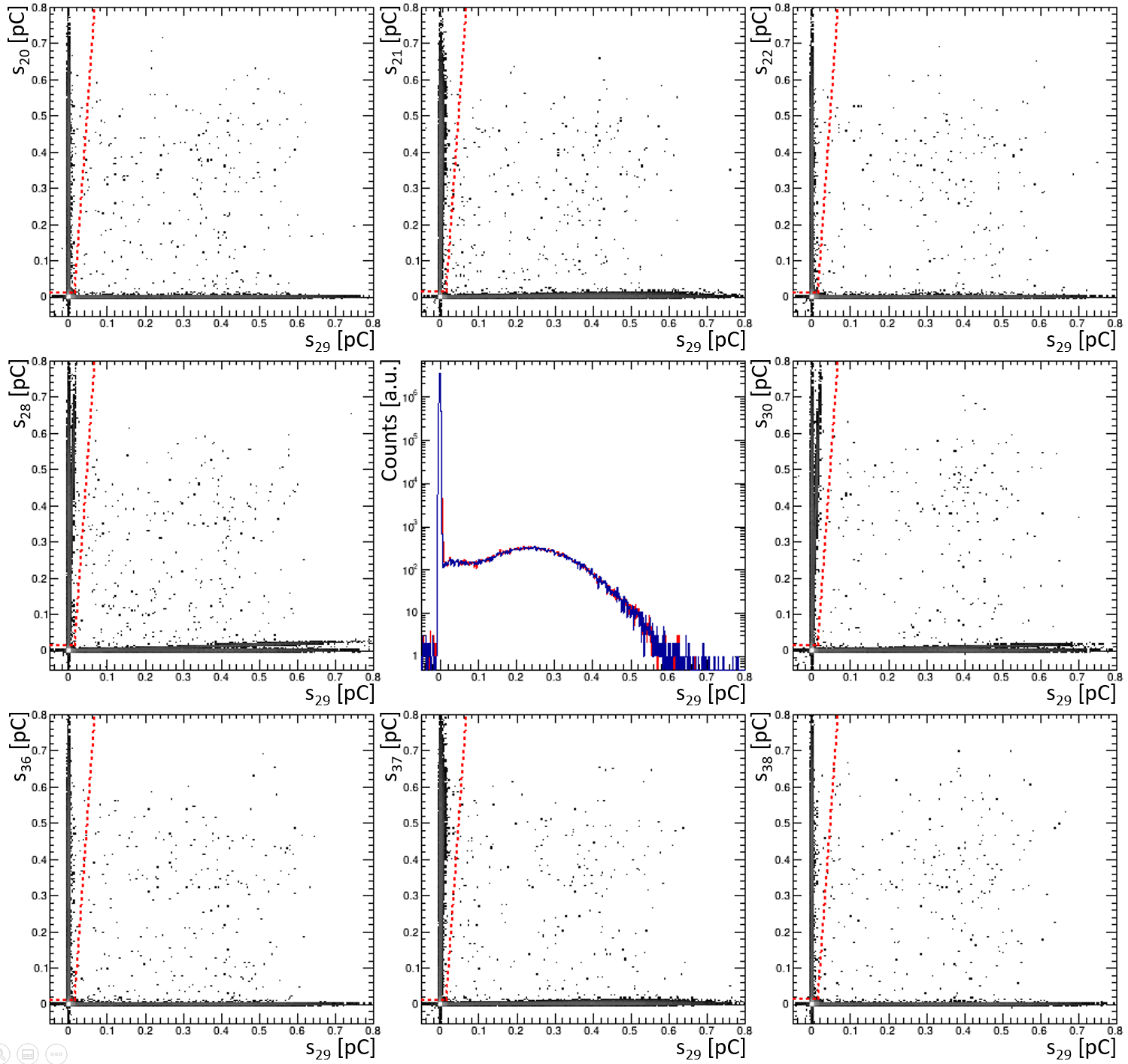}
	\caption{The charge measured in adjacent pixels is plotted as a function of the charge measured in pixel 29 for a typical H12700 MaPMT. The electronic crosstalk signature is most clearly seen in the pixels directly to the left and right of the central pixel, where a linear band of events is seen separate of the pedestal. Events which lie above the dashed (red) line in the two-dimensional plots are identified as crosstalk and are cut. The central plot shows the charge spectrum in pixel 29 before (red) and after (blue) removal of the crosstalk events.}
	\label{fig:H12700neighbors}
\end{figure*}

\begin{figure*}[h!bt]
    \centering
	\includegraphics[width=0.9\linewidth]{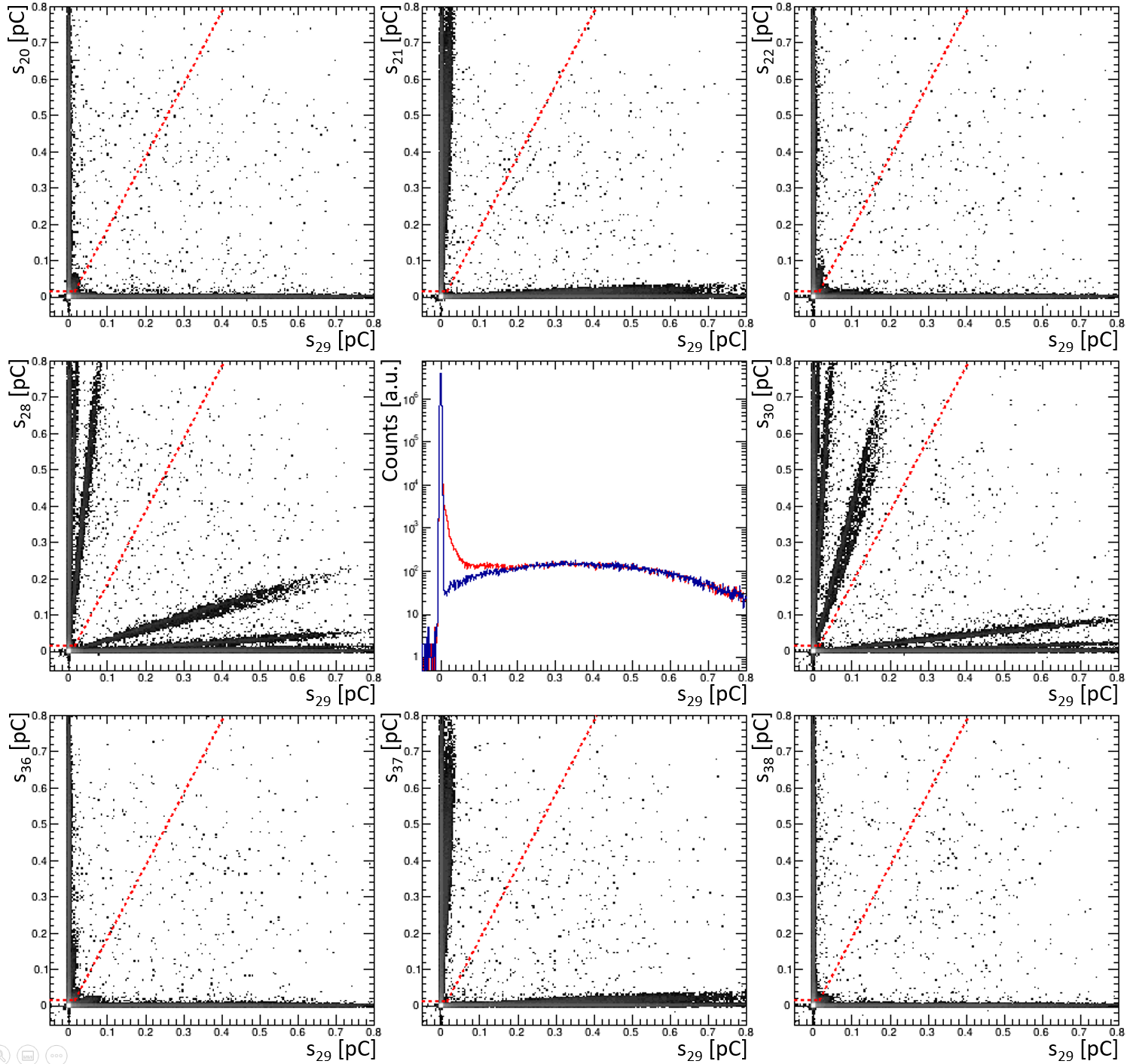}
	\caption{The charge measured in adjacent pixels is plotted as a function of the charge measured in pixel 29 for a typical H8500 MaPMT. The electronic crosstalk signature is most clearly seen in the pixels directly to the left and right of the central pixel, where a linear band of events is seen separate of the pedestal. Events which lie above the dashed (red) line in the two-dimensional plots are identified as crosstalk and are cut. The central plot shows the charge spectrum in pixel 29 before (red) and after (blue) removal of the crosstalk events.}
	\label{fig:H8500neighbors}
\end{figure*}


Using this masking scheme, we collected data with different pixels unmasked and measured the fraction of events with crosstalk in the neighboring pixels. 
Fig.~\ref{fig:H12700_ct_ratio} shows these fractions for each of the neighboring pixels of 4 different unmasked pixels. The numbers in black represent the fraction of electronic crosstalk events in the neighboring pixels, while the numbers in blue represent the fraction of optical crosstalk events.
The selection criteria for the electronic crosstalk events was that the charge measured in the unmasked pixel was larger than 25~fC, while the charge measured in the neighboring pixel was larger than three times the width of it\textquotesingle s pedestal distribution and less than 25~fC. 
Meanwhile, the optical crosstalk events were selected by requiring that the charge measured in the unmasked pixel was within 2$\sigma$ of the pedestal distribution, while the charge collected in the neighboring pixel was larger than 25~fC. Due to imperfect alignment of the masks and light leakage, there is some fraction of events where a photon is incident on one of the masked pixels. However, as observed in the red histograms of Fig.~\ref{fig:H12700pinhole}, the fraction of these events is small, and we estimate this contributes about 10$\%$ uncertainty to the numbers reported in Fig.~\ref{fig:H12700_ct_ratio}.

To properly characterize the single photoelectron spectrum for each pixel, one needs to either add a description of the crosstalk into the computational model for the SPE response, or one can attempt to identify and remove these crosstalk events from the data. A simple procedure was developed and implemented to attempt the latter option. Because the amplitude of the crosstalk is linearly dependent on the amplitude of the photo-induced signal, the crosstalk events appear as linear bands in the plots showing the measured charge in one pixel as a function of the measured charge in a neighboring pixel. Figure ~\ref{fig:H12700neighbors} and Fig.~\ref{fig:H8500neighbors} show these two-dimensional plots for all pixels that neighbor pixel 29 for one H12700 MaPMT and one H8500 MaPMT, respectively. The data shown in these plots were taken with the entire face of the MaPMTs illuminated by the laser light. From these two plots it is obvious that the strength of the crosstalk is vastly different between the H12700 and H8500 MaPMTs. On average, the amplitude of the crosstalk in an H12700 MaPMT is only about 2-3$\%$ of the main signal, whereas the crosstalk amplitude in an H8500 MaPMT can be as large as 50$\%$ of the main signal. As we will discuss later, this fact makes it more difficult to address the crosstalk for the H8500 MaPMTs in the mathematical description of the SPE response function.

Other noteworthy features from Fig.~\ref{fig:H12700neighbors} and Fig.~\ref{fig:H8500neighbors} are that the crosstalk signals are strongest in the pixels immediately to the right and left of the pixel where light was incident. The crosstalk bands in those pixels have the largest slope. Most of the crosstalk is contained within the 4 pixels that share an edge with the illuminated pixel, as the plots for the pixels on the corners show little correlation with the charge measured in the central pixel.

Because the crosstalk events are easily distinguished in these two-dimensional plots, a cut can be placed to remove these events from the data. The cut was applied to each pixel separately, and is a linear function of the charge measured in that pixel. Specifically, the cut placed a limit on the maximum charge measured in the neighboring pixels. If the maximum neighboring charge was above the cut value for the central pixel\textquotesingle s measured charge, then the event was tagged as crosstalk and was removed from the charge spectrum for the central pixel. This cut is shown as a dashed (red) line in Fig.~\ref{fig:H12700neighbors} and Fig.~\ref{fig:H8500neighbors}. The start of the cut line was placed 7$\sigma$ above the pedestal to avoid removing pedestal events. Although the slope of the crosstalk bands varied between pixels, the slope of the cut line used here was the same for each pixel on a given PMT. 

The main drawback of this crosstalk cut is that it removes events where both adjacent pixels  happen to have a photoelectron emitted from the same laser trigger. However, the fraction of these accidental coincidence events was low when the laser filter was used at the minimal setting, meaning at low light intensity this procedure can be used to provide the SPE spectrum free from electronic crosstalk. The charge spectra before and after the removal of the crosstalk events in this manner is compared in the central plot in Figs.~\ref{fig:H12700neighbors} and~\ref{fig:H8500neighbors}. For both the H12700 and the H8500, the crosstalk shoulder to the right of the pedestal is removed after applying this cut. 
\section{Calibration of laser photon flux}

\begin{figure}[h]
\centering
\includegraphics[width=0.5\textwidth]{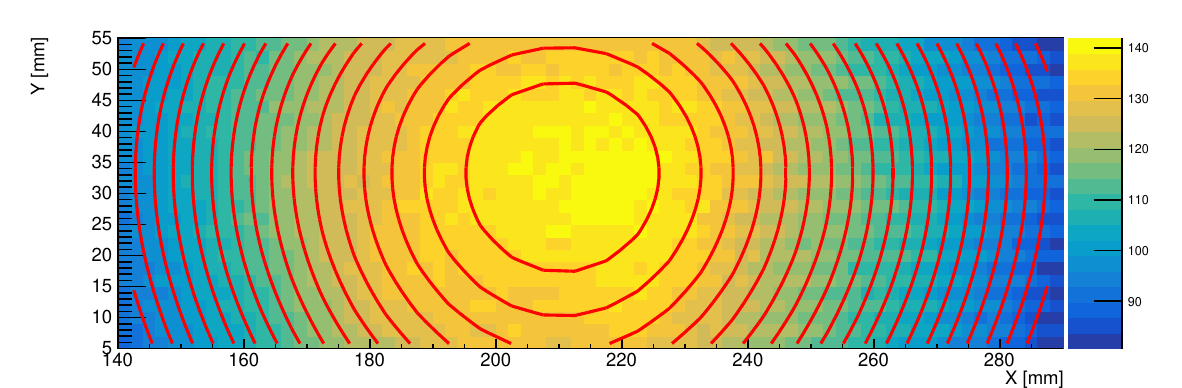}
\caption{Light intensity distribution $\frac {dN_\gamma}{dS}$, defined as the number of photons per mm$^{2}$ in one laser pulse, for a row of three MaPMTs in the laser stand.}
\label{fig:light_flux}
\end{figure}

The calibration of the absolute laser photon flux was performed with the use of the silicon photodiode Hamamatsu S2281.
The tabulated quantum efficiency of this diode at the wavelength of our laser ($\lambda=470$ nm) is 62.6\%, taken from the Hamamatsu S2281 Manual. 
The active part of the diode is a circle with a diameter of 11.3~mm, which is 100~mm$^2$. 
A KEITHLEY 6485 picoammeter was used to measure the average diode current while illuminated by the laser beam.
The noise diode current was estimated to be at the level of 0.2~pA. 
During the MaPMT characterization, the laser frequency was maintained at 20 kHz. 
For light calibration, the higher the frequency, the better the current measurement accuracy that can be achieved from the point of view of the noise level. 
The maximum frequency of our laser is 1 MHz.
However, there are additional systematic uncertainties associated with the extrapolation from one frequency to another. 
For this reason, the scan of the light field was done at the working frequency of 20 kHz. 
The measured current in the center position of the laser head was around 29.2~pA at this frequency, meaning the systematic uncertainty
of this measurement was below 1\%.  We made a detailed two-dimensional scan of the photon flux by
moving the laser head with step sizes of 2~mm in the X and Y directions along the full area where the 3 MaPMTs were located during the characterization procedure.
Normalized to one laser pulse and 1~mm$^2$ area, the number of photons with $\lambda=470$~nm  is presented in Fig.~\ref{fig:light_flux}.
The maximum value of the photon flux in the center of the light field equals 145 $\gamma$/mm$^2$/pulse.
These measurements were done without  any optical filters installed. We used neutral density calibrated optical filters with anti-reflection coating.
To check the possible filter effects, we made a measurement of the light flux for one of the filters with a tabulated attenuation of 100. 
This test was done with a frequency of 1 MHz to increase the accuracy of the current measurement. 
The ratio of the measured attenuation factor to that tabulated was determined to be 1.05$\pm 0.01$. This coefficient was applied to the map of the photon flux when used for data with optical filters. It takes into account the possible effects of rescattering or reflection of the photons by the filters.

The knowledge of the absolute number of photons hitting the photomultiplier tubes during the characterization gave us the possibility to measure the quantum efficiency of the MaPMTs for each pixel. The average number of photoelectrons, $\mu$, is proportional to the quantum efficiency:
$$
\mu=\epsilon_{QE} \int_{S_{pixel}}\frac {dN_\gamma}{dS} dS,
$$
\noindent
where $\int_{S_{pixel}}\frac {dN_\gamma}{dS} dS$ is the number of photons integrated over the pixel's area, $S_{pixel}$, and $\epsilon_{QE}$ is the quantum efficiency of the pixel.
The integration included the measured light field at the position of the pixel under study.
The parameter $\mu$ was determined during the PMT characterization. Possible photoelectron collection inefficiency was taken into account and approximated in the computational model during the calculation of $\mu$. 
\section{Computational model describing the PMT response}

The goal for using MaPMTs in RICH detectors is to achieve reliable detection of single photons in the Cherenkov light radiation cones. A single photon incident on a PMT face may knock out a single photoelectron from the PMT's photocathode with a certain probability, defined as the Quantum Efficiency (QE). The photoelectrons cascade inside the PMT to generate a typical amplified electrical signal at the anode. The amplitude distribution of the single photoelectron signal depends on the MaPMT design and high voltage applied and varies from pixel to pixel. Tests and characterization of multiple MaPMTs include measuring the SPE amplitude distributions for every pixel, finding out the appropriate amplitude thresholds, and determining the QE. To achieve this goal, we used the methods developed in Ref.~\cite{DEGTIARENKO20171}, expanded to include the new empirical method to take into account the effects of the pixel-to-pixel crosstalk in the H12700 tubes. Ref.~\cite{DEGTIARENKO20171} describes in detail the computational model used to extract and parameterize the SPE distributions from the measurements using the laser test setup. The method allows, in principle, a description of SPE functions of essentially any complexity by decomposing them into a sum of Poisson distributions with different averages. For the detailed explanations and the definition of the model parameters see Ref.~\cite{DEGTIARENKO20171}. The list of main parameters includes $\mu$, the average number of photoelectrons produced by the laser in a given pixel per test pulse, and scale, the average amplitude of the SPE distribution in pC. 
The parameter scale is directly connected with the gain (or current amplification) parameter usually given in the photomultiplier specifications. The term scale was introduced in Ref.~\cite{DEGTIARENKO20171} to handle the spectral data not necessarily normalized to the unit charge, and it is kept for compatibility. The value of scale equal to 160.2~fC corresponds to gain=$10^6$, and the value of gain may be obtained by multiplying scale (in fC) by 6241.5.
Five model parameters determine the shape of the SPE distribution, defined as a normalized sum of three Poisson distributions with different average multiplication coefficients  applied to the photoelectron on the first dynode of the PMT. The average multiplication on the first cascade ${\nu}$, or ${\nu_{average}}$ (equivalent to the {\it secondary emission ratio} as per the Hamamatsu PMT Handbook~\cite{Hamamatsu4thedition}), may be derived from these parameters. 
The parameter ${\sigma}$ describes the Gaussian shape of the pedestal function, and the parameter ${\xi}$  describes the effective cascade multiplication on the second dynode. The combination of 9 parameters describes the single-anode PMT SPE response in an ideal measurement setup with a Gaussian pedestal function. If the pedestal amplitude distribution is not exactly Gaussian, the problem of parameterizing the SPE distribution requires the addition of new parameters that take into account the distortion of the pedestal. This method was successfully implemented in~\cite{DEGTIARENKO20171} for the case of a small exponential noise contribution to the Gaussian measurement function. In the present work we use a similar ad hoc approach to parameterize and approximate the contribution of the crosstalk signals coming from the neighboring pixels to the SPE amplitude distribution. The model for the process, in agreement with the observations presented in the previous section, assumes that a portion of the signal from a neighboring pixel may be randomly added to the amplitude measured in a given pixel under investigation. Such random contributions could, in principle, depend on the neighbor. It would be very difficult to characterize all possible pair combinations separately. In the case of the H12700 MaPMTs, the signal amplitudes of the crosstalk contributions from different neighboring pixels were found to be relatively small and similar to each other, allowing us to use the single averaged spectral term for all neighbors of a given pixel. In the model every crosstalk contribution comes from a single electron in one of the neighboring pixels, their average number in one measurement $\beta$ is expected to be comparable with $\mu$, and multiple crosstalk events in one measurement happen independently. The average width of the crosstalk contribution to the measurement function from one crosstalk electron corresponds to the second new model parameter $\zeta$, and the third new parameter $\lambda$ is introduced to adjust the shape of the crosstalk contribution. The explanation of this new formalism is given in Appendix A. It requires familiarity with the formulation of the model presented in full detail in Ref.~\cite{DEGTIARENKO20171}.

\begin{figure}[hbt]
	\centering
	\includegraphics[width=0.98\linewidth, trim=80 55 105 105, clip]{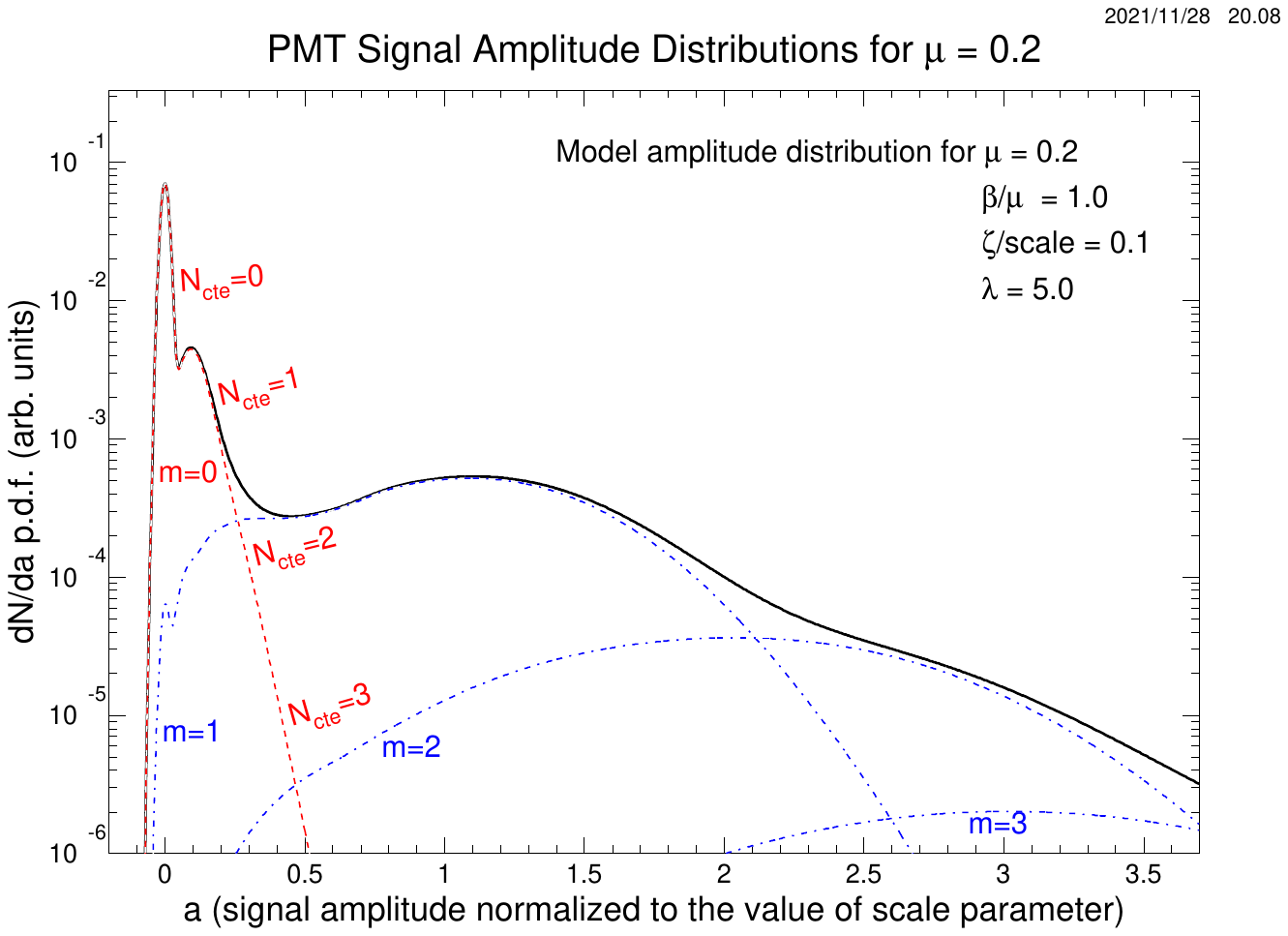}
	\caption{Model  signal  charge  distribution  (black  line)  illustrating  the parameterization for the crosstalk effects. The red line ($m=0$) corresponds to the pedestal measurement function with the additional crosstalk contribution, the blue lines ($m$ = 1, 2, 3) show the contributions from events with 1, 2, and 3 photoelectrons, with their relative probability corresponding to a Poisson distribution with an average $\mu=0.2$.}
	\label{fig:Model}
\end{figure}

The technique is illustrated in Fig.~\ref{fig:Model} showing an example of the distribution of the test events on the normalized measured charge $a$, with $a=1$ corresponding to the average charge collected from one photoelectron. The series of lines marked as $m = 1, 2, 3$ corresponds to the charge distributions in the events with the corresponding number of photoelectrons, assuming the average number of photoelectrons in the test events is $\mu = 0.2$. The red distribution corresponds to the pedestal measurement function $R_{ct}(a)$ with the added crosstalk correction. The regions in this distribution marked with $N_{cte}=0, 1, 2, 3$ correspond to the original Gaussian pedestal function and the contributions from 1, 2, and 3 crosstalk electrons. The parameters were selected for better visibility of the crosstalk effects, with $\beta$ equal to $\mu$, $\zeta$ equal to 10\% of the scale parameter, and $\lambda = 5$ to make the crosstalk Poisson peak more visible.  

The fitting procedure from Ref.~\cite{DEGTIARENKO20171} was modified to include the new three parameters in the FORTRAN routine describing the measured test spectra, bringing the total number of parameters to 12. The algorithm for the multiparametric minimization was adjusted to provide stability. The experimental verification of the fit stability and reproducibility of the results was performed using multiple measurements of the same MaPMTs in the different slots in the test setup and comparing the results. Overall confidence was assured by extracting the parameters for each MaPMT in several test conditions, varying the high voltage and the illumination conditions, and verifying the consistency of the extracted parameters. The procedure also helped us to evaluate the uncertainties of the major extracted model parameters.
\section{Characterization of MaPMTs}

\begin{figure*}[hbt] 
\centering 
  \subfloat[3 mm mask]{%
    \includegraphics[clip=true,trim=100 75 140 100,width=.49\textwidth,height=.25\textwidth]
                    {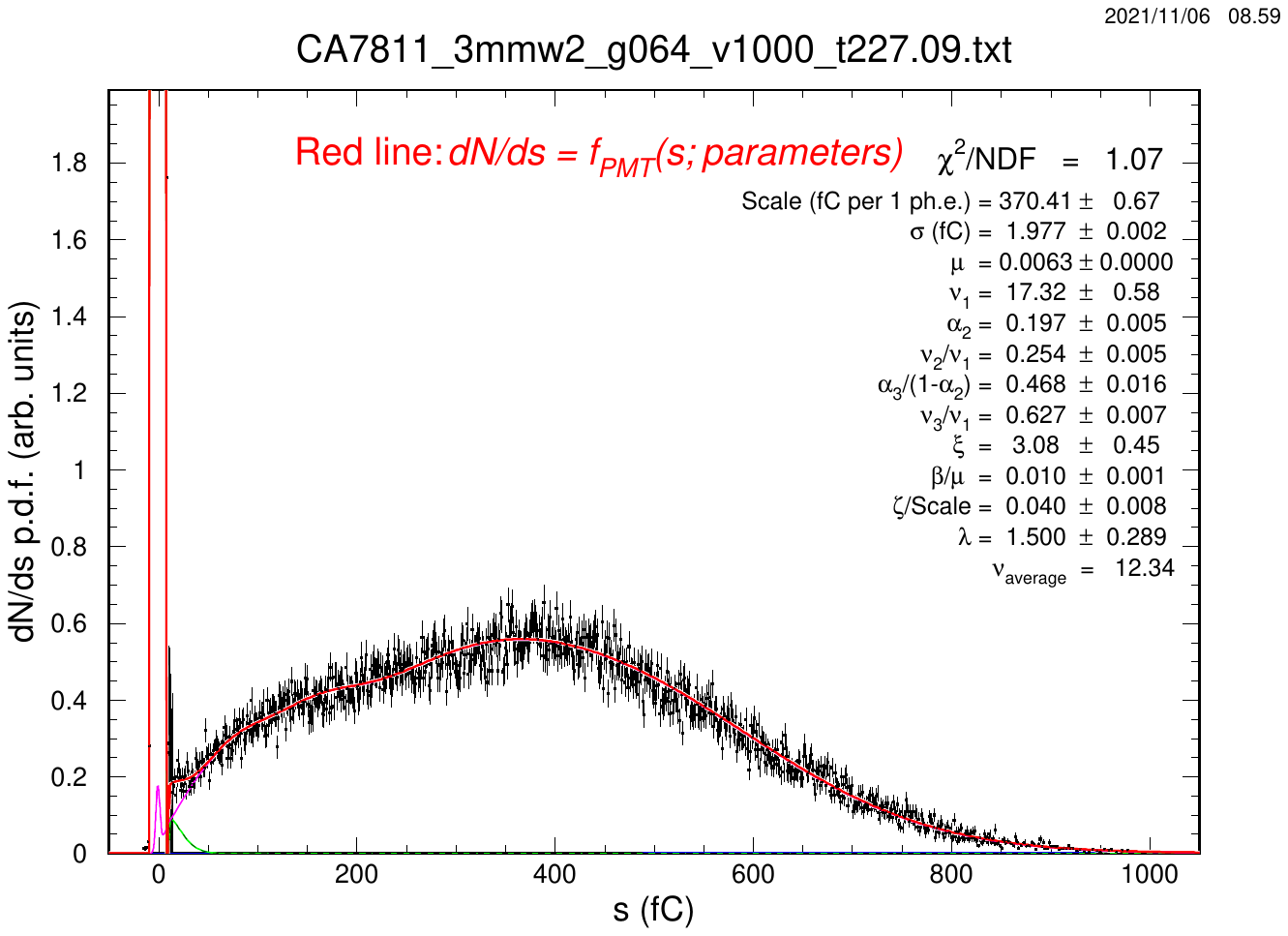}}
  \subfloat[6 mm mask]{%
    \includegraphics[clip=true,trim=100 75 140 100,width=.49\textwidth,height=.25\textwidth]
                    {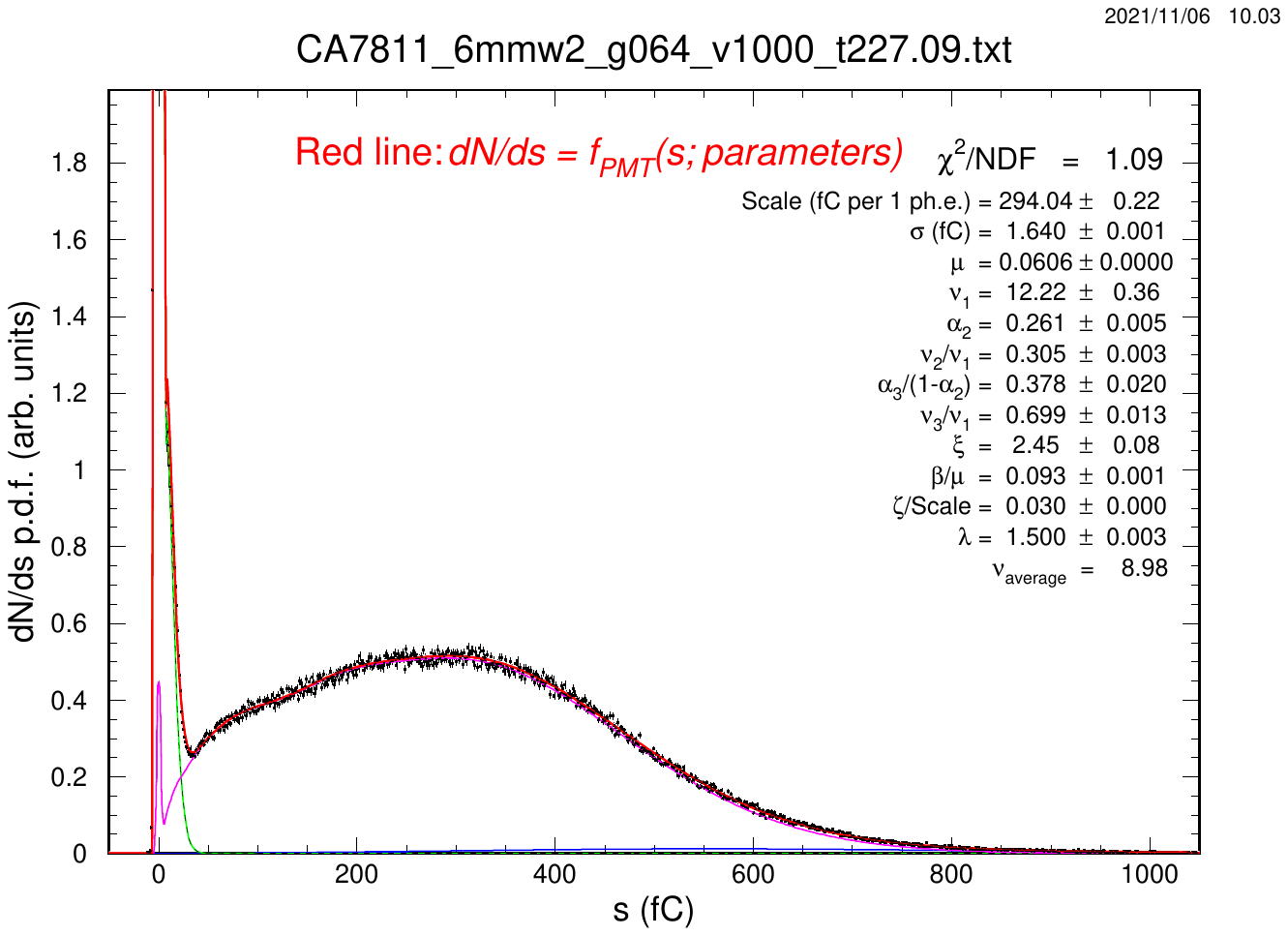}} \\
  \subfloat[No mask, crosstalk removed by software]{%
    \includegraphics[clip=true,trim=100 75 140 100,width=.49\textwidth,height=.25\textwidth]
                    {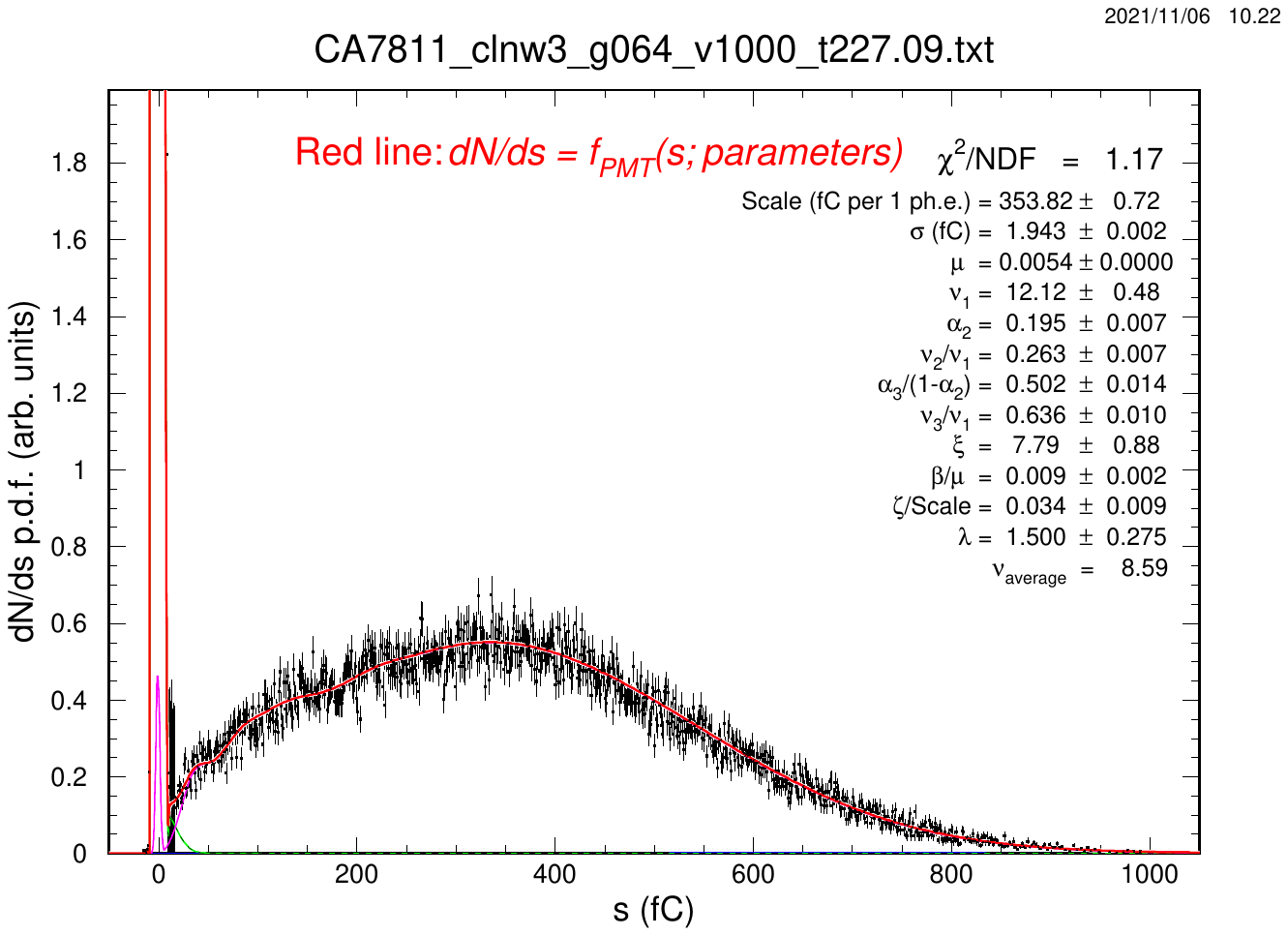}}
  \subfloat[No mask, crosstalk approximated by fit]{%
    \includegraphics[clip=true,trim=100 75 140 100,width=.49\textwidth,height=.25\textwidth]
                    {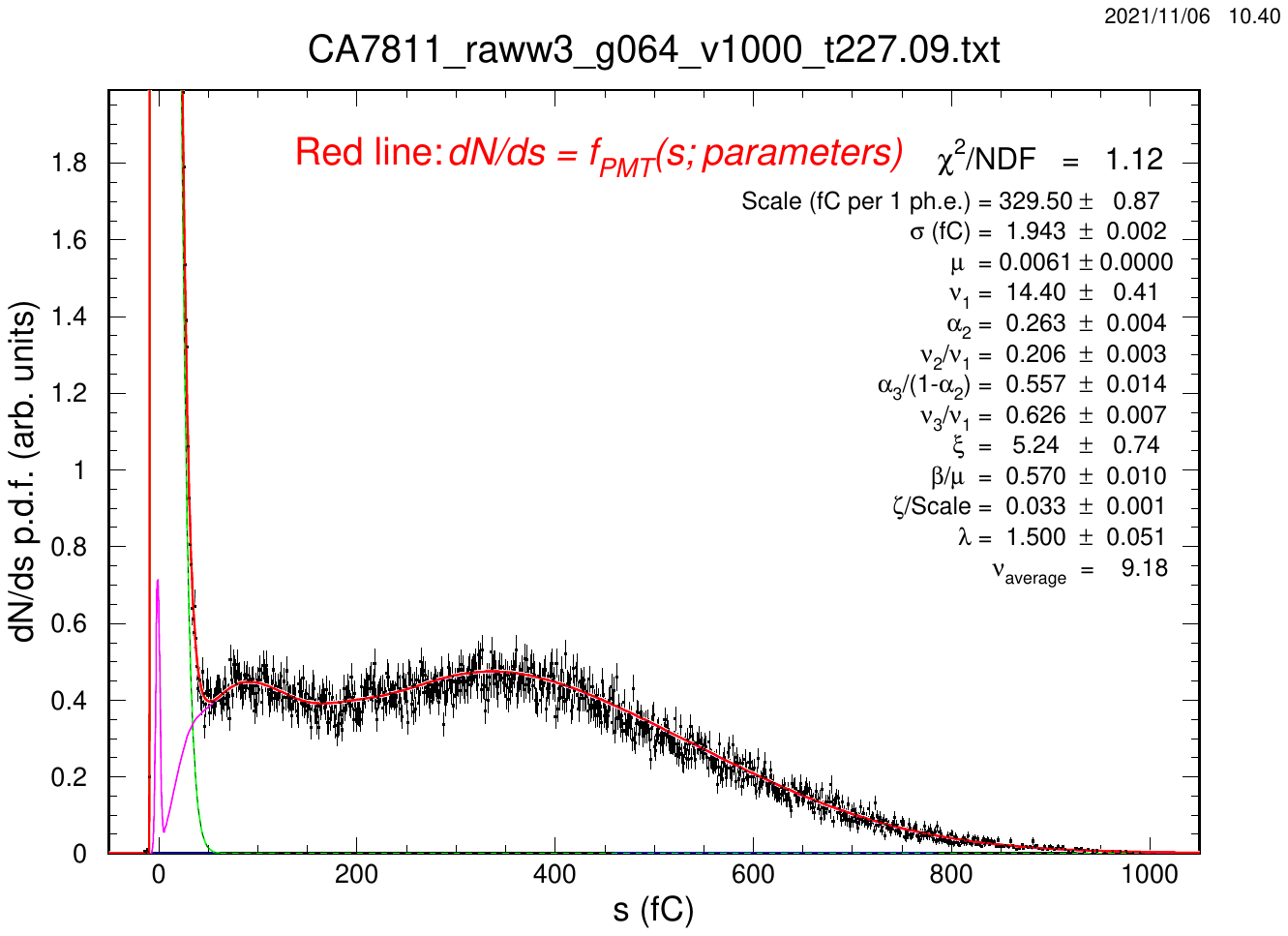}}
  \caption{Signal amplitude probability distributions for MaPMT CA7811 (H8500), pixel 9, at HV = 1000 V. The signal amlitude $s$ is in units of fC, and the measured spectra are shown as black dots with statistical errors. Red lines correspond to the parameterized model charge distributions. Green and violet lines correspond to $m=0$ and $m=1$ functions as explained in Fig.~\ref{fig:Model}. Subplots: (a) 3~mm mask; (b) 6~mm mask; (c) run with full PMT face open with the crosstalk events removed by the correlation analysis; (d) run with full MaPMT face open, with the contribution to the spectrum from the crosstalk events approximated and parameterized by the analysis algorithm. The crosstalk effects in the open configuration are too wide, the fitting algorithm cannot distinguish between the crosstalk and the SPE distribution, and the evaluated SPE function in the (d) plot differs from the ``clean'' one in the (c) plot.
    }
\label{fig:CA7811}
\end{figure*}

\begin{figure*}[hbt] 
\centering 
  \subfloat[3 mm mask]{%
    \includegraphics[clip=true,trim=100 75 140 100,width=.49\textwidth,height=.25\textwidth]
                    {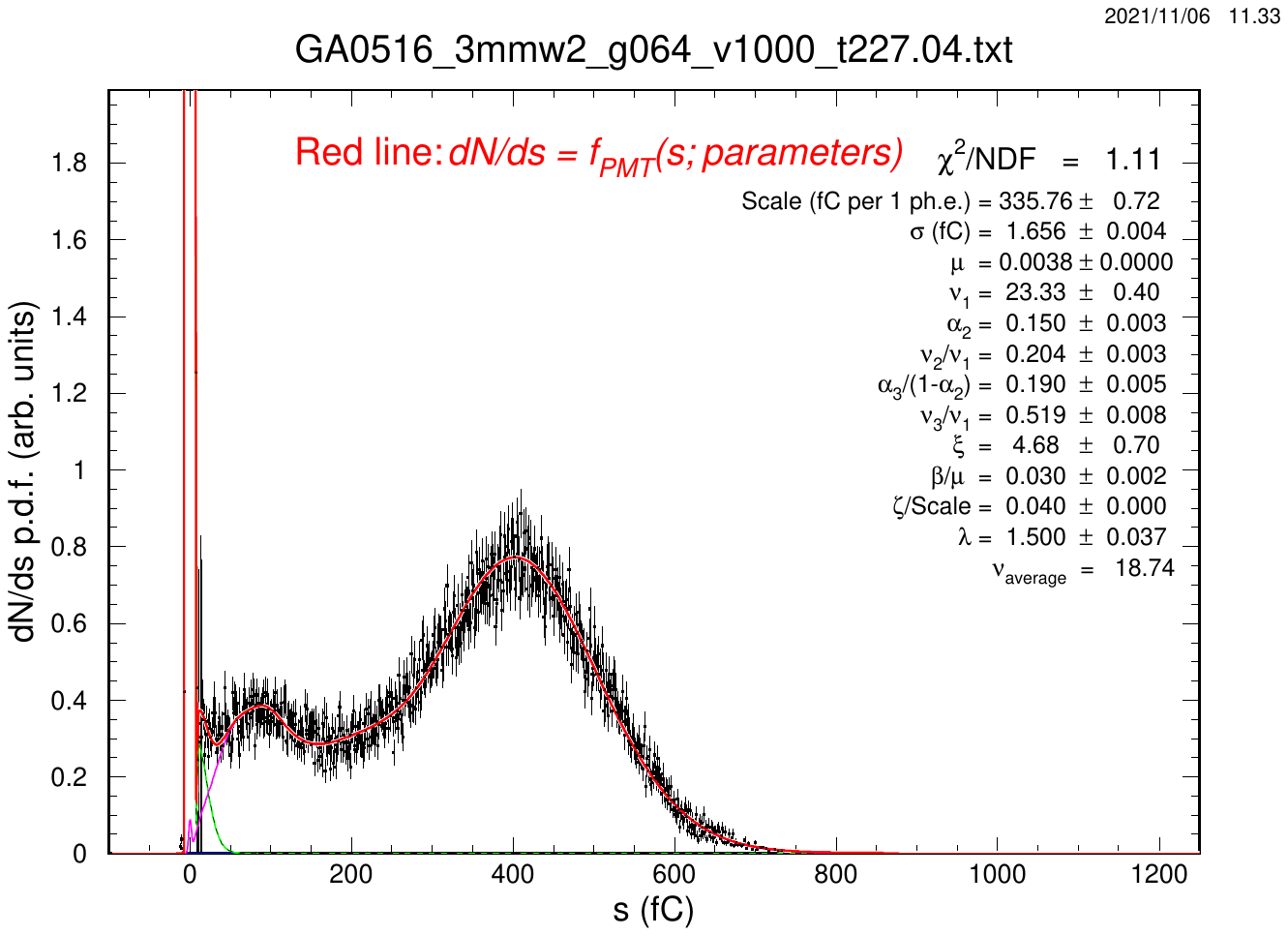}}
  \subfloat[6 mm mask]{%
    \includegraphics[clip=true,trim=100 75 140 100,width=.49\textwidth,height=.25\textwidth]
                    {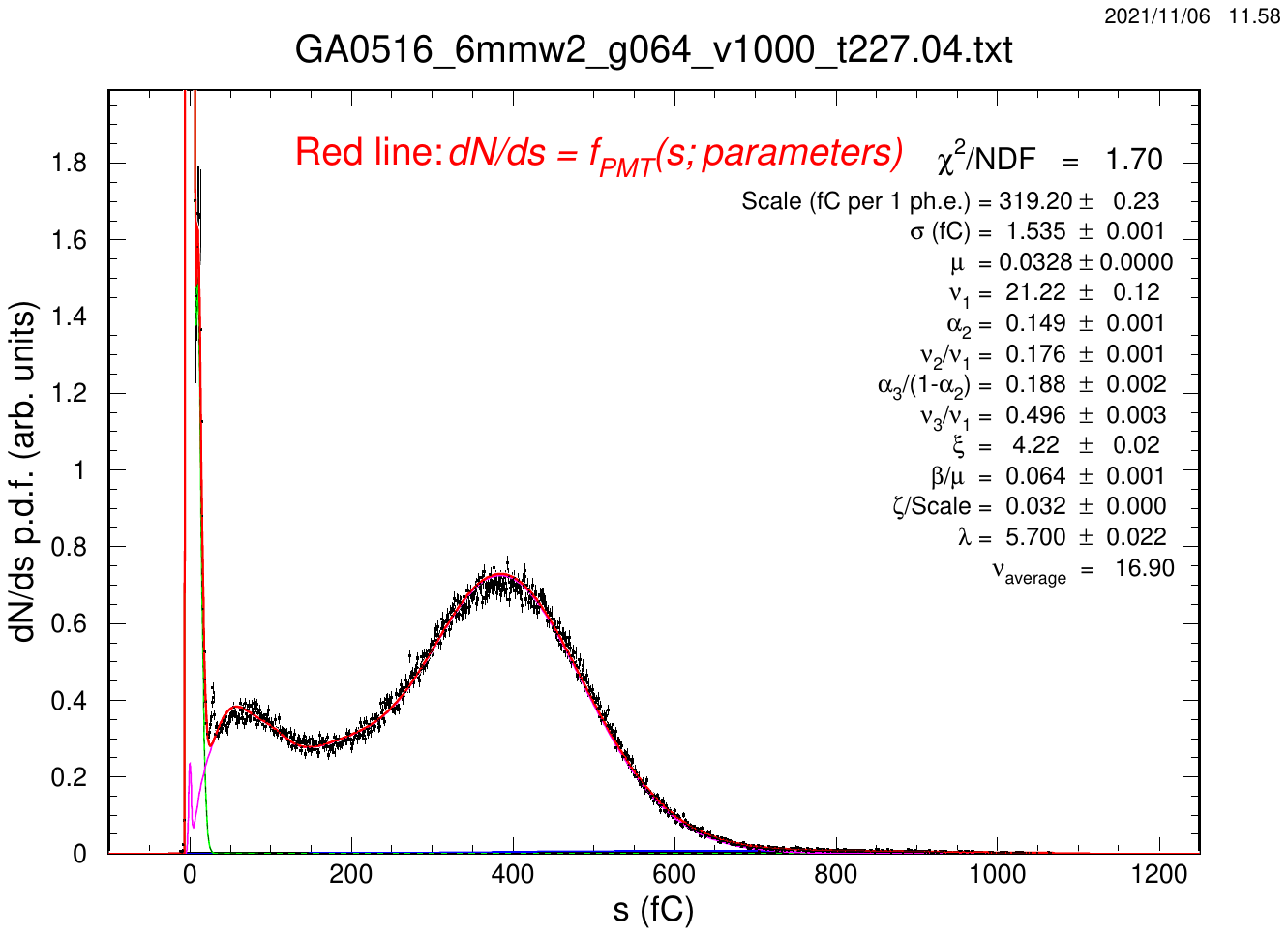}} \\
  \subfloat[No mask, crosstalk removed by software]{%
    \includegraphics[clip=true,trim=100 75 140 100,width=.49\textwidth,height=.25\textwidth]
                    {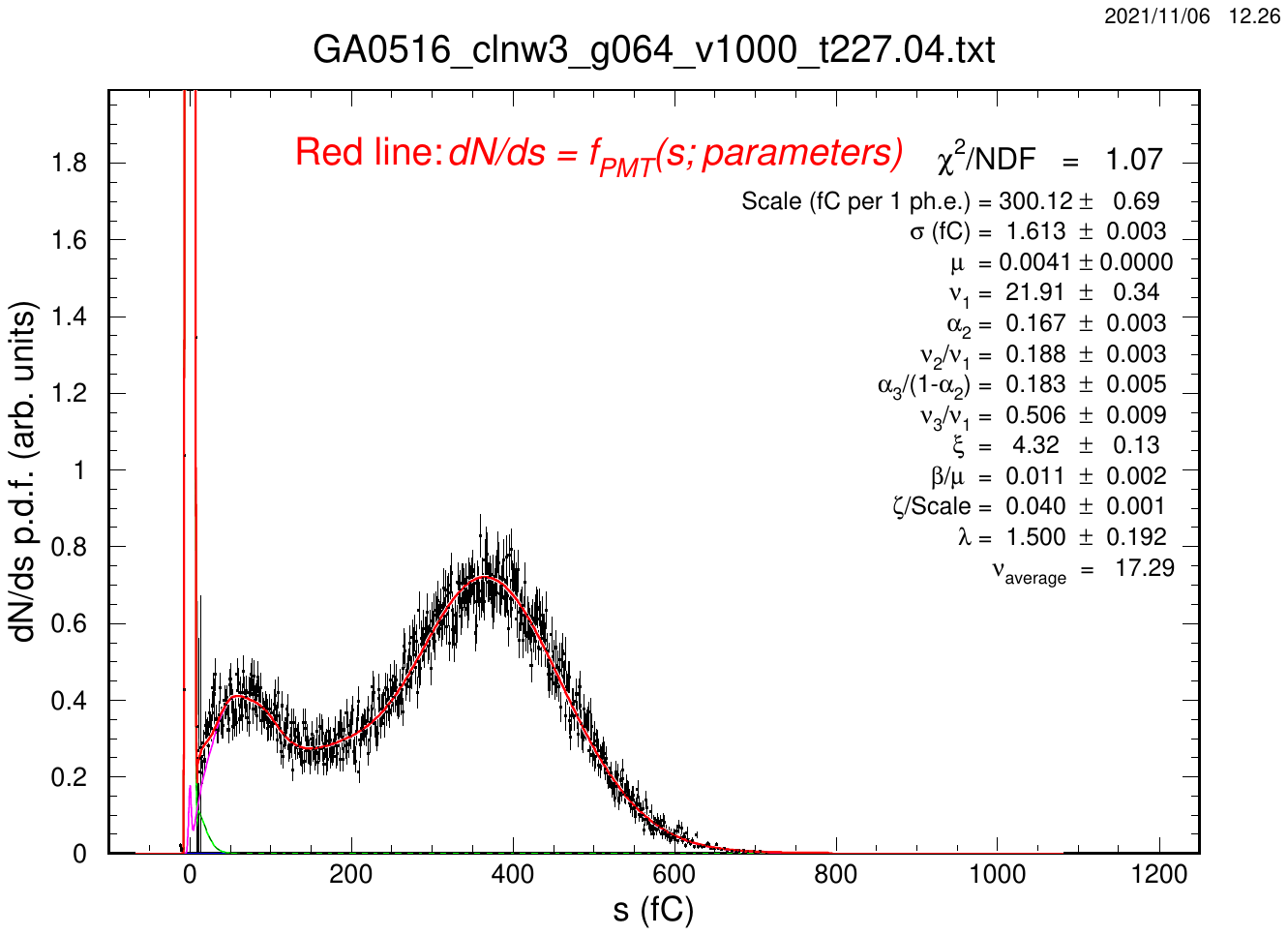}}
  \subfloat[No mask, crosstalk approximated by fit]{%
    \includegraphics[clip=true,trim=100 75 140 100,width=.49\textwidth,height=.25\textwidth]
                    {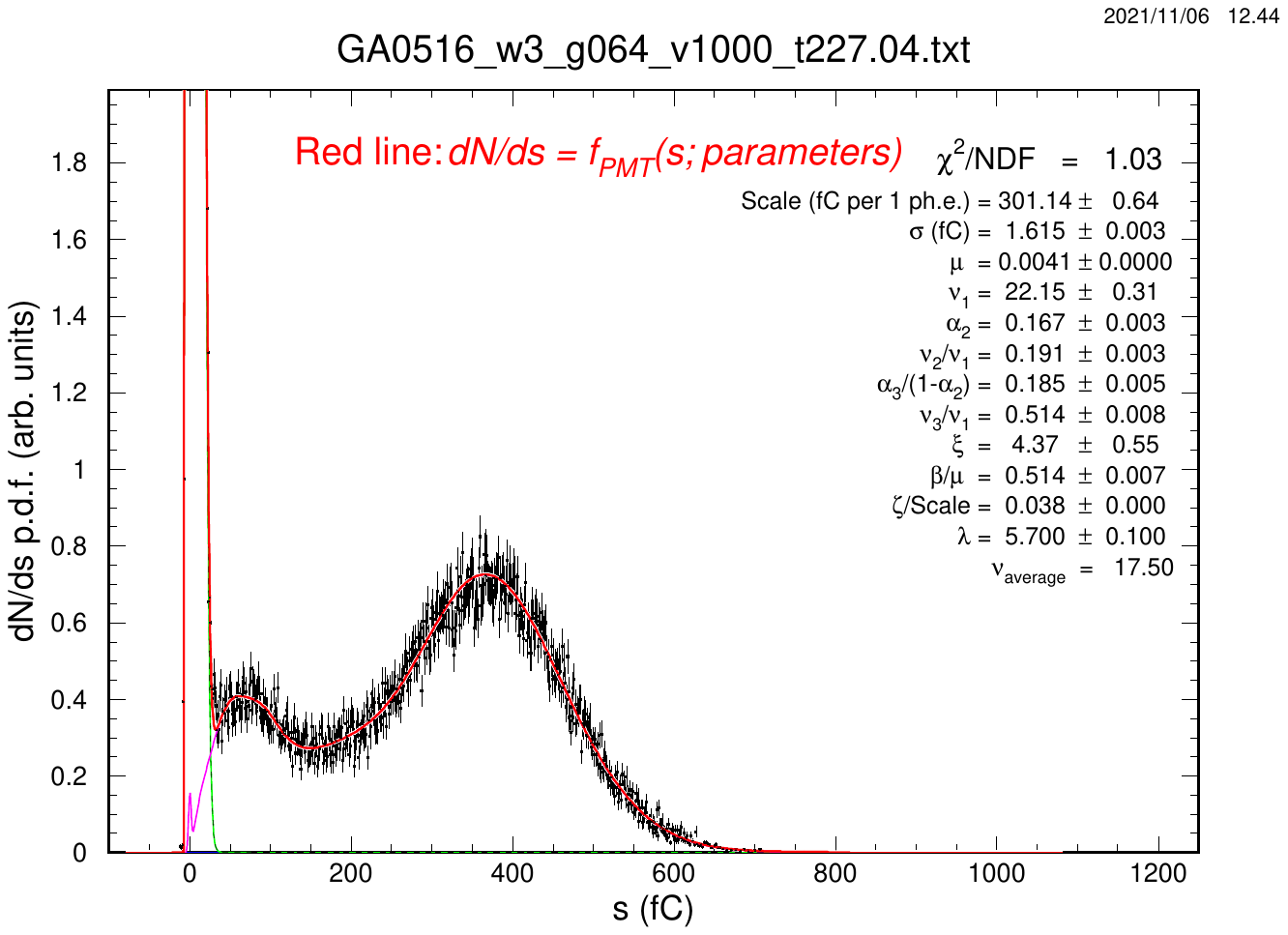}}
  \caption{Signal amplitude probability distributions for MaPMT GA0516 (H12700), pixel 4, at HV = 1000 V. Notation similar to Fig.~\ref{fig:CA7811}. Subplots: (a) 3~mm mask; (b) 6~mm mask; (c) run with full PMT face open with the crosstalk events removed by the correlation analysis; (d) run with full PMT face open with the contribution to the spectrum from the crosstalk events approximated and parameterized by the analysis algorithm.
    }
\label{fig:GA0516_1}
\end{figure*}
\begin{figure*}[h!bt] 
\centering 
  \subfloat[3 mm mask]{%
    \includegraphics[clip=true,trim=100 75 140 100,width=.49\textwidth,height=.25\textwidth]
                    {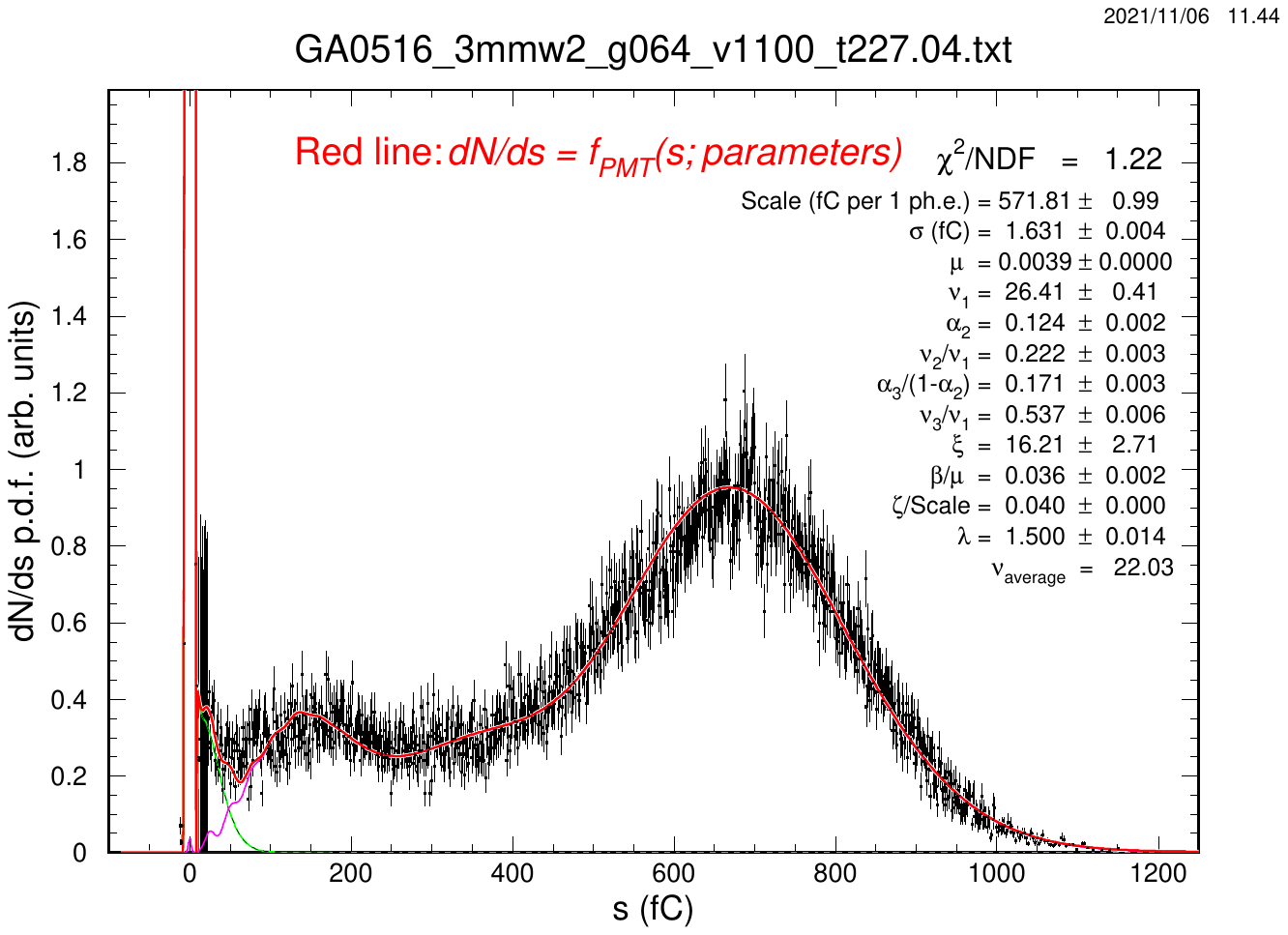}}
  \subfloat[6 mm mask]{%
    \includegraphics[clip=true,trim=100 75 140 100,width=.49\textwidth,height=.25\textwidth]
                    {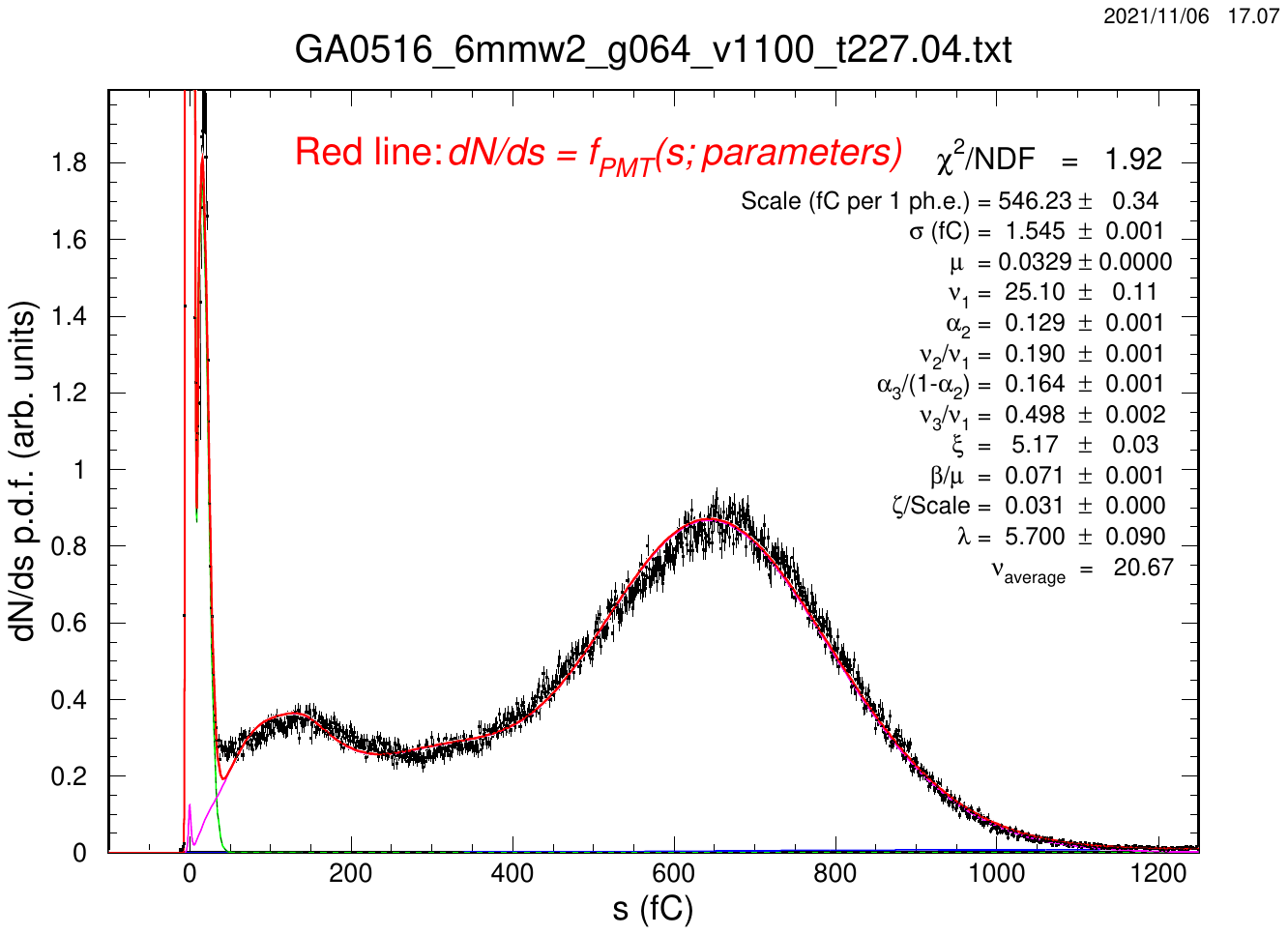}} \\
  \subfloat[No mask, crosstalk removed by software]{%
    \includegraphics[clip=true,trim=100 75 140 100,width=.49\textwidth,height=.25\textwidth]
                    {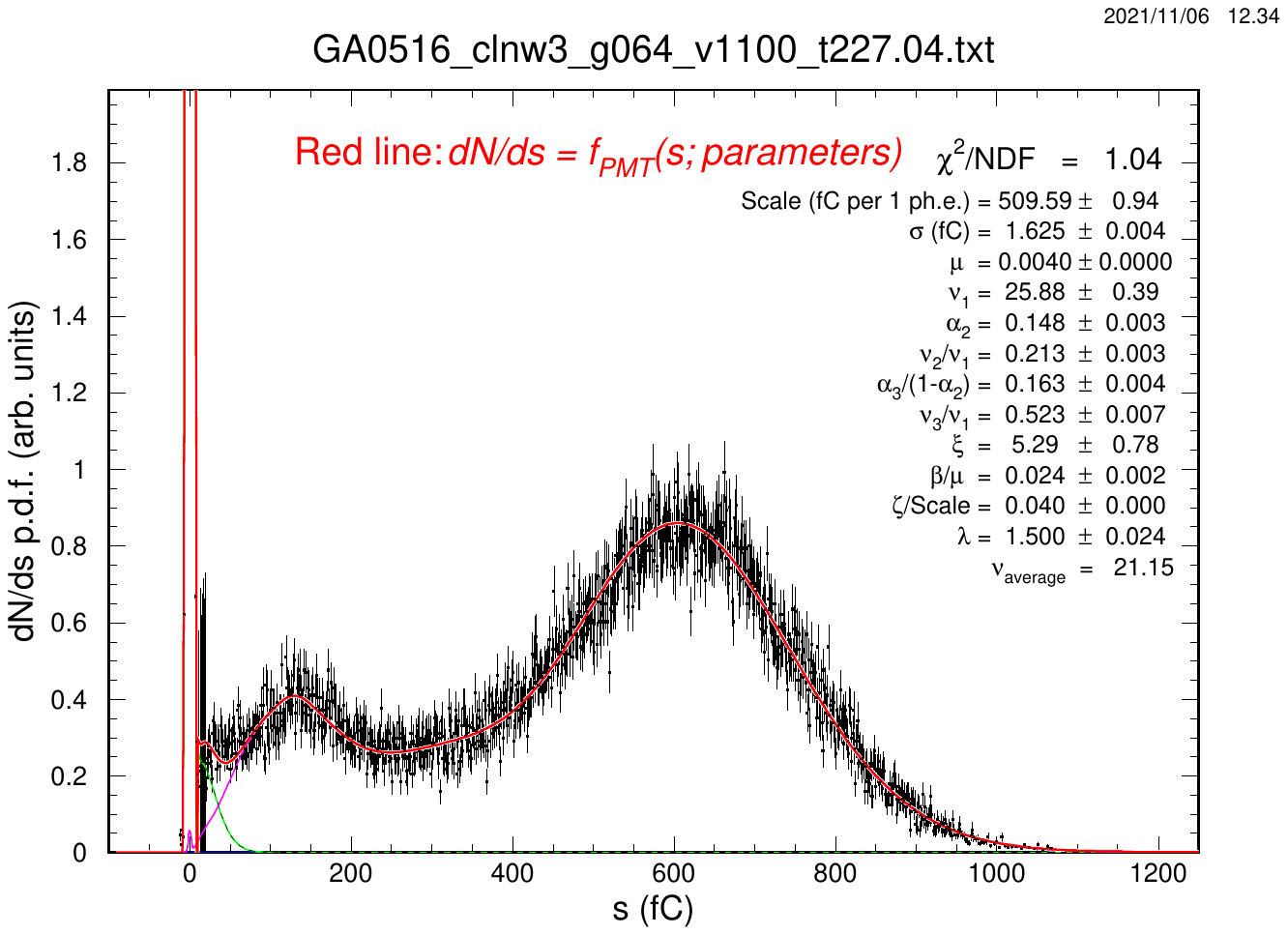}}
  \subfloat[No mask, crosstalk approximated by fit]{%
    \includegraphics[clip=true,trim=100 75 140 100,width=.49\textwidth,height=.25\textwidth]
                    {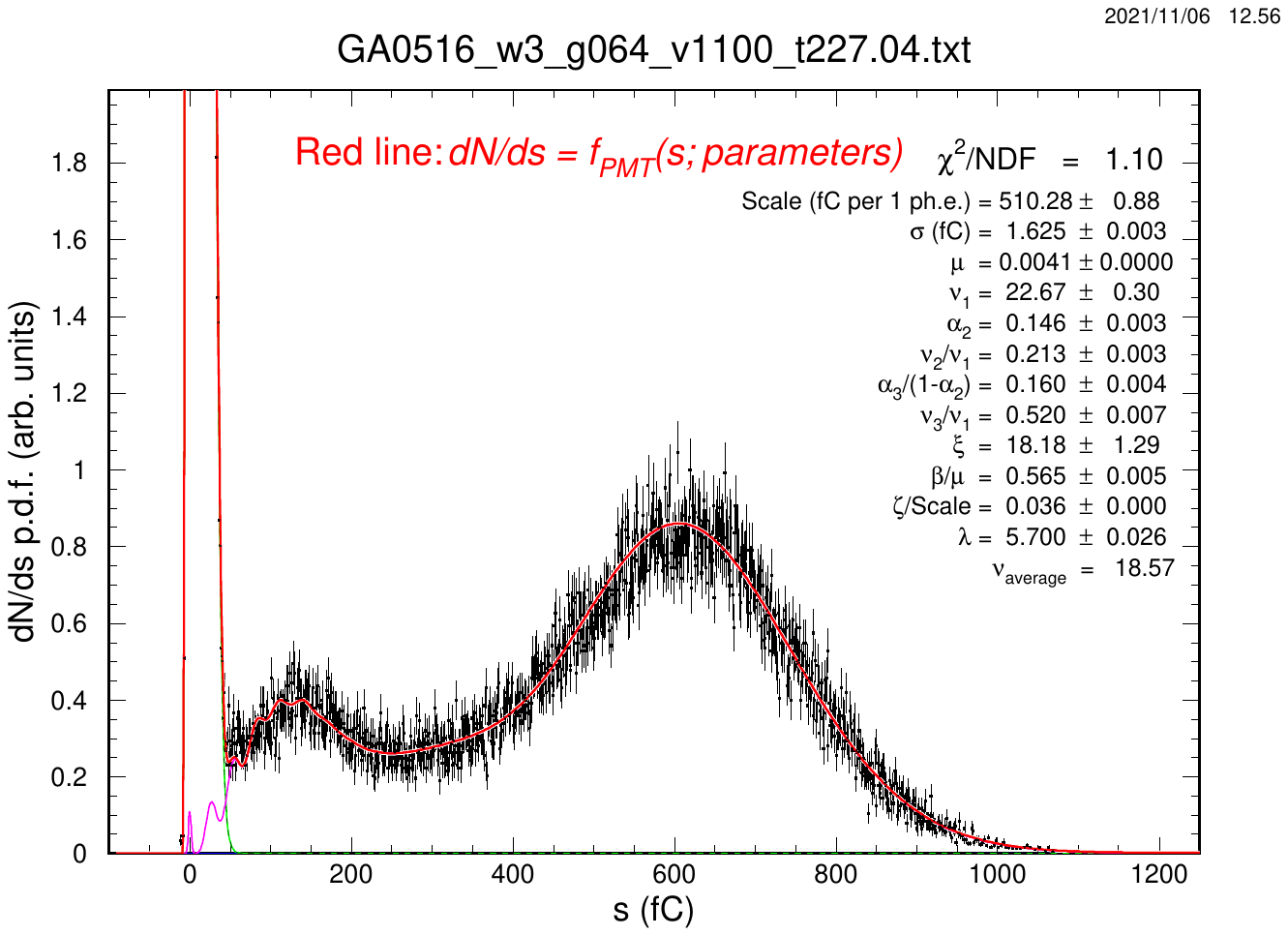}}
  \caption{Same as Fig.~\ref{fig:GA0516_1}, but with all the data taken at HV = 1100~V. 
    }
\label{fig:GA0516_2}
\end{figure*}
\begin{figure*}[h!bt] 
\centering 
  \subfloat[6 mm mask, at HV = 1000 V]{%
    \includegraphics[clip=true,trim=100 75 140 100,width=.49\textwidth,height=.24\textwidth]
                    {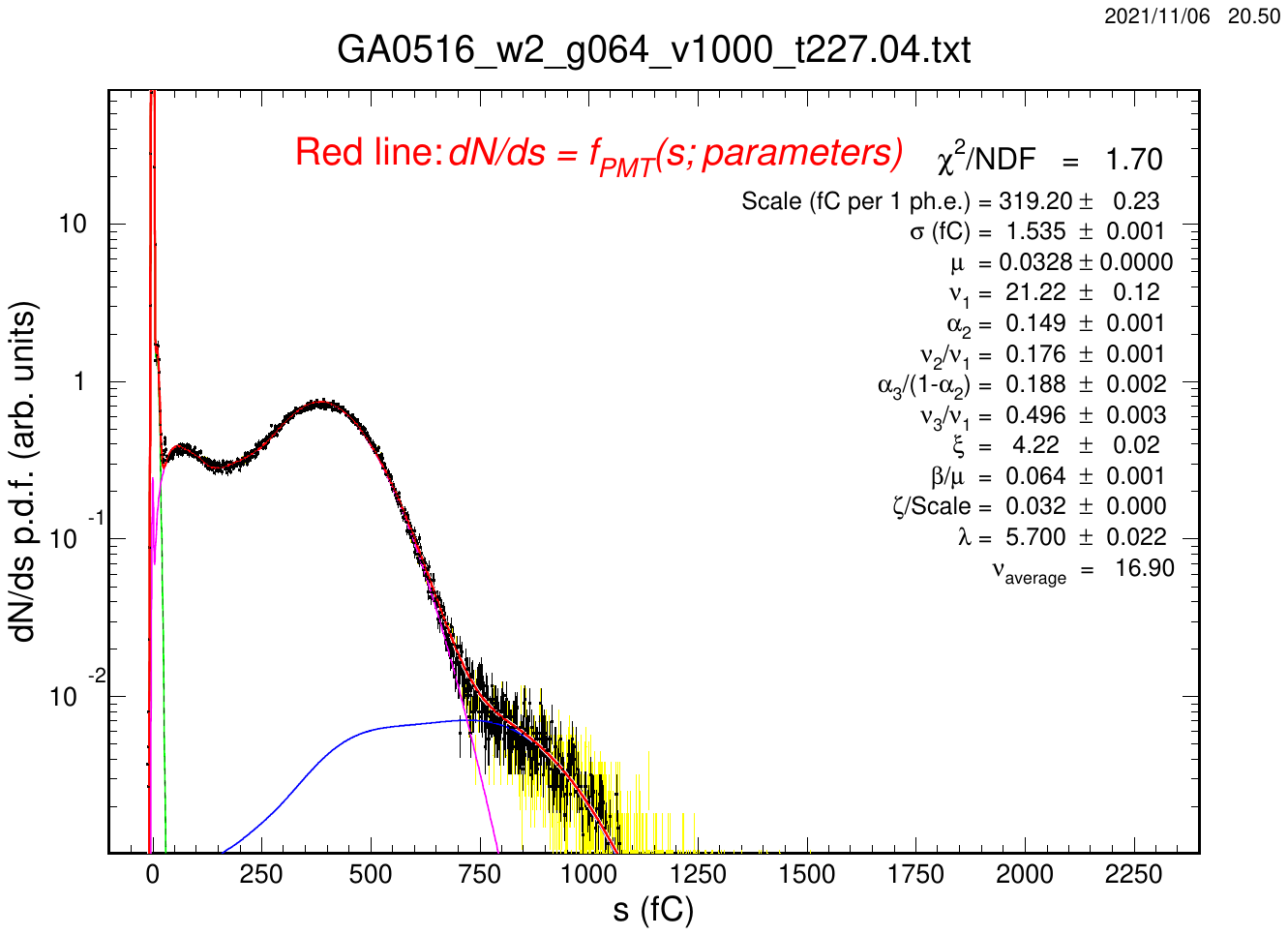}}
  \subfloat[6 mm mask, at HV = 1100 V]{%
    \includegraphics[clip=true,trim=100 75 140 100,width=.49\textwidth,height=.24\textwidth]
                    {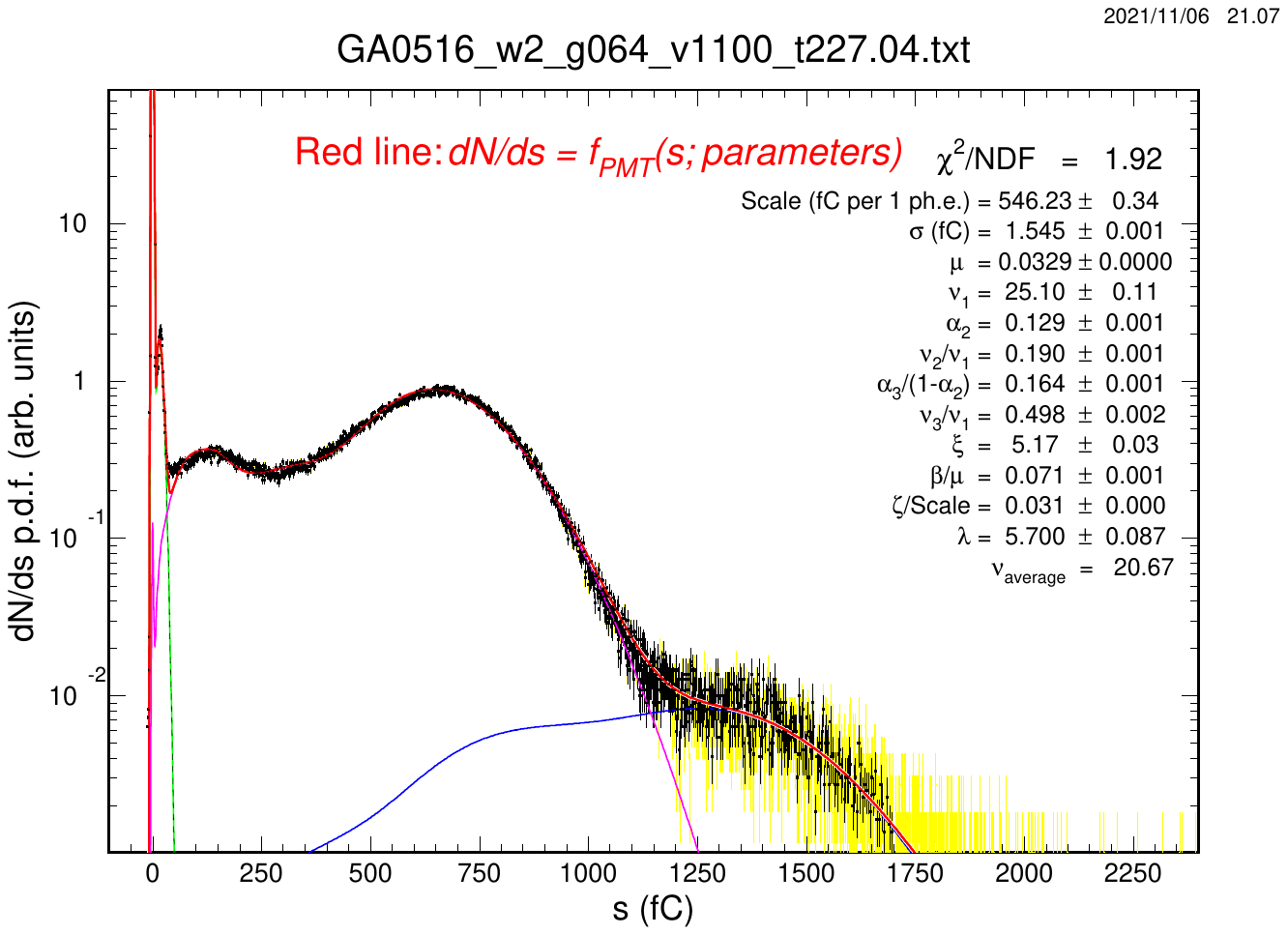}} \\
  \subfloat[No mask, at HV = 1000 V]{%
    \includegraphics[clip=true,trim=100 75 140 100,width=.49\textwidth,height=.24\textwidth]
                    {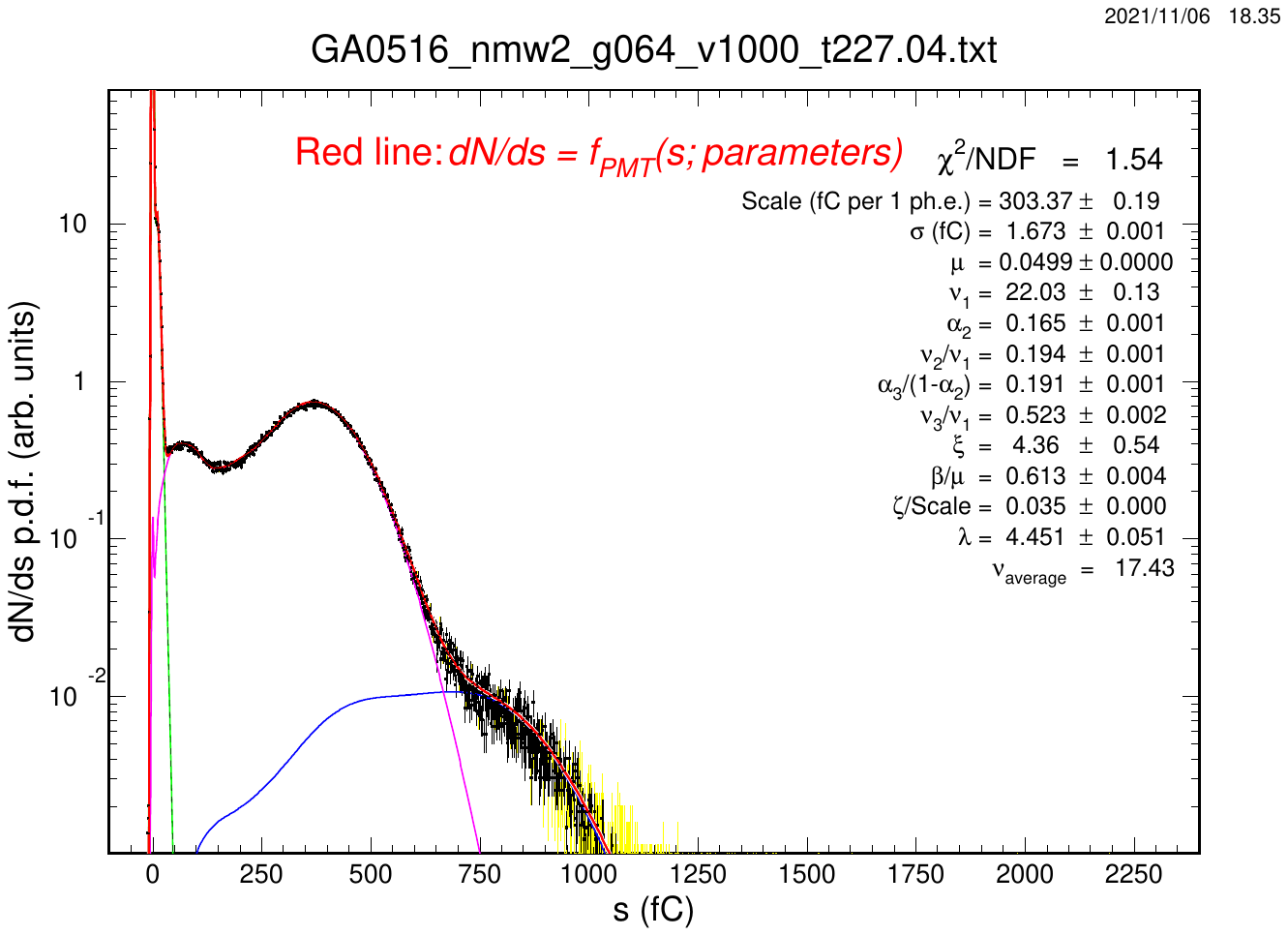}}
  \subfloat[No mask, at HV = 1100 V]{%
    \includegraphics[clip=true,trim=100 75 140 100,width=.49\textwidth,height=.24\textwidth]
                    {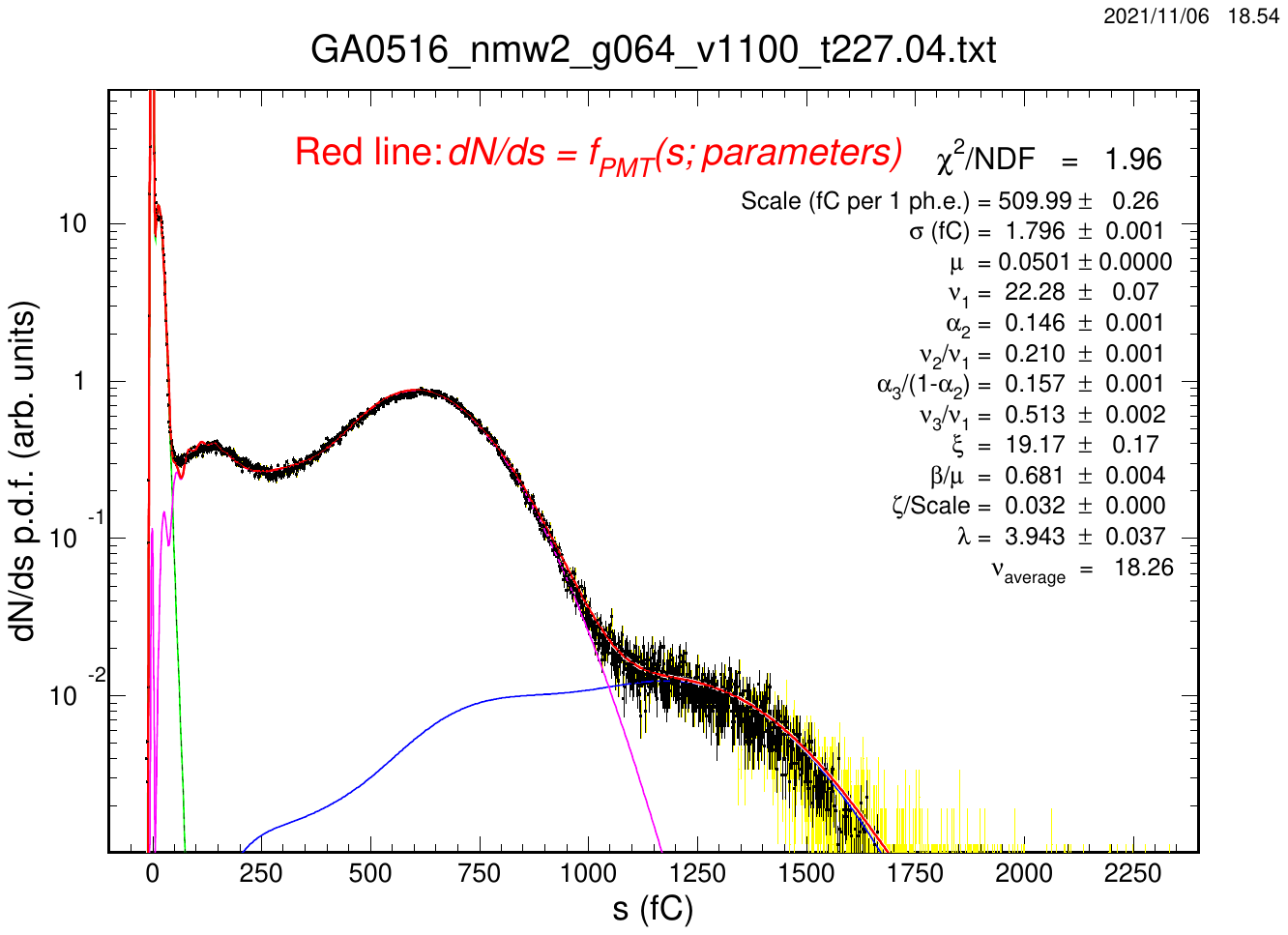}}
  \caption{Signal amplitude probability distributions for PMT GA0516 (H12700), pixel 4, medium light intensity, at HV = 1000 V ((a) and (c)) and at HV = 1100 V ((b) and (d)). Notation similar to Fig.~\ref{fig:CA7811}. To avoid statistical instabilities in the fitting procedure bins with low statistics at high amplitudes (shown by the yellow histogram) were combined and averaged to provide better Gaussian spread (black points with errors).  Subplots: (a) and (b) run with 6 mm mask covering the full PMT face except pixel 4; (c) and (d) run with full PMT face open with the contribution to the spectrum from the crosstalk events approximated and parametrized by the analysis algorithm. Contributions to the spectra are shown by colors: red is the single photoelectron, blue - two or more photoelectrons, green-black dashed line shows the measurement function including the pedestal Gaussian and the crosstalk contribution. 
    }
\label{fig:GA0516_3}
\end{figure*}
\begin{figure*}[h!bt] 
\centering 
  \subfloat[6 mm mask, at HV = 1000 V]{%
    \includegraphics[clip=true,trim=100 75 140 100,width=.49\textwidth,height=.24\textwidth]
                    {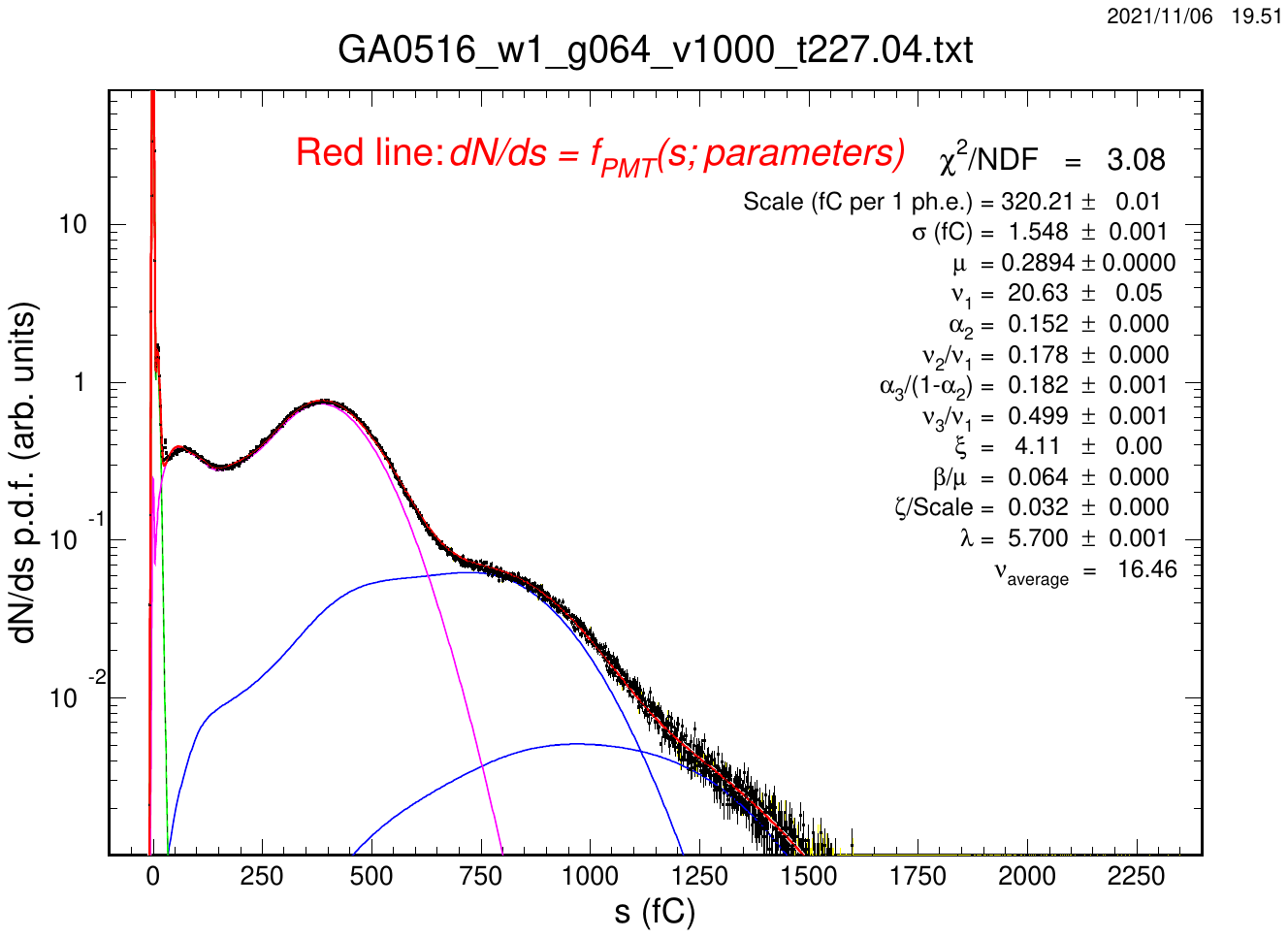}}
  \subfloat[6 mm mask, at HV = 1100 V]{%
    \includegraphics[clip=true,trim=100 75 140 100,width=.49\textwidth,height=.24\textwidth]
                    {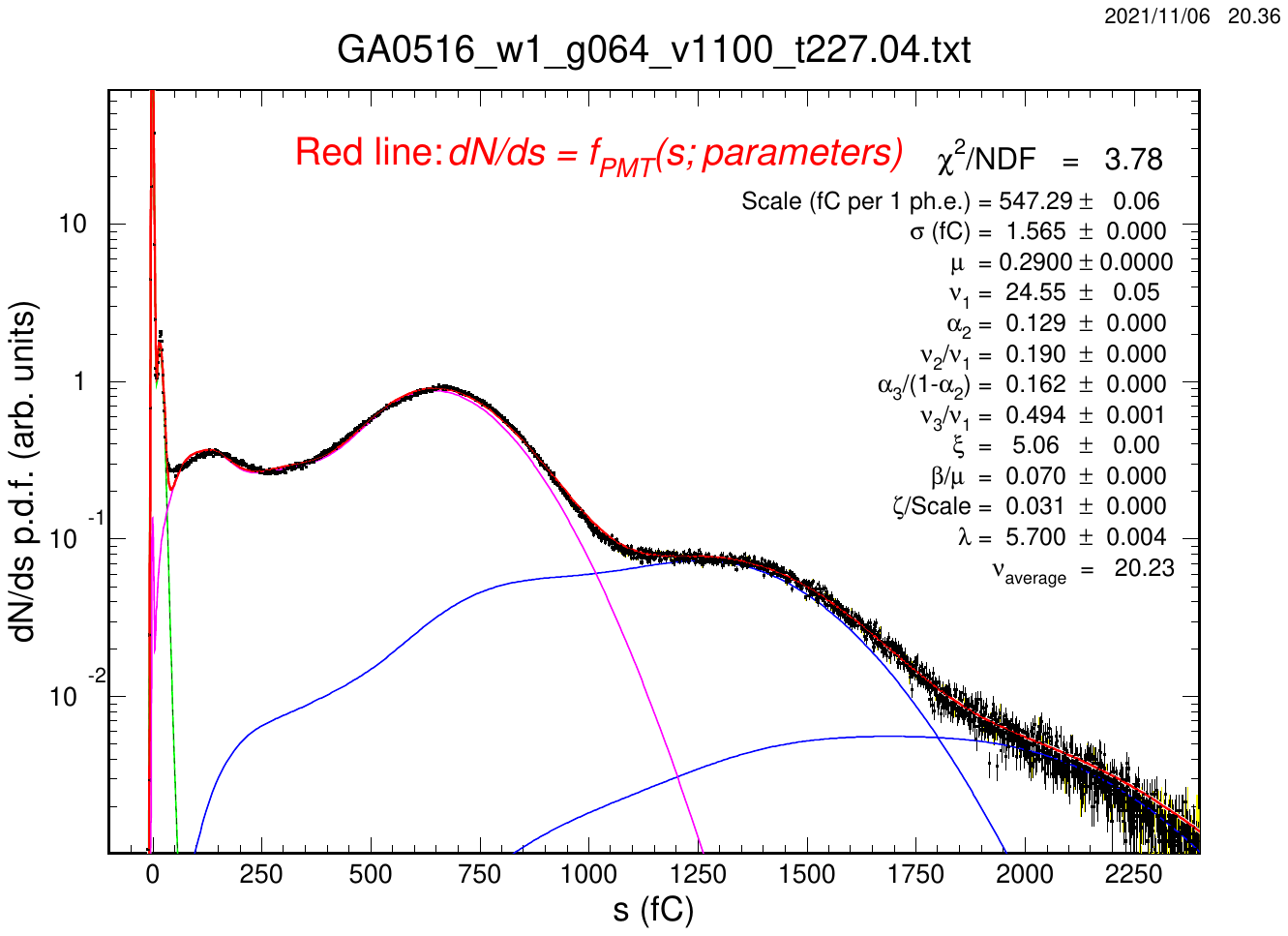}} \\
  \subfloat[No mask, at HV = 1000 V]{%
    \includegraphics[clip=true,trim=100 75 140 100,width=.49\textwidth,height=.24\textwidth]
                    {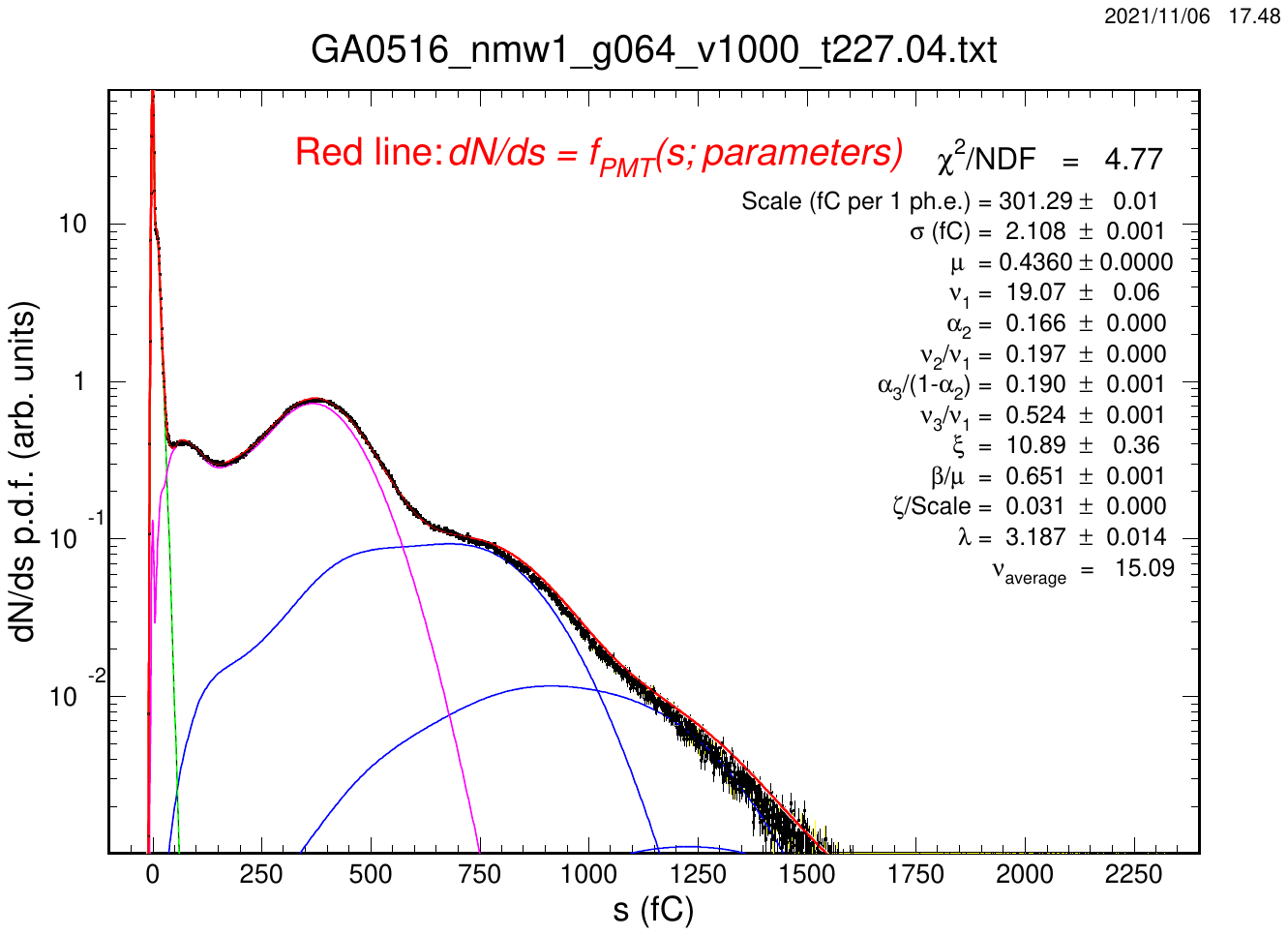}}
  \subfloat[No mask, at HV = 1100 V]{%
    \includegraphics[clip=true,trim=100 75 140 100,width=.49\textwidth,height=.24\textwidth]
                    {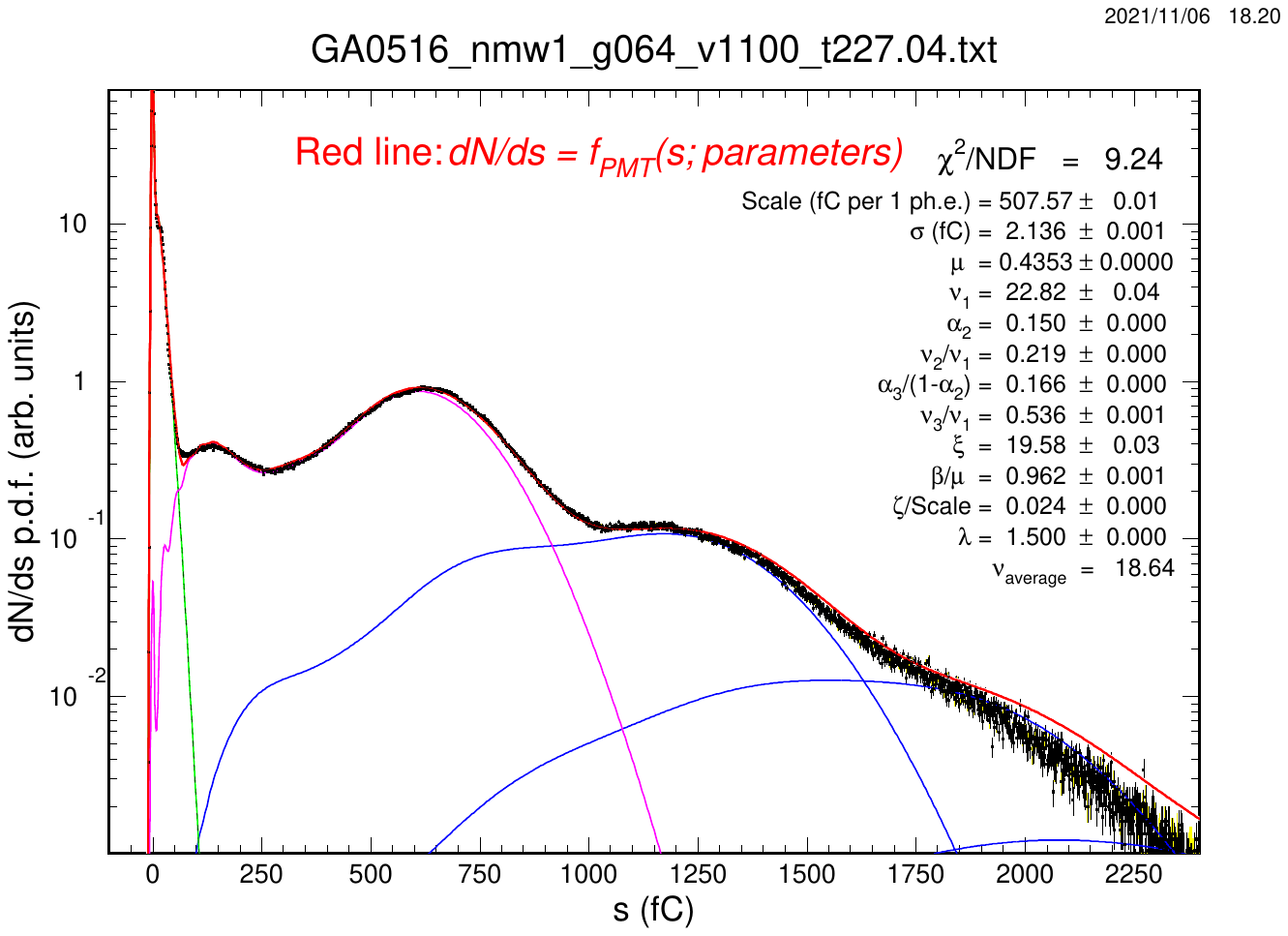}}
  \caption{Same as Fig.~\ref{fig:GA0516_3}, but at the light intensity approximately 10 times higher. Green, purple, and violet lines correspond to m = 0, m = 1, and m = 2, 3,... functions as explained in Fig.~\ref{fig:Model}. Higher formal $\chi^2/NDF$ values are due to very high number of events in the plots.
    }
\label{fig:GA0516_4}
\end{figure*}

\begin{figure*}[!ht]
	\centering
	\includegraphics[width=1.0\textwidth,height=.7\textwidth]{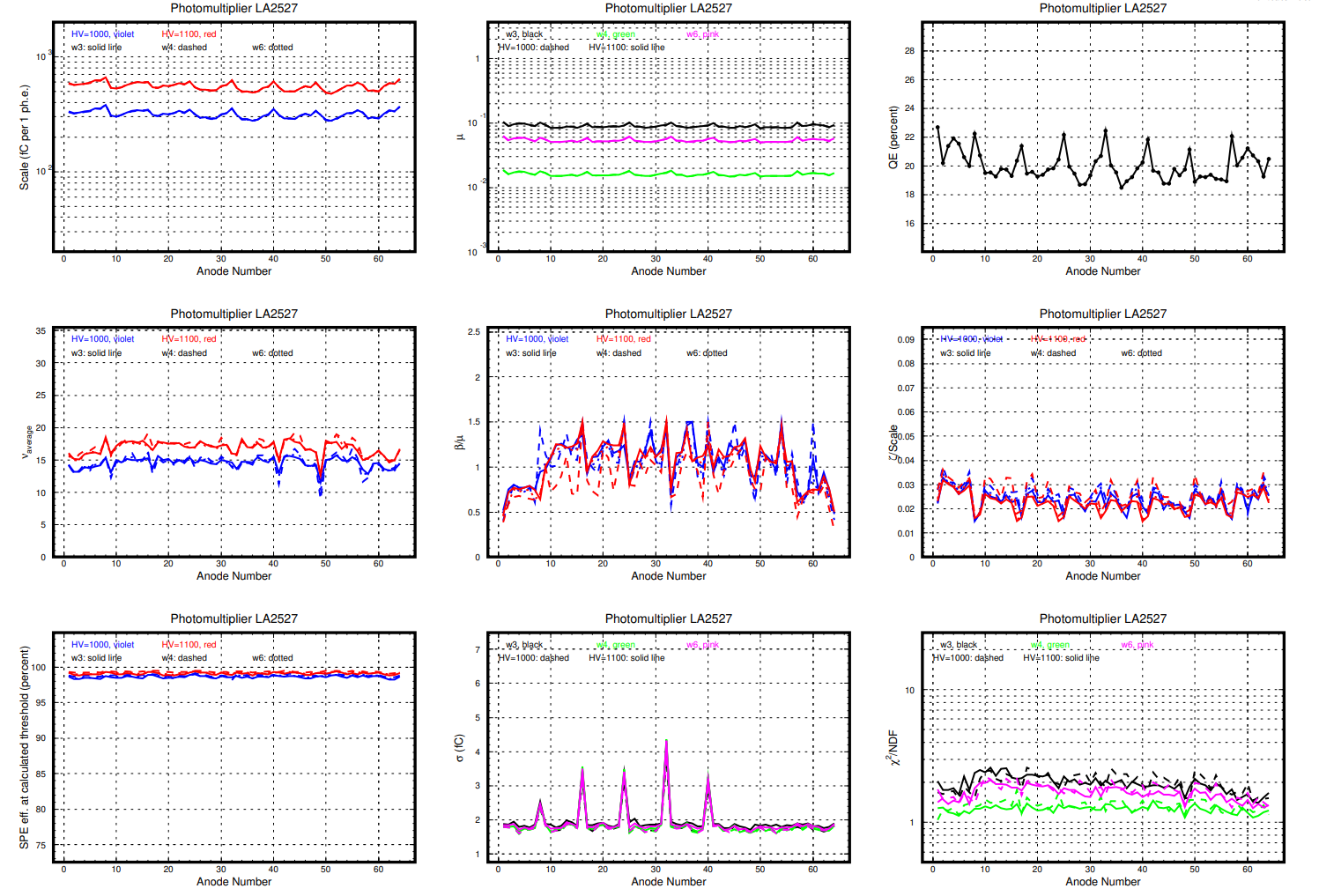}
	\caption{Illustration of the ``MaPMT passport'' plots for one of the MaPMTs, LA2527 (H12700). The standard six measurements included runs at three illumination settings (wheel positions 3, 4, and 6), each at two operating high voltage values (1000~V and 1100~V). The formal statistical errors from the minimization routine are too small to be visible in the plots. The systematical errors are evaluated comparing independent measurements of each pixel at different conditions, not shown in the plot and discussed further in the text.
	}
	\label{fig:LA2527_passport}
\end{figure*}
\begin{figure*}[!ht]
	\centering
	\includegraphics[width=1.0\textwidth,height=.7\textwidth]{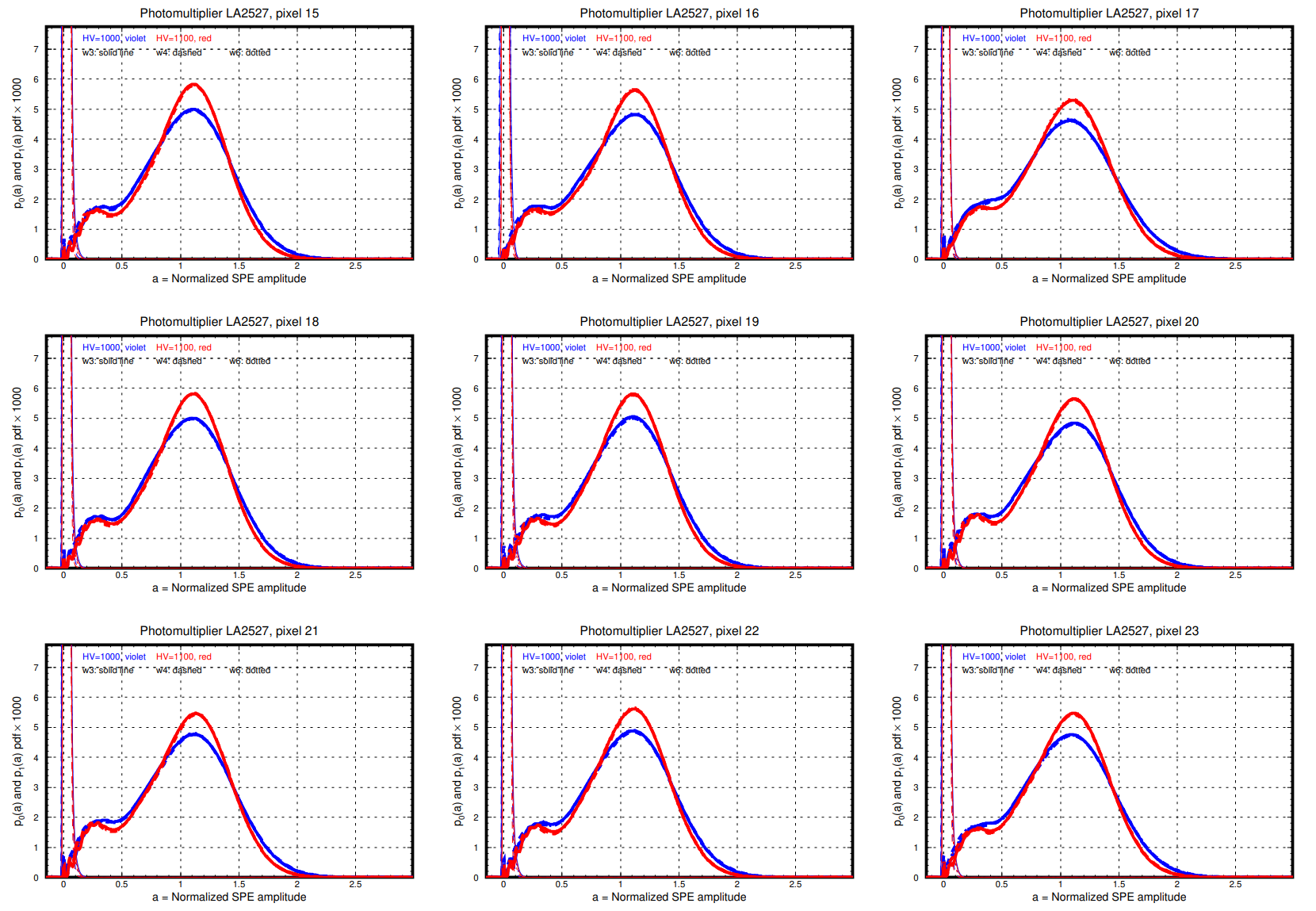}
	\caption{Illustration of the ``MaPMT passport'' plots for one of the PMTs, LA2527 (H12700), continued. The standard six measurements included runs at three illumination settings (wheel positions 3, 4, and 6), each at two operating high voltages (1000~V, and 1100~V). Shown are the calculated SPE probability distribution functions $p_1(a)$, defined by the fit parameters resulting from the independent fitting procedures for each of the six settings. The blue color corresponds to the three sets at HV = 1000~V, and red - to the sets at HV = 1100~V. The parameters of the independent fits at three different illuminations result in very stable SPE shapes, practically indistinguishable in the plot. The measurement functions $p_0(a)$ are shown as peaks around the pedestal at $a=0$ with the left sharp edge width corresponding to $\sigma$, and the right edge determined by the crosstalk.}
	\label{fig:LA2527_passport_spectra}
\end{figure*}

As a demonstration of the characterization procedure for the MaPMTs, Figs.~\ref{fig:CA7811}-\ref{fig:GA0516_4} show the measured signal amplitude probability distributions for one H8500 MaPMT pixel (CA7811, pixel 9) and one H12700 MaPMT pixel (GA0516, pixel 4) under various conditions, as well as their respective fit results. 
Figure ~\ref{fig:CA7811} and Fig.~\ref{fig:GA0516_1} illustrate the effect that the electronic crosstalk from neighboring pixels has on the measured SPE fit parameters. 
We collected two sets of data intended to reduce the contribution of crosstalk from neighboring pixels. In the first (as described in Section 4) we used a black sheet of paper to mask all pixels on a single MaPMT and punctured a 3~mm hole over the pixel of interest (see Fig.~\ref{fig:CA7811}a). 
However, with this setup, one cannot fully characterize the unmasked pixel, as there is some dependence of the measured signal on the location of the incident photon. 
To provide full coverage of a single pixel\textquotesingle s surface, another set of measurements was taken with a 6~mm x 6~mm square hole cut out over a single pixel. 
With this configuration, the full face of the pixel of interest was illuminated, while the neighboring pixels remained mostly covered by the black paper. 
However, there is still a non-negligible contribution from crosstalk with this configuration, due to imperfect alignment of the masks. 
This can be clearly seen in Fig.~\ref{fig:CA7811}b which shows the signal amplitude distribution with this 6~mm x 6~mm square hole cut out over pixel 9. 
One can see the contribution of the crosstalk appearing as a shoulder to the pedestal, albeit smaller than the crosstalk shoulder seen in Fig.~\ref{fig:CA7811}d where the full face of the MaPMT was illuminated. 

The resulting SPE fit parameters for Figs.~\ref{fig:CA7811}a-d indicate the inability of the model to fully describe the crosstalk in the H8500 MaPMTs. 
Most notably, in the data sets where the full-face of the MaPMT was illuminated (see Figs.~\ref{fig:CA7811}c-d) the $scale$ parameter changes by almost $7\%$ when the crosstalk is removed by the offline correlation analysis procedure compared to when it is kept in the data. 
Because the $scale$ parameter gives the average charge measured per photoelectron, it should be independent of the crosstalk. 
In contrast, we observe that the crosstalk in the H12700 MaPMTs can indeed be well described by the updated model, as is evident by comparing the fit parameters for Figs.~\ref{fig:GA0516_1}c-d. 
All parameters are consistent between the two fits, despite the fact that the crosstalk was removed by the offline analysis prior to performing the fit for Fig.~\ref{fig:GA0516_1}c. 
This result exemplifies the ability of the model to extract the SPE parameters from the measured signal amplitude distributions in a crosstalk-independent manner. 

The sample comparison between typical H8500 and H12700 MaPMTs as shown in Figs.~\ref{fig:CA7811} and \ref{fig:GA0516_1} generally confirms our decision to switch to H12700 as the MaPMT of choice for the RICH detector. In the previous study (Ref.~\cite{DEGTIARENKO20171}), using a different electronics front-end and data acquisition system, we observed that the values of the $\nu_{average}$ parameters were generally much smaller for H8500 than for the H12700, leading to a significant improvement of the expected efficiency of the H12700 MaPMTs to SPE events. In the previous study the amplitude resolution was not good enough to uncover the additional difference between the two models: the crosstalk spectra are significantly wider in the H8500, decreasing the expected SPE efficiency further, as compared to H12700. Wide crosstalk distributions in the H8500 overlap noticeably with the shapes of the model SPE functions and do not allow the model to isolate them, while for the H12700 MAPMTs the separation between the crosstalk and SPE distributions is reliable.  

The same sets of data were taken with the H12700 MaPMT high voltage set to 1100~V to compare with the results of Fig.~\ref{fig:GA0516_1} which were taken at 1000~V. 
The resulting amplitude probability distributions and fits are shown in Fig.~\ref{fig:GA0516_2}. 
As expected, both the $scale$ and $\nu_{average}$ parameters are larger when the high voltage is increased to 1100~V, while the parameters describing the crosstalk, $\beta/\mu$ and $\zeta/scale$, are fairly consistent. 
Furthermore, by comparing Fig.~\ref{fig:GA0516_2}c and Fig.~\ref{fig:GA0516_2}d, we observe the same desirable characteristic that the SPE fit parameters are consistent with or without the offline removal of the crosstalk events from the data even at a larger high voltage setting.

Finally, Fig.~\ref{fig:GA0516_3} and Fig.~\ref{fig:GA0516_4} show the signal amplitude probability distributions for the same pixel on MaPMT GA0516 at higher illumination intensities. 
Specifically, Fig.~\ref{fig:GA0516_3} shows the results with new light intensity for high voltage settings 1000~V and 1100~V, both with the full MaPMT face illuminated, and with the 6~mm x 6~mm square hole mask cutout applied. 
Comparing Fig.~\ref{fig:GA0516_3}c to Fig.~\ref{fig:GA0516_1}d (full-face illumination, 1000~V), the $\mu$ parameter is almost a factor of 10 larger for the data collected with the new light intensity, but the characteristic parameters for the SPE response are consistent. 
The same can be said by comparing to the signal amplitude probability distribution in Fig.~\ref{fig:GA0516_4}c, which was measured at higher illumination.
Even at roughly 100 times the light intensity, the resulting $scale$ parameter is consistent to the one measured at low light intensity. Such consistency brings about the confidence in the bulk model approximation results, their independence on the pixel-to-pixel variability of the measurement conditions, and allows evaluation of the systematical errors, as it will be discussed further in the text.

Figure~\ref{fig:LA2527_passport} shows an example of the ``passport'' plots obtained for a single MaPMT - in this case, an H12700 MaPMT labeled LA2527. 
Each plot shows different parameters extracted from the fits to the signal amplitude probability distributions vs. the pixel number, resulting in 64 data points per curve.
In all plots (excluding the top-right plot), the fit results are compared for the data taken with wheel positions 3, 4, and 6, and high voltages 1000~V and 1100~V (6 different configurations in total).
The wheel positions 4, 6 and 3 correspond to the increasing relative light intensities of 0.18:0.60:1.
As expected, the $scale$ and $\nu_{average}$ parameters are independent of the light intensity, but change with the applied high voltage. 
This is due to the increased amplification at each dynode at higher applied voltages. The values of the extracted $scale$ parameters are identical when obtained in the independent experiments with different light intensity. Similarly the independence of extracted $\mu$ parameters on the value of high voltage applied can be used in evaluating the consistency of the measurement and the systematic error. 
The $\beta/\mu$ and $\zeta/scale$ parameters that describe the crosstalk from neighboring photoelectrons remain somewhat consistent between the different experimental configurations. 
However, the $\beta/\mu$ passport plot shows the dependence of the relative probability of crosstalk on pixel location. For example, the first 8 and last 8 pixels all have significantly lower $\beta/\mu$ parameters. These pixels are along the edge of the MaPMT and therefore have (at least) one fewer neighboring pixel than those in the center of the MaPMT.
Consequently, the $\beta$ parameter for the amplitude probability distributions in these pixels is lower. 

The measurement of the absolute photon flux on each pixel was discussed in Section 5. 
The stability of the light flux was demonstrated by running the same PMT many times during the characterization.
The QE is obtained for each pixel by relating the light flux measurement to the average number of photoelectrons measured per laser pulse, $\mu$, which is extracted separately for each pixel as a parameter of the fit to the signal amplitude probability distribution. The resulting QE distribution is shown in the top-right plot of Fig.~\ref{fig:LA2527_passport}. These results indicate that on average the QE for each pixel of the H12700 MaPMTs is about 21$\%$ for incident photons with wavelength 470~nm. Generally, we observe significant pixel-to-pixel spread of various characterization parameters in every MaPMT, within the specifications. We believe the spread is inevitable in the manufacturing process.

The lower-right plot illustrates the quality of the SPE fit by showing the standard $\chi^2/NDF$ values for every fit, calculated for all bins in the measured spectrum with amplitudes above threshold. The accumulated number of events in each measured spectrum was very high and it is hard to expect an ideal model description with $\chi^2/NDF = 1$. The statistical quality of the fit was reasonably good for all measured spectra.

One final remark from the plots included in Fig.~\ref{fig:LA2527_passport} is that the SPE efficiency shown in the lower-left plot is slightly larger at 1100~V than at 1000~V. The efficiency was defined as the percent of SPE events above the threshold, which, in turn, was defined as the amplitude at which the number of events in the SPE distribution below the threshold was equal to the number of events in the crosstalk spectrum above it. The higher voltage leads to increased separation between the SPE spectra and the pedestal, corresponding to larger values of $\nu_{average}$, and thus increasing the efficiency. 

Figure~\ref{fig:LA2527_passport_spectra} shows the extracted SPE functions for 9 pixels on the same MaPMT, again for all 6 configurations. The probability distributions are given as a function of the normalized charge amplitude, $a$. The functions extracted from the data measured at 1100~V are noticeably more narrow around the peak than the data collected at 1000~V, in agreement with the previously noticed differences between the values of $\nu_{average}$ and the efficiency at the different high voltages. The plots also illustrate the pedestal measurement functions around $a=0$, including the crosstalk contributions. The pedestal functions and the SPE functions measured independently at three illumination settings visibly overlap, and thus illustrate the stability of the fitting procedure and validate the applicability of the model in its function to objectively extract the MaPMT characteristics.

\section{Results}

This section reports on the study of 399 H12700 MaPMTs, acquired for the CLAS12 RICH2 detector upgrade. Each of them was tested in the same conditions by groups of six mounted in the MAROC tiles and irradiated simultaneously. The test procedure included six different setup conditions: two sets of applied high voltage (1000~V and 1100~V), and three laser light intensity settings at wheel positions 3, 4, and 6. The data were accumulated and pre-processed to make the non-linearity corrections and to convert the amplitudes into units of electric charge. After that the data were transferred to the ``parameterization factory'' computer workstation in which every accumulated spectrum was automatically analyzed and approximated with the 12-parameter fitting function, as was explained earlier. Each MaPMT was issued a ``passport'' document listing the fit parameters for every measurement for all 64 anodes, showing the extracted SPE functions, and the parameter dependencies on pixel number, as illustrated in Figs.~\ref{fig:LA2527_passport} and \ref{fig:LA2527_passport_spectra}. The most important parameters extracted from the analysis for every pixel were i) $scale$, which measured the average charge collected at the anode from the single photoelectron events, ii) the average multiplicity $\mu$ of the photoelectrons per laser pulse, which can be converted to the quantum efficiency of the pixel when normalized to the calibrated incoming light in the pulse, iii) the calculated optimal threshold value for the separation of the single photoelectron events from the pedestal (including the crosstalk background), and iv) the corresponding estimate of the photodetection efficiency based on that value. The parameters of interest are also the characteristics of the photomultiplier, such as i) the gain on the first dynode evaluated in the model, ii) the amplitude width, and iii) the intensity of the crosstalk signal. The pedestal $\sigma$ parameter characterizes the quality of the MAROC measurement channel.

The six independent measurements in different conditions were used to verify the self-consistency of the results, using the model approximation features allowing the $scale$ parameter to be measured at various light conditions, ideally providing the same value, and similarly allowing the $\mu$ parameter (and hence the quantum efficiency) to be measured at various high voltages, also providing the same value. These features may be found in each of the ``MaPMT passports'', and they are also further illustrated in the following figures. \begin{figure}[h!]
	\centering
	\includegraphics[width=0.98\linewidth,trim=0 12 50 35,clip]{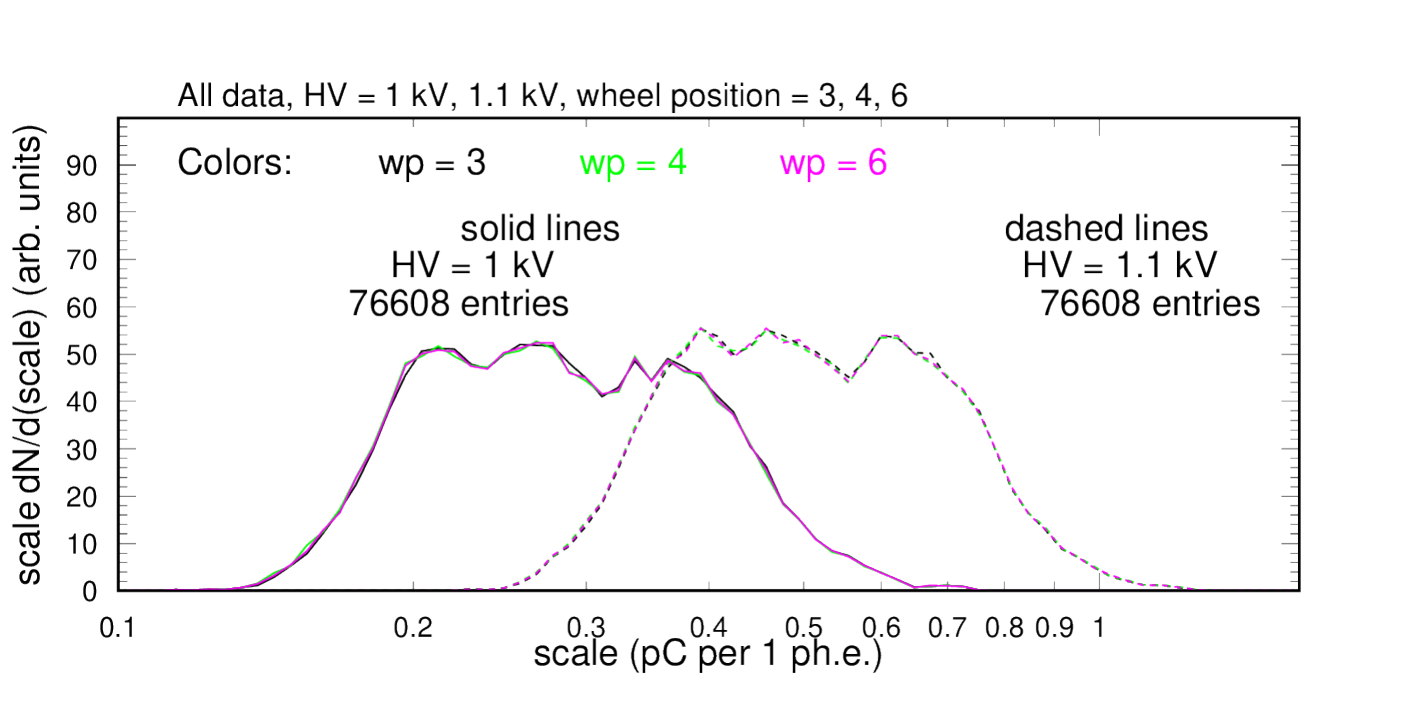}
	\caption{Distribution of $scale$ (average charge per photoelectron) as determined by the fitting procedure for a set of 399 PMTs. All measured pixels contributed to the plots. Distributions measured at HV = 1000~V are shown by the solid lines, and those at HV = 1100~V by the dashed lines. The three colors correspond to the three different illuminations (essentially on top of each other).
	}
	\label{fig:pglobal_sc}
\end{figure}
Figure~\ref{fig:pglobal_sc} shows the distribution of the $scale$ parameter for the whole data set, separately for different high voltages and illumination settings. The distributions are clearly identical if obtained in different illuminations, and the change in high voltage is seen as an approximate multiplication of the $scale$ parameter by a factor about 2 when switching from 1000~V to 1100~V. Logarithmic $x$ scale in the plot helps to see the multiplication as a shift on the plot, roughly preserving the shape of the distribution. 

\begin{figure}[h!]
	\centering
	\includegraphics[width=0.98\linewidth, trim=0 12 50 35, clip]{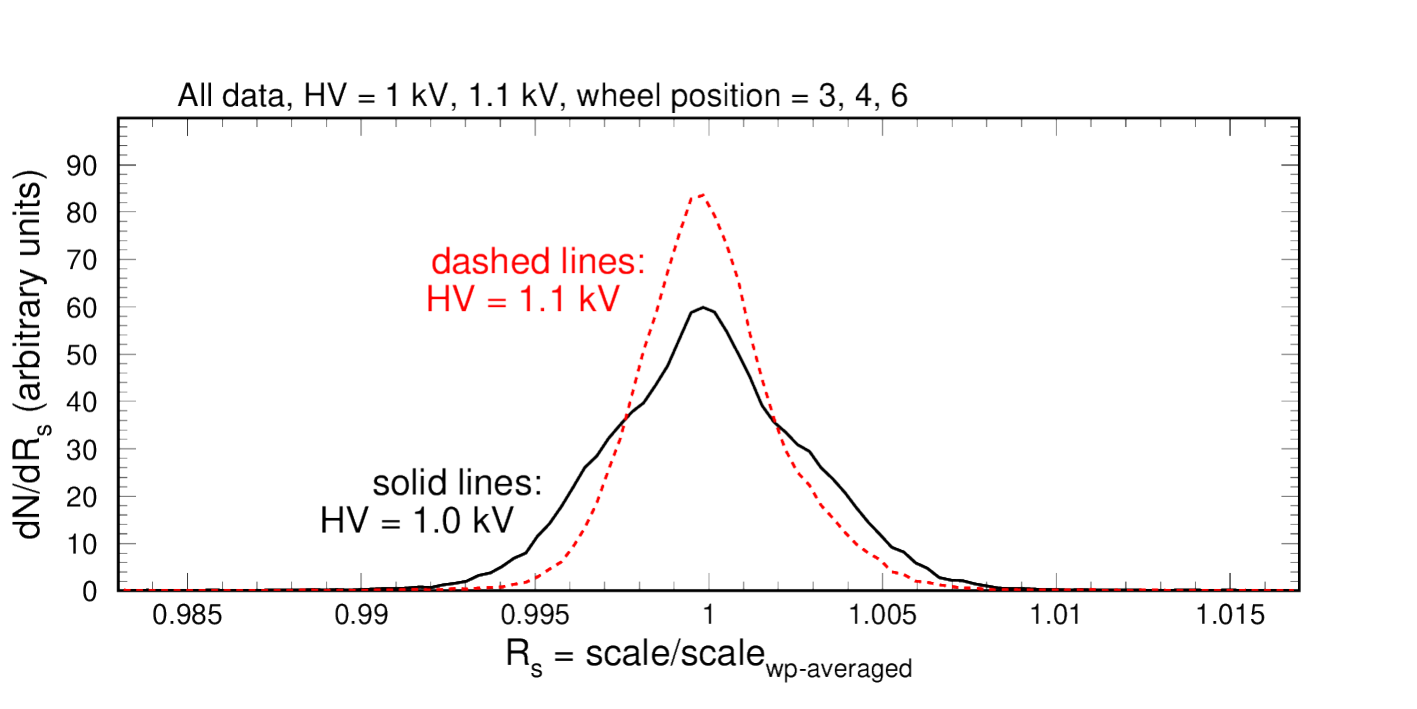}
	\caption{Parameter $scale$ normalized to its average value over the three different illumination settings (wheel positions 3, 4, and 6).}
	\label{fig:pglobal_Rs}
\end{figure}
The stability and consistency of the fitting procedure is illustrated in Fig. \ref{fig:pglobal_Rs} in which every measured $scale$ parameter is normalized to the value of $scale$ averaged over the three measurements on the same pixel at the three different illuminations. The value of the ratio $R_{\mathrm{s}}$ serves as an estimate of the statistical uncertainty of the $scale$ evaluation procedure, and is approximately within 0.75\% for the tests at 1000~V, and within 0.5\% at 1100~V 

\begin{figure}[h!]
	\centering
	\includegraphics[width=0.98\linewidth, trim=0 10 15 10, clip]{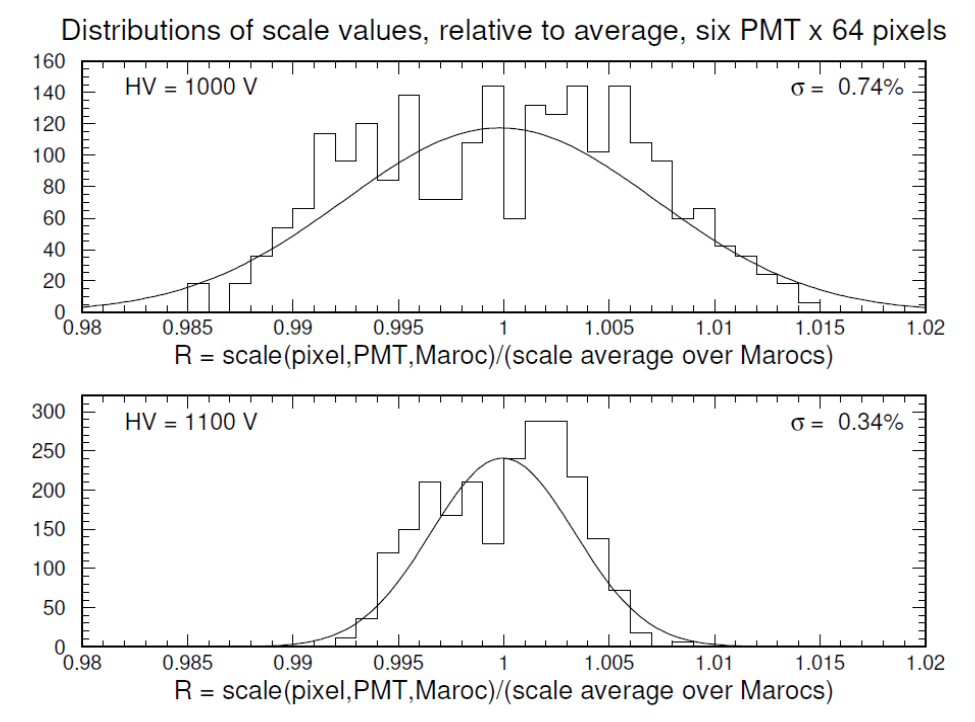}
	\caption{Evaluated precision of the scale parameter measurement for the two high voltage settings.}
	\label{fig:R_scale_maroc_avg}
\end{figure}

In the bulk measurements, one MaPMT was measured in one MAROC location. To be confident that different MAROC locations do not systematically contribute to the differences between the MaPMTs, we compared all six locations by making the standard sets of measurements using six MaPMTs in six runs in which every MaPMT occupied each of the six MAROC positions in turn, and compared the extracted parameters for every pixel made six times in the different locations. One of the results of such a comparison is shown in Fig.~\ref{fig:R_scale_maroc_avg}. 
The histograms show the distributions of the ratios of the measured $scale$ parameter to the average of its values measured in the six MAROC locations. The spreads observed are different for the runs at 1000~V and at 1100~V, and the values are comparable to the spreads observed in Fig.~\ref{fig:pglobal_Rs}. Thus we conclude that switching the location of the MaPMT in the test setup did not cause significant systematic uncertainties in the measured parameters. Similar studies were performed for the other extracted parameters. The observed stability of the extracted quantum efficiencies during these tests, and also comparisons of measurements of quantum efficiency on the same MaPMT made few months apart, indicated to the short- and long-term stability of the laser light source yield at a very good level within the range of statistical errors in the evaluated $\mu$ parameter.

\begin{figure}[h!]
	\centering
	\includegraphics[width=0.98\linewidth,trim=0 12 50 35,clip]{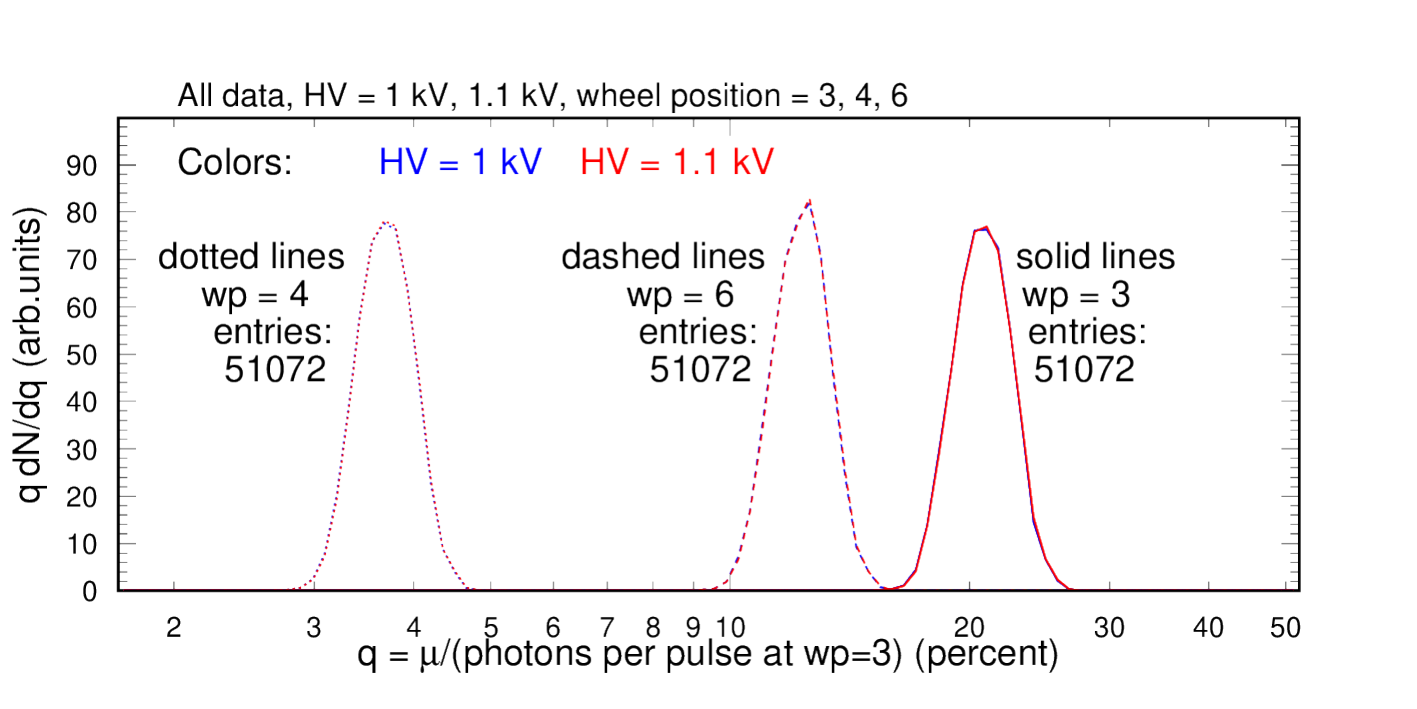}
	\caption{Distribution of $\mu$ in all wheel positions divided by the calibrated number of photons per pulse at wheel position 3. All measured pixels contributed to the plots. Distributions measured at HV = 1000 V are shown in blue, the ones at HV = 1100 V in red, practically indistinguishable in the plot. The three line styles (dotted, dashed, and solid) correspond to different illuminations. For the data collected at wheel position 3, this ratio is the quantum efficiency of the individual pixels.}
	\label{fig:pglobal_qe_all}
\end{figure}
Figure~\ref{fig:pglobal_qe_all} shows a pattern similar to Fig.~\ref{fig:pglobal_sc} for the $\mu$ parameter, with the difference that $\mu$ essentially does not depend on high voltage, but it is proportional to the light intensity. The plot shows that the distributions at different high voltages are on top of each other at a given light intensity but shift in log scale when the light intensity changes. In the plot, the parameter $\mu$ is shown normalized to the number of photons coming to each pixel in the ``wheel position 3'' setting, to provide the associated value of quantum efficiency. The overall averaged quantum efficiency measured in this work at the wavelength of 470 nm is close to the values given in the manufacturer's specifications for the H12700 MaPMTs \cite{H12700}. The average value of QE for all measured pixels is slightly above 20\%, with the pixel-to-pixel spread of about 30\%, to be compared with the average QE number quoted by Hamamatsu at about 21\%. 

\begin{figure}[h!]
	\centering
	\includegraphics[width=0.98\linewidth, trim=0 12 50 35, clip]{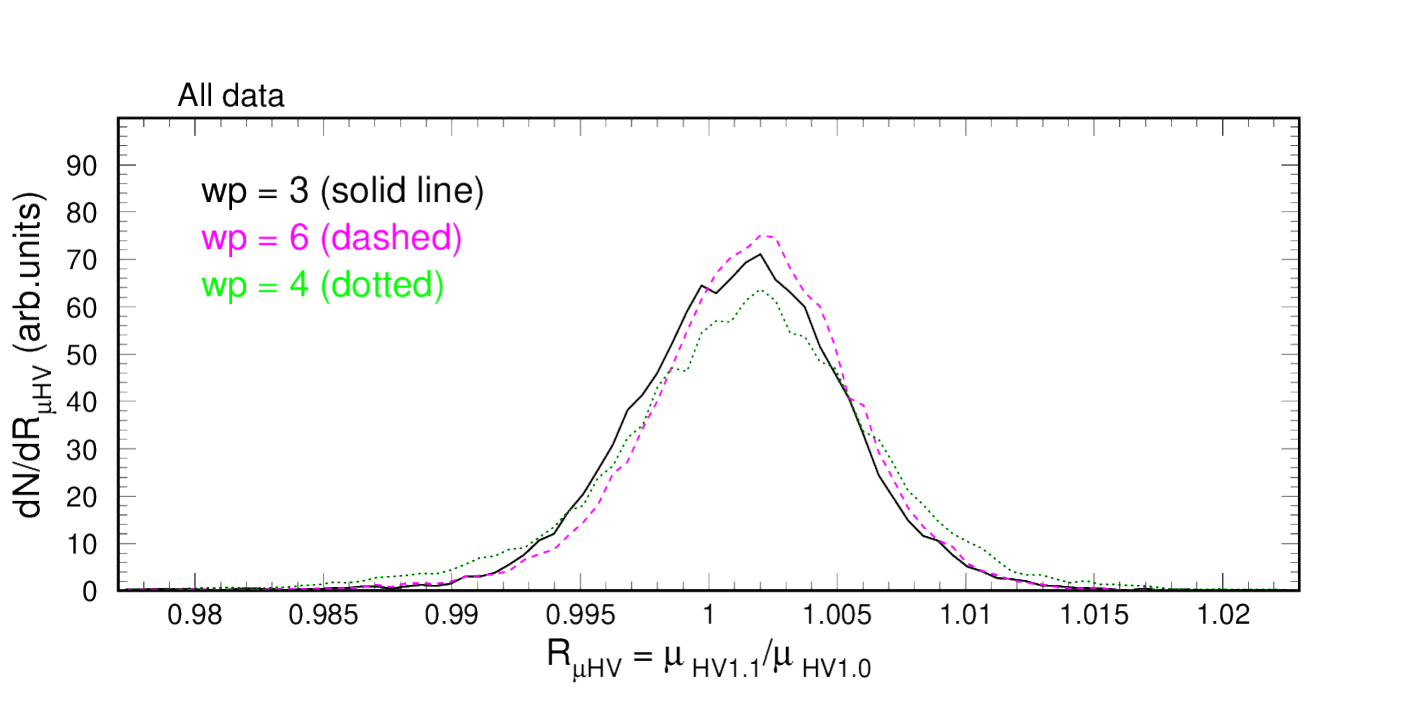}
	\caption{The ratio of the $\mu$ parameters from the fit results at HV = 1100 V to the results at HV = 1000 V.}
	\label{fig:pglobal_mHV}
\end{figure}
Figure~\ref{fig:pglobal_mHV} illustrates the stability of the evaluated $\mu$ parameter measured at different values of high voltage. As we had only two settings, the plot shows the distributions of the ratios $R_{\mu\mathrm{HV}} = \mu_{\mathrm{HV1.1}}/\mu_{\mathrm{HV1.0}}$ of the values of $\mu$ measured at 1100~V to the values at 1000~V. The width of the distribution around $R = 1$ may characterize the statistical uncertainty in the measurement of $\mu$. The plot shows that the relative $\mu$ spread is approximately within 1\% of the value. In first approximation, the quantum efficiency is not expected to be dependent on the high voltage applied to a MaPMT. However, the distributions show slight systematic shifts in the ratio, indicating a small dependence of quantum efficiency on the high voltage applied, with a slope of about 0.2\% per 100 V change. Practically the change is insignificant and within the statistical uncertainties, however, there might be some attempts to explain it assuming, for example, that the larger electric field at the cathode region may improve the probability of photoelectron knock out, or improve the collection probability of the photoelectrons at the first dynodes.

\begin{figure}[h!]
	\centering
	\includegraphics[width=0.98\linewidth, trim=0 12 50 35, clip]{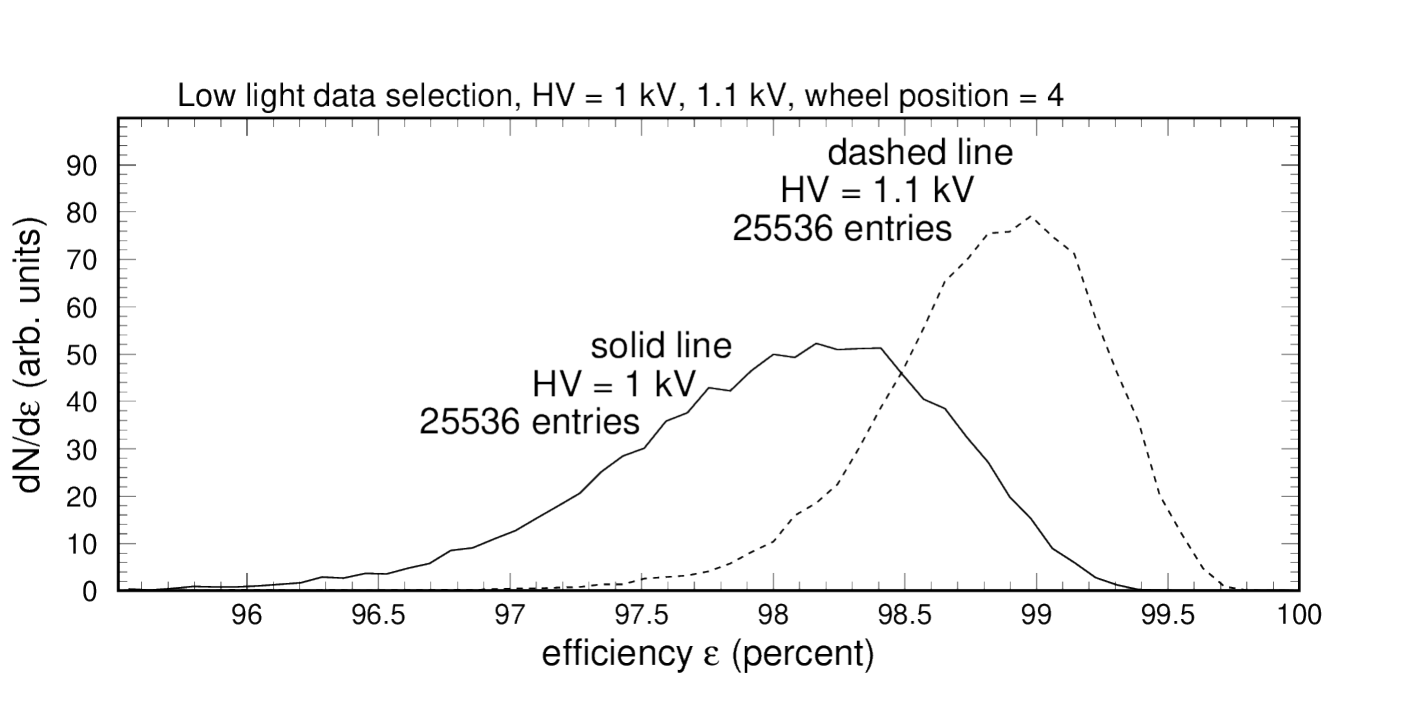}
	\caption{Distribution of the measured efficiency for all pixels at wheel position 4.}
	\label{fig:pglobal_eff}
\end{figure}
Figure~\ref{fig:pglobal_eff} shows the estimated values of the photodetection efficiency based on the calculated optimal threshold value for the separation of the single photoelectron events from pedestal (including the crosstalk background). The calculation for every pixel was performed for the measurements at the lowest illumination settings at wheel position 4, when both parameters $\mu$ and $\beta$ are small and the probability of having two crosstalk electrons in one event was negligible. Such a condition imitates the real operations of the MaPMTs in the RICH detector in the best way, as the number of photons from one relativistic particle is expected to be small. The figure also illustrates the generally very high (above 96\%) single photon efficiency of all tested H12700 MaPMTs at the planned operational high voltage value of 1000~V. The efficiency is improved significantly at 1100~V, with the value of inefficiency decreasing by approximately a factor of 2 in these conditions.

\begin{figure}[h!]
	\centering
	\includegraphics[width=0.98\linewidth, trim=0 12 50 35,clip]{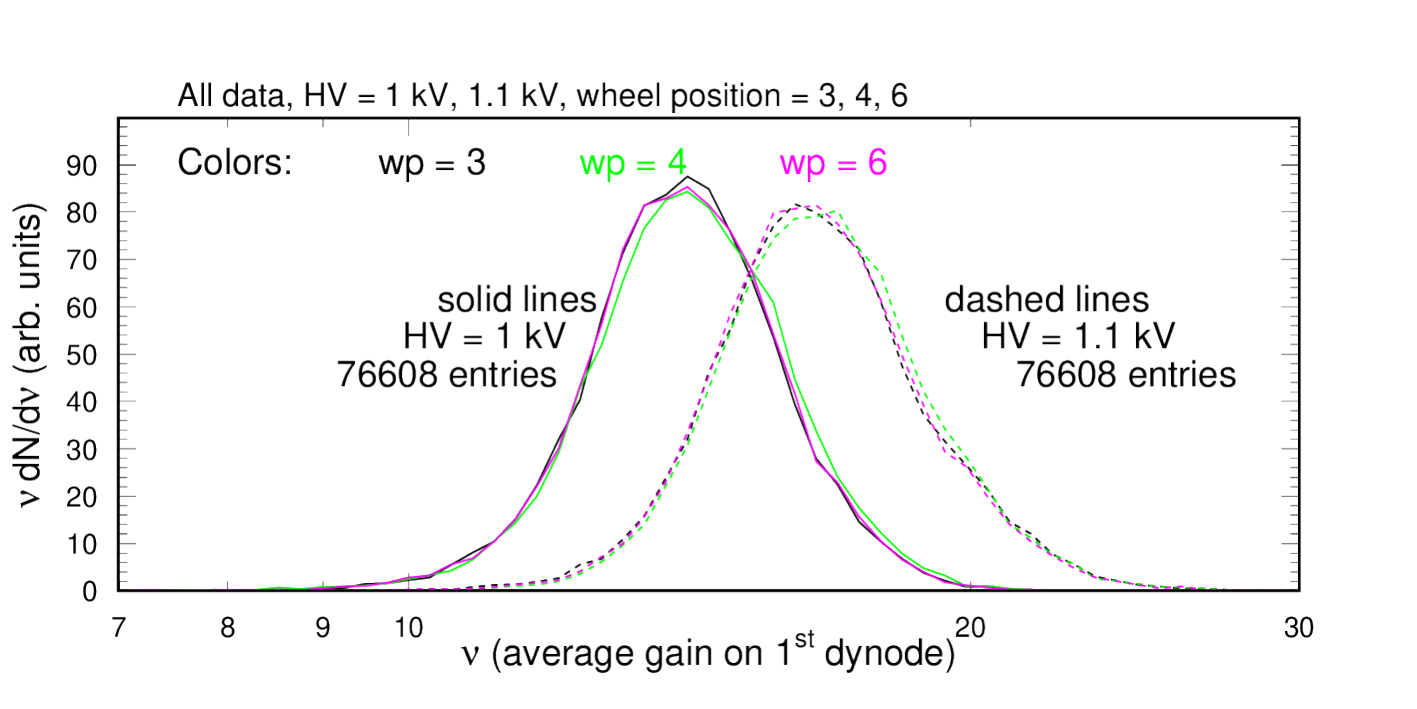}
	\caption{Distribution of $\nu$ (average gain on first dynode) as determined by the fitting procedure for a set of 399 PMTs.}
	\label{fig:pglobal_nu}
\end{figure}
The efficiency improvements at larger high voltage are correlated with the observed increases of the average degree of multiplication of the photoelectrons on the first dynodes of the MaPMTs. The average gain $\nu$ is evaluated in the model using the five parameters describing the shapes of the SPE amplitude distributions. The average gain $\nu$ is clearly dependent on the energy acquired by the photoelectron traveling from the photocathode to the first dynode. The spread in this parameter over the whole data set is noticeable, but the systematic increase at 1100~V is quite prominent, as shown in Fig.~\ref{fig:pglobal_nu}. This figure further illustrates the consistency and stability of the fitting procedure as the distributions built for different illuminating conditions are very close to each other.

\begin{figure}[h!]
	\centering
	\includegraphics[width=0.98\linewidth, trim=0 12 50 35, clip]{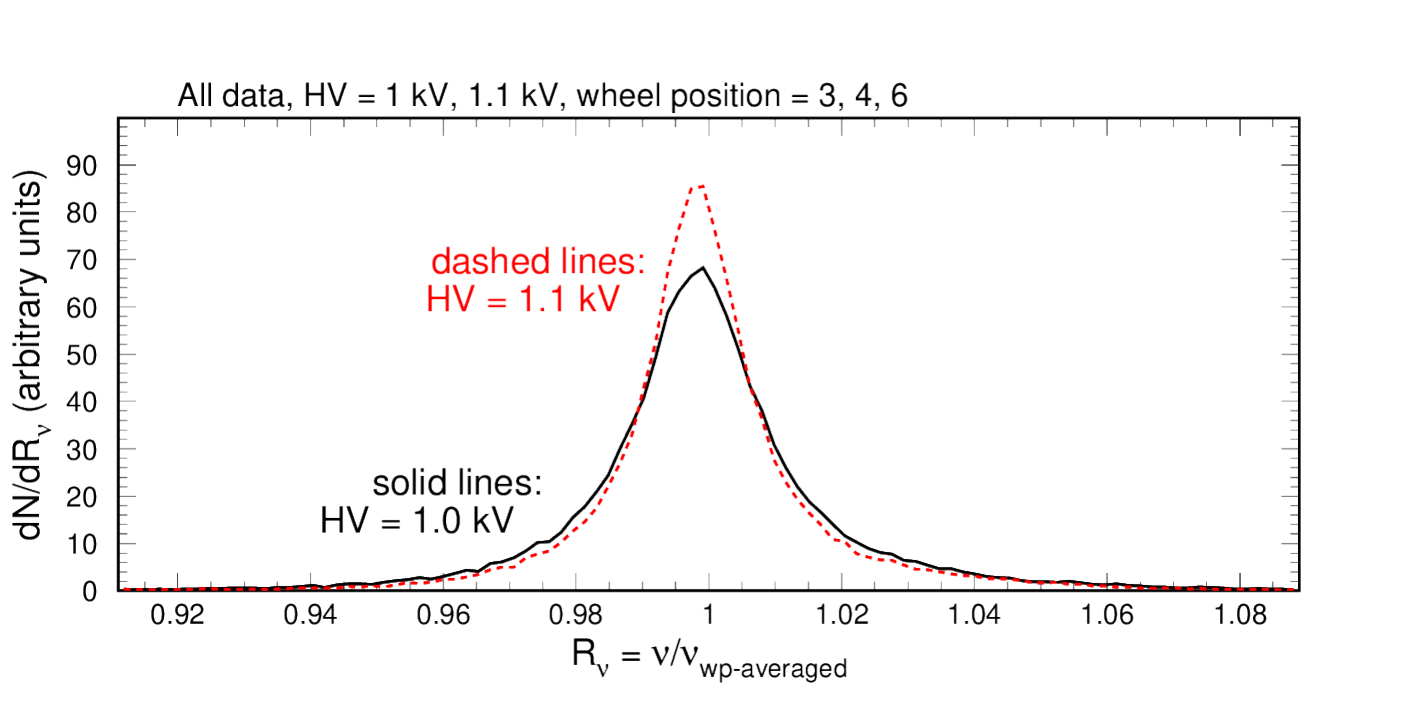}
	\caption{Parameter $\nu$ normalized to its average value over the three different illumination settings (wheel positions 3, 4, and 6).}
	\label{fig:pglobal_Rn}
\end{figure}
Figure~\ref{fig:pglobal_Rn} is similar to Fig.~\ref{fig:pglobal_Rs}, showing the measured $\nu$ parameters normalized to the value of $\nu$ averaged over the three measurements on the same pixel at the three different illuminations. The value of the ratio $R_{\nu}$ serves as an estimate of the statistical uncertainty of the $\nu$ evaluation procedure, and is approximately within 5\%. The distribution is visibly non-Gaussian as $\nu$ is a complicated function of five variable signal shape parameters in the fit. There is a small difference between the distributions at different high voltage settings.

\begin{figure*}[t!]
	\centering
	\begin{subfigure}[c]{0.48\linewidth}
		\centering
		\includegraphics[width=\linewidth]{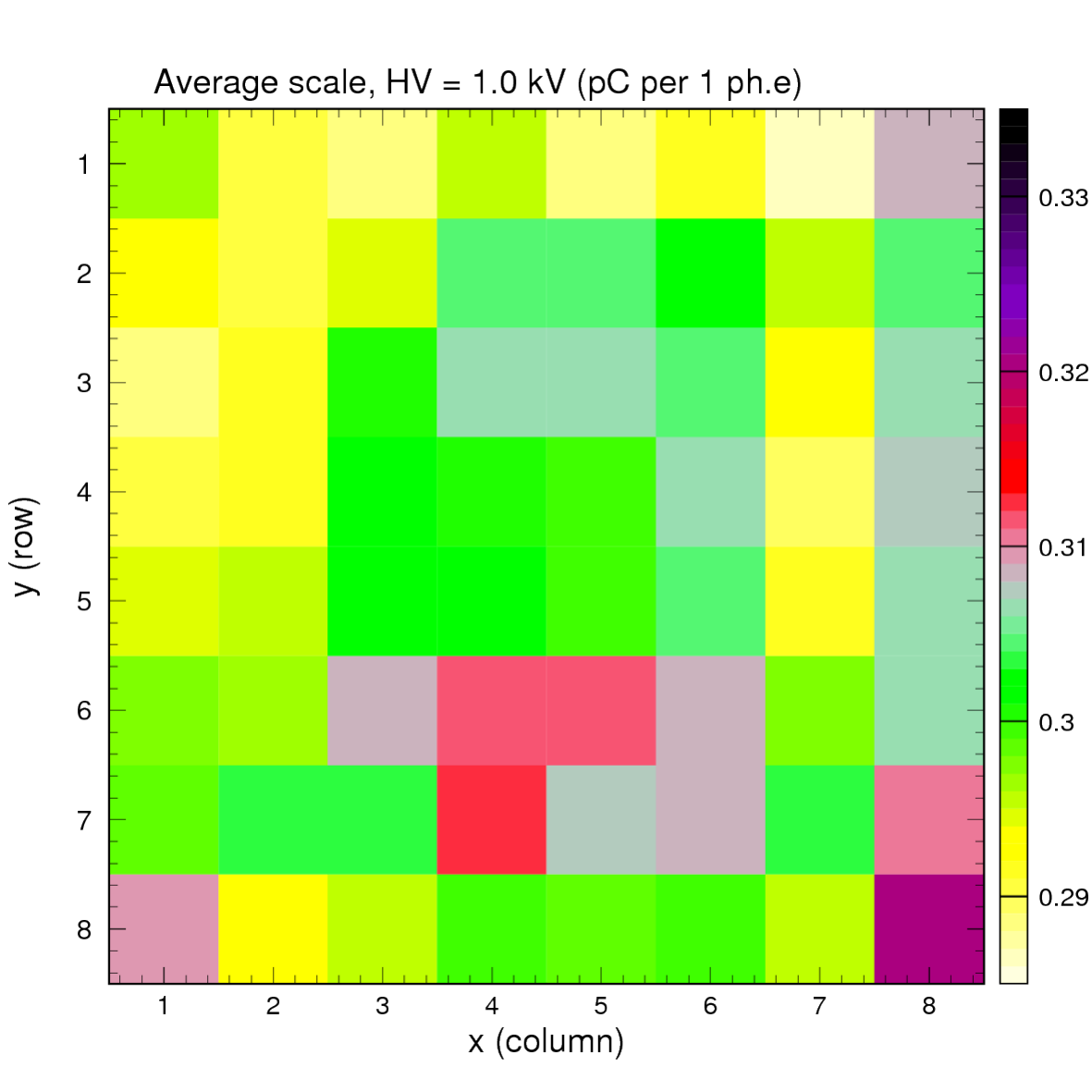}
		\caption{Scale, HV = 1.0 kV (pC per 1 photoelectron)}
		\vspace{0mm}
	\end{subfigure}
	\begin{subfigure}[c]{0.48\linewidth}
		\centering
		\includegraphics[width=\linewidth]{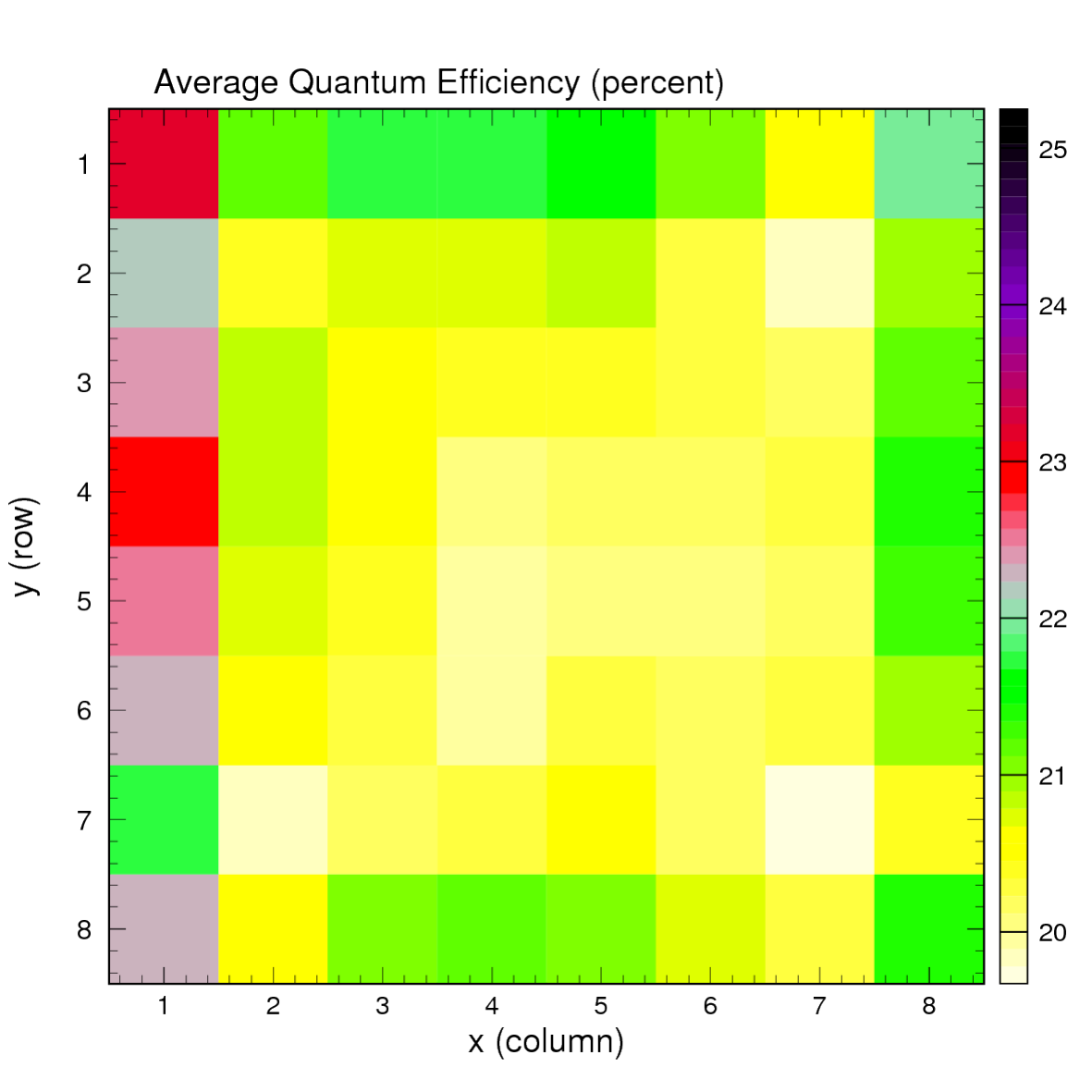}
		\caption{Quantum Efficiency (percent)}
		\vspace{0mm}
	\end{subfigure}
	\vspace{3mm}
	\begin{subfigure}[c]{0.48\linewidth}
		\centering
		\includegraphics[width=\linewidth]{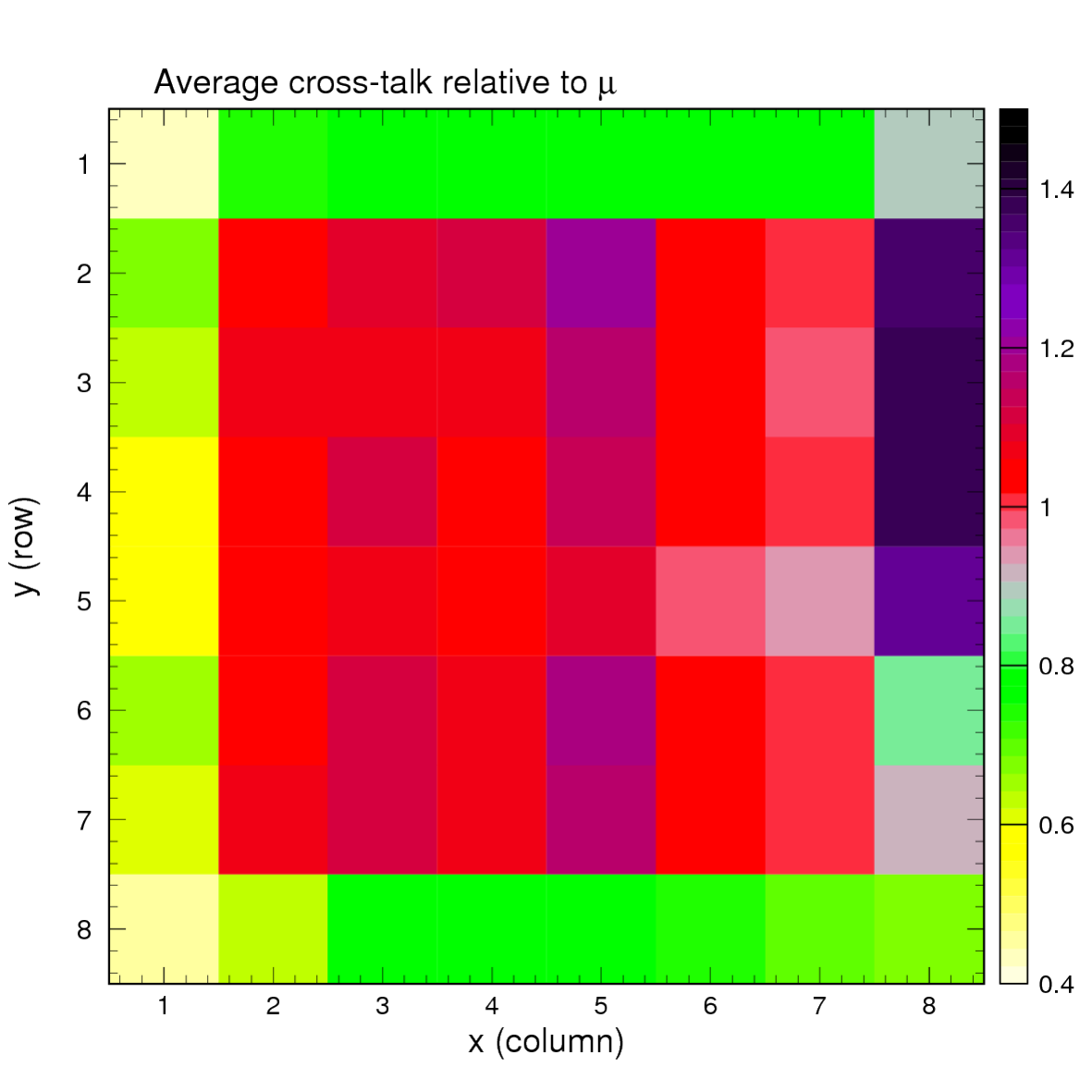}
		\caption{Crosstalk relative to $\mu$}
		\vspace{0mm}
	\end{subfigure}
	\begin{subfigure}[c]{0.48\linewidth}
		\vspace{3mm}
		\centering
		\includegraphics[width=\linewidth]{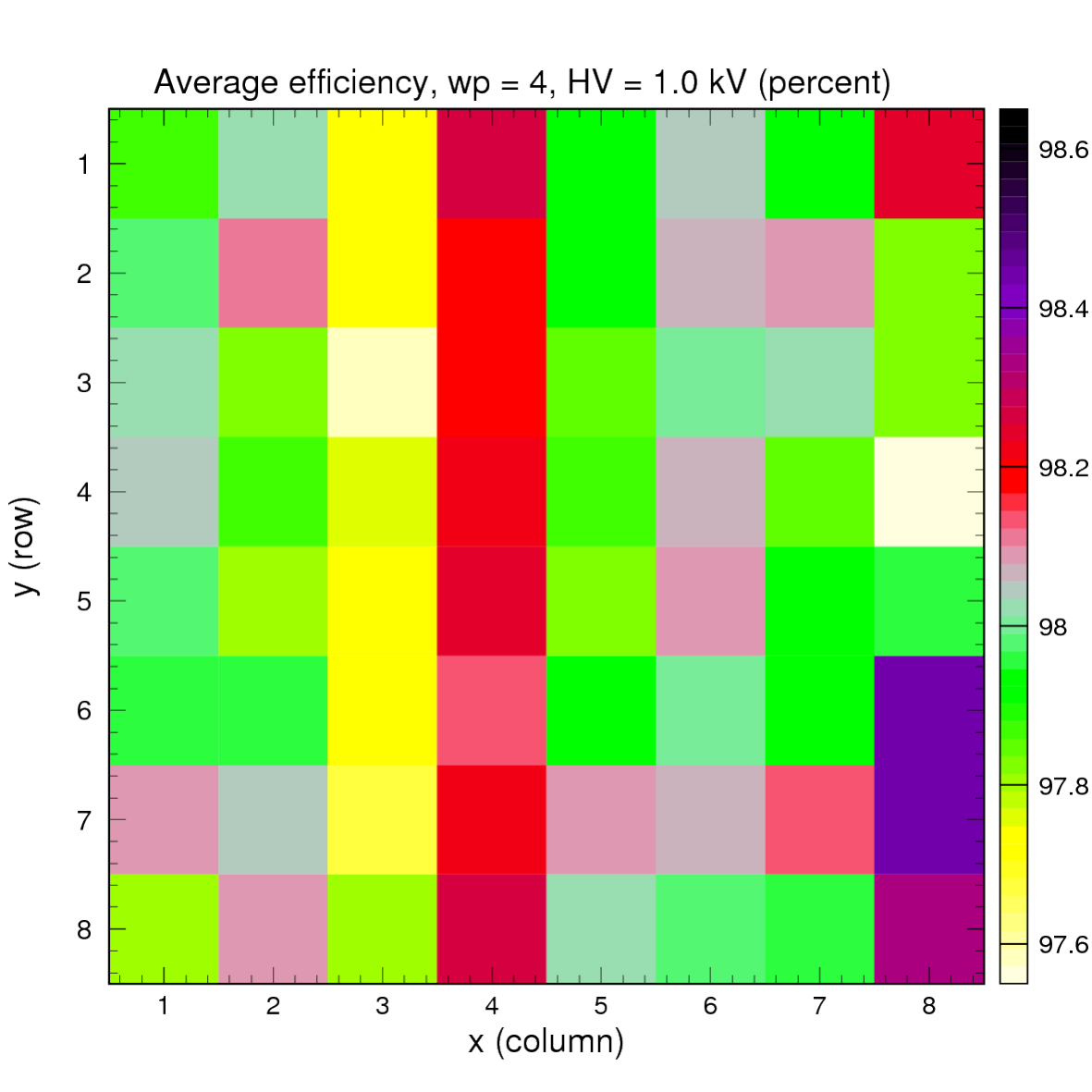}
		\caption{Efficiency, wheel position 4, HV = 1.0 kV (percent)}
		\vspace{0mm}
	\end{subfigure}
	\caption{Two dimensional plots showing the average (a) scale, (b) quantum efficiency, (c) crosstalk relative to $\mu$, and (d) efficiency as a function of pixel location. The results are averaged for the full set of 399 Hamamatsu H12700 MaPMTs. The pixel numbers increment from left to right, top to bottom, with pixel \#1 in the top left corner.}
	\label{fig:2d_avg_fit_results}
\end{figure*}

Figure~\ref{fig:2d_avg_fit_results} illustrates the dependencies of several major parameters on the pixel number for the full set of MaPMTs studied, including the average amplitude of the single photon amplitude scale, quantum efficiency, the relative probability of the crosstalk events $\beta/\mu$, and the evaluated efficiency. Generally, the set exhibits a very good uniformity of the average parameters, much smaller than the spreads observed between pixels in a single MaPMT or between the tubes. The Quantum Efficiency is slightly higher at the edges of the MaPMT and still higher at the corners (larger areas of the border pixels are taken into account in the QE calculation). The crosstalk probability pattern is consistent with the hypothesis that it is dependent on the number of neighbors: it is smaller at the edges, and still smaller in the corners of the MaPMT. The four outliers in pixels 16, 24, 32, and 40 are most likely due to the feature of all MAROC boards used, exhibiting significantly wider pedestals in these pixels, hiding the crosstalk under the pedestal Gaussian and causing the fitting procedure to fail to fit the crosstalk properly. The average efficiency pattern shows somewhat better values in columns 4 and 8 (with the exception of the same four outliers), likely correlated with the widths of the crosstalk contributions and the parameters of the average gain on the first dynode $\nu$.

The parameter database accumulated as the result of this work was used for the selection of the MaPMTs for installation in the RICH detector, and for the optimization of the future run parameters, such as the tube placement selection, as well as setting the values of operating high voltage, electronics gains, and thresholds in the detector.

The data also provide the opportunity to evaluate the spread of such parameters in the mass production of the MaPMT devices as the channel gains, quantum efficiencies, SPE spectral shapes, and parameters of the crosstalk, - across the face of each tube, and across the whole set. The results show that the quality of MaPMT mass production at Hamamatsu is high and satisfies our needs in good quality single photoelectron detection.

\section{Conclusion}

As a part of CLAS12 RICH detector upgrade at Jefferson Lab, we have conducted a mass study of 399 H12700 MaPMTs from Hamamatsu, with the goal to evaluate every tube and characterize every pixel in terms of their gain, quantum efficiency, crosstalk contribution, and optimized threshold for detecting single Cherenkov photons. The dedicated test setup included a precision picosecond laser, 
gears for the  positioning of the laser beam in the setup,
RICH detector front-end electronics, and fully automated data acquisition and control systems. The non-linearity of the data acquisition, the ADC-to-charge conversion calibration parameters of every channel, and the absolute calibration of the number of laser photons reaching every pixel in every event were measured in special separate experiments. The bulk measurements consisted of six expositions of every group of six MaPMTs at three levels of low light and two applied high voltages, 1000~V, and 1100~V. The systematic uncertainties dependent on the MaPMT placement in the group of six were evaluated and found to be within the final parameter uncertainties.

In a set of dedicated detailed studies we observed and quantified the pixel-to-pixel signal crosstalk using a two-dimensional amplitude distribution analysis. Using several representative MaPMTs of both types we found that the H8500 model is characterized by quite significant amplitude spectral contributions to a given pixel from its neighbors in the matrix, with such crosstalk contributions reaching up to 50\% of the spectral amplitude. At the same time, the crosstalk in H12700 MaPMTs was generally less than about 3-5\%. Methods of separating and taking into account the crosstalk contributions to the amplitude distributions from any pixel were developed, using the two-dimensional analysis, and also approximating and evaluating the contributions based on the spectral shape using the computational model. The first approach is applicable to all MaPMTs studied, but it is labor intensive and works correctly only in the conditions of extremely low light in the tests. The second approach works well for the H12700 MaPMTs and was used for the bulk measurements.

The accumulated amplitude spectra were corrected to the non-linearity of the data acquisition and converted to the calibrated total charge distributions. The recently published state-of-the-art computational model, describing photon detector response functions measured in conditions of low light, was extended to include the successful description of the crosstalk contributions to the spectra from the neighboring pixels. The updated model was used to parameterize and extract the SPE response functions of every pixel, and characterize its properties such as gain, quantum efficiency, and crosstalk, and to determine the optimal signal threshold values to evaluate its efficiency to Cherenkov photons. The stability and reproducibility of the extracted parameter values were verified by the comparison of the six independent measurements of each pixel, allowing us to evaluate the uncertainties in the measurements of the major model parameters. One of the extracted parameters, the average multiplication of a photoelectron on the first dynode $\nu$ was found significantly larger on the H12700 compared to the H8500 MaPMTs. That difference corresponds to the resulting difference between the SPE efficiency of the two models.  That observation, together with much smaller crosstalk contributions, generally confirms our early decision to switch to the H12700 as the MaPMT of choice for the RICH detector.

The database of extracted parameters has been used for the final selection and arrangement of the MaPMTs in the new RICH detector, and for determining their optimal operation parameters, such as operating high voltage, gain, and threshold of the front-end electronics. A good model description of the measured amplitude distributions from MaPMT pixels, including the crosstalks, will allow using the parameterization in the Monte Carlo simulations of the detector. The results show that the quality of the H12700 MaPMT mass production at Hamamatsu is high, satisfying our needs in the good position-sensitive single photoelectron detectors.

\section*{Acknowledgements}
This material is based upon work supported by the U.S. Department of Energy, Office of Science, Office of Nuclear Physics under contract DE-AC05-06OR23177, and in part by DOE Grant No. DE-FG02-04ER41309, DOE Grant No. DE-FG02-03ER41231 and NSF Award no: 2012413. 
We sincerely thank 
Bishnu Karki,
Charles Hanretty,
Aiden Boyer,
Chris  Cuevas,
Matteo Turisini, and
Carl Zorn
for their technical support and help in taking and analyzing the experimental data. We would
also like to thank 
Patrizia Rossi, 
Marco Contalbrigo, 
Marco Mirazita,
Kyungseon Joo,
and Zhiwen Zhao
for the fruitful discussions, continuous support, and interest in this work. Special thanks to Danial Carman for proofreading the article and making valuable comments.		  

\newpage
\newpage
\appendix
\section{}
\label{AppendixA}

In the case of the H12700 MAPMTs, the signal amplitudes of the crosstalk contributions from different neighboring pixels were found to be relatively small and similar to each other, allowing us to use in the model a single average spectral term for all neighbors of a given pixel. Each crosstalk contribution comes from a single electron in one of the neighboring pixels, their average number in one measurement $\beta$ is expected to be comparable with $\mu$, and multiple crosstalk events in one measurement happen independently. That means that the probability of observing $i$ crosstalk contributions in one event is distributed according to a Poisson distribution
\begin{equation}
\label{CTPoisson}
 P(i;\beta) = \frac{\beta^{i} e^{-\beta}}{i!}.
\end{equation}
Poisson-like shapes of the general SPE distribution functions suggest a shape of the crosstalk contribution in the form of a Poisson distribution, scaled to represent the portion of the charge generated in the neighboring pixel, transferred to the pixel studied. The representation of such a distribution for one crosstalk electron takes the form
\begin{equation}
\label{C1Poisson}
 C_1(j) = P(j;\lambda) = \frac{\lambda^{j} e^{-\lambda}}{j!},
\end{equation}
where $j$ is a non-negative integer, corresponding to the amplitude values $a_j = j \zeta / \lambda $, relating the discrete Poisson scale to the set of $a$ values, such that the average crosstalk contribution to the measurement function from one crosstalk electron was equal to the value of the $\zeta$ parameter (the average $\langle j \rangle$ in Eq.~(\ref{C1Poisson}) equals to $\lambda$).

The corresponding distributions for the events with $i$ crosstalk electrons then take the form of convolution powers, which can be explicitly calculated in the case of Poisson distributions: 
\begin{equation}
\label{CiPoisson}
 C_i(j) = C_1^{*i}(j) = P(j;i\lambda).
\end{equation}

Thus, similar to Eq.~(13) in Ref.~\cite{DEGTIARENKO20171}, the discrete distribution can be represented as a function of the normalized amplitude $a$ in the form of the infinite sum of correspondingly weighted delta-functions, one per each value of $j \geq 0$:
\begin{equation}
\label{deltaf}
  D_{ct}(a)= \sum\limits_{j=0}^{\infty}
  \delta \left (a - \frac{j\zeta}{\lambda} \right )
   \sum\limits_{i=0}^{\infty} P(i;\beta) C_i(j) .
\end{equation}
The convolution of this distribution with the Gaussian measurement function (sigma equal to $\sigma_a$) will result in a continuous function similar to Eq.~(15) in Ref.~\cite{DEGTIARENKO20171}:
\begin{equation}
\label{RCmodel}
  R_{ct}(a)= \sum\limits_{j=0}^{\infty} \frac{1}{\sqrt{2 \pi} \ \sigma_a} 
  \exp{\left [- \frac{(a - j \zeta / \lambda)^{2}}{2 
\ \sigma_a^{2}} \right ]}
   \sum\limits_{i=0}^{\infty} P(i;\beta) C_i(j) .
\end{equation}

The new function $R_{ct}(a)$, parametrically dependent on $\sigma_a$, $\beta$, $\zeta$, $\lambda$, describes the effective measurement function applied to every signal. The recorded signals are the results of the convolution with this function. In particular, in the events with no photoelectrons ($m=0$), the pedestal distribution takes the form of $R_{ct}(a)$. For a given set of parameters the function $R_{ct}(a)$ is evaluated numerically in the model implementation and then used in the calculations as described in Ref.~\cite{DEGTIARENKO20171}, by replacing the measurement function $R(a)$ with $R_{ct}(a)$ in convolution with the $D(a)$ function in Eq.~(14) in Ref.~\cite{DEGTIARENKO20171}. The function $D(a)$ as defined in Eq.~(13), Ref.~\cite{DEGTIARENKO20171}, much like the function $D_{ct}(a)$ in Eq.~(\ref{deltaf}) in this work, represents an infinite set of delta-functions, and the convolution calculation just needs the values of the tabulated function $R_{ct}(a)$ in all the final sums. The new equivalent for Eq.~(16) in Ref.~\cite{DEGTIARENKO20171} is thus
\begin{equation}
\label{GCmodel}
  G_{ct}(a,n;\sigma_a,\beta,\zeta,\lambda)= R_{ct}(a-n/\nu;\sigma_a,\beta,\zeta,\lambda).
\end{equation}
The new function $G_{ct}(a,n;\sigma_{\mathrm{eff}},\beta,\zeta,\lambda)$ is then used to replace the function $G(a,n;\sigma_{\mathrm{eff}})$ in the final model equation, Eq.~(36) in Ref.~\cite{DEGTIARENKO20171}, keeping the same form. The change is that instead of being a standard Gaussian, the measurement function is now distorted by the crosstalk contribution, requiring three extra parameters to approximate the data.

\end{document}